\documentclass[aps,pra,twocolumn,reprint,amsmath,amssymb,nofootinbib,superscriptaddress,floatfix,eqsecnum]{revtex4-1}
\usepackage{graphicx}
\usepackage{amssymb,bbm,amsmath,braket,indentfirst,bm}
\allowdisplaybreaks 
\usepackage[pdftex]{color} 
\usepackage[pdftex,colorlinks,citecolor=blue,linkcolor=blue,urlcolor=blue]{hyperref}
\usepackage{color}
\usepackage{ulem} 
\usepackage[paperwidth=8.5in,paperheight=11in,centering,hmargin=2cm,vmargin=2.5cm]{geometry} 

\newcommand{\pri}[1]{\ensuremath #1^{\prime}}

\renewcommand\({\ensuremath \left(} \renewcommand\){\ensuremath
  \right)} \renewcommand\[{\ensuremath \left[}
  \renewcommand\]{\ensuremath \right]}
\def\:={\,\raisebox{0.85pt}{.}\hspace{-2.78pt}\raisebox{2.85pt}{.}\!\!=\,}
\def\=:{\,=\!\!\raisebox{0.85pt}{.}\hspace{-2.78pt}\raisebox{2.85pt}{.}\,}

\newcommand{\T}{\mathsf{T}}

\begin{document}

\title{Wire constructions of Abelian topological phases 
in three or more dimensions}

\author{Thomas~Iadecola} \affiliation{Physics Department, Boston
  University, Boston, Massachusetts 02215, USA}

\author{Titus~Neupert} \affiliation{Princeton Center for Theoretical
  Science, Princeton University, Princeton, New Jersey 08544, USA}

\author{Claudio~Chamon} \affiliation{Physics Department, Boston
  University, Boston, Massachusetts 02215, USA}

\author{Christopher~Mudry} \affiliation{Condensed Matter Theory Group,
  Paul Scherrer Institute, CH-5232 Villigen PSI, Switzerland}

\date{\today}

\begin{abstract}
Coupled-wire constructions have proven to be useful tools to
characterize Abelian and non-Abelian topological states of matter in
two spatial dimensions.  In many cases, their success has been
complemented by the vast arsenal of other theoretical tools available
to study such systems.  In three dimensions, however, much less is
known about topological phases.  Since the theoretical arsenal in this
case is smaller, it stands to reason that wire constructions, which
are based on one-dimensional physics, could play a useful role
in developing a greater microscopic understanding of three-dimensional
topological phases.  In this paper, we provide a comprehensive
strategy, based on the geometric arrangement of commuting projectors
in the toric code, to generate and characterize coupled-wire
realizations of strongly-interacting three-dimensional topological phases.  
We show how this method can be used to construct pointlike and linelike
excitations, and to determine the topological degeneracy.  We also point
out how, with minor modifications, the machinery already developed in
two dimensions can be naturally applied to study the surface states of
these systems, a fact that has implications for the study of surface topological
order. Finally, we show that the strategy developed for the construction of
three-dimensional topological phases generalizes readily to arbitrary
dimensions, vastly expanding the existing landscape of coupled-wire theories.
Throughout the paper, we discuss $\mathbb{Z}^{\,}_m$ topological
order in three and four dimensions as a concrete example of this approach, but
the approach itself is not limited to this type of topological order.
\end{abstract}

\maketitle

\tableofcontents

\section{Introduction}
\label{sec: Introduction}

The experimental discovery of the integer and fractional quantum Hall
effects excited enormous interest in the study of topological states
of matter in two dimensional space.  Strongly interacting states of
matter distinguished by the presence of excitations with fractional
quantum numbers or nontrivial boundary modes have attracted particular
attention from theorists.  Over time, a vast arsenal of theoretical
tools has been developed to study such systems, from the microscopic
(e.g., numerical techniques to study lattice models with topologically
ordered ground states) to the macroscopic (e.g., topological quantum
field theories).

Wire constructions, which were first undertaken for the integer%
~\cite{Poilblanc87,Yakovenko91,Lee94}, and later the fractional%
~\cite{Kane02,Teo14,Mong14,Klinovaja14a,Meng14a,Klinovaja14b,Sagi14,Neupert14,Meng14b,Klinovaja15,Santos15,Sagi15a},
quantum Hall effect, are conveniently poised midway between these two
extremes.  The approach in this case is to model a topological phase
by starting from an anisotropic theory of decoupled gapless quantum
wires, and then introducing local couplings between the wires to
produce a gapped state of matter with an isotropic low-energy
description.  This approach has the virtue of yielding the edge
theory, itself that of a Luttinger liquid, directly, and of providing
means to construct the low-lying quasiparticle excitations of the bulk
quantum liquid.  Furthermore, because wire constructions make use of
well-understood techniques in one-dimensional physics, such as Abelian 
(or non-Abelian) bosonization, one can construct analytically tractable
theories of states of matter that might not otherwise admit a
controlled analytical description.

In recent years, wire constructions have also been used
to study fractional topological insulators
(FTIs)~\cite{Neupert14,Sagi14,Santos15} and spin
liquids~\cite{Meng15a,Gorohovsky15}, and also to develop an
extension~\cite{Neupert14} of the ten-fold way for noninteracting
fermions~\cite{Altland97,Schnyder09,Kitaev09,Ryu10} 
to strongly-correlated systems.

Since the prediction~\cite{Fu07} and
discovery~\cite{Hsieh08,Hsieh09,Chen09} of three-dimensional
$\mathbb{Z}^{\,}_{2}$ topological insulators (TIs), there has been a
growing interest in understanding topological states of matter in
three spatial dimensions. In addition to generalizing these
time-reversal invariant $\mathbb{Z}^{\,}_{2}$ topological insulator to
the strongly-interacting regime%
~\cite{Maciejko10,Swingle11,Maciejko14}, there has been an effort to
derive effective field theories describing the bulk of such TIs, and
to determine the bulk-boundary correspondence in such theories that
yields the hallmark single Dirac cone on the two-dimensional surface%
~\cite{Cho11,Chan13,Tiwari14,Ye15a,Chen15b}.  
Further work has undertaken efforts to understand broader features 
of three-dimensional topological states of matter, 
such as the statistics of pointlike and
linelike excitations~\cite{Wang14,Lin15}.  
For example, it has been
shown that certain three-dimensional topological phases can only be
distinguished by the mutual statistics among three linelike
excitations~\cite{Lin15}.

Another major direction of work concerns three-dimensional systems
whose surfaces are themselves two-dimensional topological states of
matter. The simplest example of this phenomenon occurs on the surface
of a $\mathbb{Z}^{\,}_{2}$ TI when time-reversal symmetry is locally
broken by a magnetic field on the surface, in which case a
half-integer surface quantum Hall effect
develops~\cite{Fu07b,Qi08,Xu14,Yoshimi15}.  
Further theoretical work has shown that generic 
three-dimensional topological phases, including
but not limited to the fermionic $\mathbb{Z}^{\,}_{2}$ TI, 
can exhibit more exotic surface topological phases 
that cannot exist with the same realization of symmetries 
for local Hamiltonians in purely two-dimensional space.  
This family of surface phenomena is known
as surface topological order%
~\cite{Keyserlingk13,Vishwanath13,Wang13a,Wang13b,Burnell14,Chen14,Metlitski15,Mross15}.
Several recent works~\cite{Mross15,Sahoo15} have approached the
question of surface topological order by applying the
quasi-one-dimensional physics of wire constructions, although it
appears that this approach necessitates the use of an unusual
``antiferromagnetic'' time-reversal symmetry rather than the 
usual (physical) realization of reversal of time,
which acts on-site.  It is possible that a fully
three-dimensional wire construction could remedy this peculiarity,
although such a description is still lacking.

Layer constructions, in which planes of two-dimensional topological
liquids are stacked on top of one another and coupled, were used to
construct the single surface Dirac cone of the three-dimensional
$\mathbb{Z}^{\,}_{2}$ TI~\cite{Hosur10} and to study surface
topological order~\cite{Jian14}.  Wire constructions of
three-dimensional topological states of matter have also recently been
undertaken, yielding Weyl semimetals~\cite{Vazifeh13,Meng15b} and a
class of fractional topological insulators~\cite{Sagi15b}.  However,
in all three cases, different methods are used to develop the wire
constructions themselves, and little effort has been made to extend
these constructions beyond the specific problem at hand in each
example.  In order to attack the most distinctive aspects of
topological states of matter in three dimensions, such as surface
topological order, it is therefore necessary to develop a framework
that lends itself readily to a variety of approaches with minimal
modifications.

In this paper, we provide a comprehensive strategy to design wire
constructions of strongly-interacting Abelian topological states of
matter in three dimensions.  The strategy that we present is to start
with decoupled quantum wires placed on the links of a two-dimensional
square lattice, and then to couple the wires with many-body
interactions associated with each star and plaquette of the lattice.
In this way, each interaction term that couples neighboring wires can
be viewed as corresponding to one of the commuting projectors that
enters Kitaev's toric code Hamiltonian~\cite{Kitaev03}.  
This correspondence simplifies the application of a criterion, first
proposed by Haldane, to ensure that these interaction terms do not
compete, and are sufficient in number to gap out all gapless modes in
the array of quantum wires when periodic boundary conditions are
imposed along all three spatial directions.

When all interaction terms satisfy this criterion, the Hamiltonian is
frustration-free, and taking the strong-coupling limit produces a
gapped three-dimensional state of matter.  With this done, one can
proceed to characterize this state of matter in terms of its pointlike
and linelike excitations, as well as their statistics, and calculate
the topological degeneracy, if any, of the ground-state manifold.
The class of three-dimensional models studied in this work
features a topological degeneracy given by 
$|\text{det }\varkappa|^{3}$, 
where the integer-valued matrix $\varkappa$ contains
information about the mutual statistics of pointlike and linelike
excitations in the theory.  This is in close analogy with the
$K$-matrix formalism developed for two-dimensional topological
states of matter~\cite{Wen92}.  When periodic boundary conditions
are relaxed by the presence of two-dimensional terminating surfaces,
we further show that gapless surface states result.  One can apply the
coupled-wire techniques already developed in two dimensions to study
the various gapped surface states that can be produced by introducing
interwire hoppings or interactions on the surface, provided that the
added terms are compatible with the interactions in the bulk.

In addition, we show that the above strategy for constructing
three-dimensional Abelian topological states of matter can be readily
extended to arbitrary dimensions, vastly expanding the existing scope
of the coupled-wire approach.  Indeed, much as it is possible to
define higher-dimensional versions of the toric code on hypercubic
lattices (see, e.g., Ref.~\cite{Mazac12}), one can arrange a set of
decoupled quantum wires on a $d$-dimensional hypercubic lattice and
couple them with interactions defined on stars and plaquettes of this
lattice.  Applying Haldane's compatibility criterion, one can show
that these interactions produce a gapped $(d+1)$-dimensional state of
matter, whose excitations and topological properties can be
investigated much as in the three-dimensional case.

The structure of this paper is as follows.  
In Sec.~\ref{sec: Three-dimensional wire constructions}, we develop
in detail the strategy discussed above for constructing 
three-dimensional topological phases from coupled wires.
In Sec.~\ref{subsec: decoupled wires d=2}, 
we establish the basic notation used to describe the array of
decoupled quantum wires.  
In Sec.~\ref{subsec: 2D coupling discussion}, 
we present Haldane's compatibility
criterion and a class of many-body interactions between wires that satisfy it.  
(This class is mainly chosen for
analytical expedience, and is not the only class of interactions 
that can be constructed according to our strategy.)
In Sec.~\ref{subsec: Fractionalization in the coupled wire array}, 
we show how to use the interacting arrays
of quantum wires defined in Secs.~\ref{subsec: decoupled wires d=2} and 
\ref{subsec: 2D coupling discussion} 
to study states of matter with fractionalized excitations.  
In particular, we show how to construct pointlike and linelike excitations, 
and determine their statistics, as well as the
topological ground state degeneracy.  
Next, in Sec.~\ref{subsec: Zm example} we exemplify our strategy with
perhaps the simplest type of topological order in three dimensions, namely 
$\mathbb{Z}^{\,}_{m}$ topological order.
Furthermore, we investigate the surface states of these 
$\mathbb{Z}^{\,}_{m}$-topologically-ordered states of matter,
and find that they are unstable to interwire hoppings. 
Additionally, a surface fractional quantum Hall effect with
Hall conductivity $[(2e)^2/h] \times (1/2m)$ 
can develop at the expense of breaking time-reversal symmetry on the surface.
We also discuss how these observations
regarding surface states can be extended to the more general class 
of interwire interactions introduced in
Sec.~\ref{subsec: 2D coupling discussion}.

Next, in Sec.~\ref{sec: Higher-dimensional wire constructions}, 
we outline the generalization of our results to arbitrary dimensions.  
In Sec.~\ref{subsec: Review of toric codes in arbitrary dimensions}, 
we discuss how to define $d$-dimensional hypercubic arrays of quantum
wires that are analogous to the square array of quantum wires used to
construct three-dimensional topological states.  Then, in
Sec.~\ref{subsec: Generalizing the results of Sec. ...}, we generalize
the results of Sec.~\ref{subsec: 2D coupling discussion} regarding the
definitions of appropriate interwire couplings and their compatibility
in the strong-coupling limit.  Finally, in 
Sec.~\ref{subsec: Example: 4D toric codes}, 
we provide an example of this generalization by
constructing $\mathbb{Z}^{\,}_{m}$-topologically-ordered states of
matter in four dimensions, and constructing their pointlike, linelike,
and membranelike excitations, before concluding in 
Sec.~\ref{sec: Conclusion}.

\section{Three-dimensional wire constructions}
\label{sec: Three-dimensional wire constructions}

In this section, a method to construct arrays of coupled wires realizing
topological phases of matter in three-dimensional space is presented.
We begin by defining a class
of gapless theories describing decoupled wires, before moving on to a
discussion of interwire couplings.  In particular, we provide a set of
algebraic criteria that are sufficient to determine whether the theory
is gapped when periodic boundary conditions are imposed.

\subsection{Decoupled wires}
\label{subsec: decoupled wires d=2}

We consider a two-dimensional array of $2N$ quantum wires, 
labeled by Latin indices $j=1,\dots,2N$, 
placed on the links of a two-dimensional square
lattice embedded in three-dimensional Euclidean space.  Each quantum
wire is assumed to be gapless and \textit{nonchiral}, and therefore
to contain $2M$ gapless degrees of freedom, labeled by Greek indices
$\alpha=1,\dots,2M$.  We take the wires (of length $L$) to lie along
the $z$-direction, and the square lattice to lie in the $x$-$y$ plane.
We will impose periodic boundary conditions in all directions ($x$, $y$,
and $z$) until further notice.  The set of decoupled quantum wires is
described by the quadratic Lagrangian
\begin{subequations}
\label{L0 d=2}
\begin{align}
\hat{L}^{\,}_{0}=
\frac{1}{4\pi} 
\int\limits^{L}_{0}
\mathrm{d}z
\Bigg[
\(\partial^{\,}_{t}\hat{\Phi}\)^{\T}
\!
\mathcal{K}
\(\partial^{\,}_{z}\hat{\Phi}\)
- 
\(\partial^{\,}_{z}\hat{\Phi}\)^{\T}
\!
\mathcal{V}
\(\partial^{\,}_{z}\hat{\Phi}\)
\Bigg]
\label{L0 d=2 a}
\end{align}
where
\begin{align}
\begin{split}
\hat{\Phi}(t,z)&\:=
\Big(
\hat{\phi}^{\,}_{1,1}(t,z)\ 
\dots\ 
\hat{\phi}^{\,}_{1,2M}(t,z) 
\mid 
\\
&\qquad
\dots
\mid  
\hat{\phi}^{\,}_{2N,1}(t,z)\ 
\dots\ 
\hat{\phi}^{\,}_{2N,2M}(t,z) 
\Big)^{\T}
\end{split}
\end{align}
is a vector that collects the $2M$ scalar fields 
$\hat{\phi}^{\,}_{j,\alpha}(t,z)$
defined in each of the $j=1,\ldots,2N$ wires.  
We use vertical bars as a visual
aid to separate degrees of freedom defined in different wires.  
The block-diagonal $4MN$-dimensional matrix
\begin{align}
\mathcal{K}\:= 
\mathbbm{1}^{\,}_{2N}\otimes K,
\end{align}
where $\mathbbm{1}^{\,}_{2N}$ is the unit matrix of dimension $2N$ and
$K$ is a $2M\times 2M$ symmetric matrix with integer entries,
yields the equal-time commutation relations
\begin{align}
\[
\partial^{\,}_{z}\hat{\phi}^{\,}_{j,\alpha}(z),
\hat{\phi}^{\,}_{\pri{j},\pri{\alpha}}(\pri{z})
\]&=
\mathrm{i}\,2\pi\, 
\delta^{\,}_{j \pri{j}}\, 
K^{-1}_{\alpha\pri{\alpha}}\, 
\delta(z-\pri{z}).
\label{eq: K-matrix commutator}
\end{align}
We will omit the explicit time dependence of the fields from now on.
Finally, the block-diagonal $4MN\times 4MN$ matrix
\begin{align}
\mathcal{V}\:= 
\mathbbm{1}^{\,}_{2N}\otimes V,
\end{align}
where the $2M\times 2M$ matrix $V$ is real, symmetric, and
positive-definite.  The matrix $V$ is set by microscopics within each
wire, and will usually be taken to be a diagonal matrix in this work.
However, the matrix $K$, which enters the commutation relations
\eqref{eq: K-matrix commutator}, contains crucial data that define the
fundamental degrees of freedom in a wire.  The final data necessary to
complete the definition of the theory describing the 
two-dimensional array of decoupled quantum wires
is the $4MN$-dimensional ``charge-vector''
\begin{align}\label{charge vector definition}
\mathcal Q \:= 
\begin{pmatrix}
Q & \mid & Q & \mid & \dots  & \mid & Q 
\end{pmatrix}^{\T}.
\end{align}
\end{subequations}
The $2M$-dimensional integer vector $Q$ collects the $U(1)$ electric
charges associated with the scalar fields $\hat{\phi}^{\,}_{j,\alpha}$,
$\alpha=1,\dots,2M$.

The theory defined by Eqs.~\eqref{L0 d=2} can be viewed as an
effective low-energy description of 
a two-dimensional array
of decoupled physical quantum wires containing
fermionic or bosonic degrees of freedom.  

For fermions, each wire $j=1,\ldots,2N$
contains $M$ flavors of \textit{chiral} scalar fields
$\hat{\phi}^{\,}_{j,\alpha^{\,}_{R}}$ and $\hat{\phi}^{\,}_{j,\alpha^{\,}_{L}}$, where
$\alpha^{\,}_{R,L}=1,\dots,M$ label right- and left-moving degrees of
freedom, respectively.  
These fields obey the chiral equal-time commutation relations
\begin{subequations}
\label{eq: def fermionic K}
\begin{align}
\[
\partial^{\,}_{z}\hat{\phi}^{\,}_{j,\alpha^{\,}_{R}}(z),
\hat{\phi}^{\,}_{\pri{j},\alpha^{\prime}_{R}}(\pri{z})
\]&=
+\mathrm{i}\,2\pi\,
\delta^{\,}_{j \pri{j}}\, 
\delta^{\,}_{\alpha^{\,}_{R}\alpha^{\prime}_{R}}\, 
\delta(z-\pri{z}) ,
\nonumber\\
\[
\partial^{\,}_{z}\hat{\phi}^{\,}_{j,\alpha^{\,}_{L}}(z),
\hat{\phi}^{\,}_{\pri{j},\alpha^{\prime}_{L}}(\pri{z})
\]&=
-\mathrm{i}\,2\pi\, 
\delta^{\,}_{j \pri{j}}\, 
\delta^{\,}_{\alpha^{\,}_{L}\alpha^{\prime}_{L}}\, 
\delta(z-\pri{z}) ,
\nonumber\\
\[
\partial^{\,}_{z}\hat{\phi}^{\,}_{j,\alpha^{\,}_{R}}(z),
\hat{\phi}^{\,}_{\pri{j},\alpha^{\,}_{L}}(\pri{z})
\]&=0,
\label{eq: def fermionic K a}
\end{align}
and therefore, for fermions, the $2M\times 2M$ matrix $K$ entering
Eq.~\eqref{eq: K-matrix commutator} is given by
\begin{align}
K^{\,}_{\mathrm{f}}\:= 
\bigoplus^{M}_{\alpha=1}\, 
\begin{pmatrix}
+1&0
\\
0&-1
\end{pmatrix}.
\label{eq: def fermionic K b}
\end{align}
We further adopt the convention that the charge-vector 
\begin{align}
Q^{\,}_{\mathrm{f}}\:=
\begin{pmatrix}
1 & \dots & 1
\end{pmatrix}^{\T},
\label{eq: def fermionic K c}
\end{align}
\end{subequations}
in units where the electron charge $e$ is set to unity,
for a fermionic wire with $2M$ channels. Treating an array of fermionic
quantum wires within Abelian bosonization, as we do here, further
requires the use of Klein factors, which are needed in order to assure
that fermionic vertex operators (defined below) defined in different wires
anticommute with one another.  These Klein factors can be subsumed into the
equal-time commutation relations for the scalar fields
$\hat{\phi}^{\,}_{j,\alpha^{\,}_{R}}$ and $\hat{\phi}^{\,}_{j,\alpha^{\,}_{L}}$. 
This can be done by integrating both sides of 
Eqs.~\eqref{eq: def fermionic K a} 
over all $z$ and fixing the arbitrary constant
of integration to be the Klein factor necessary to ensure the appropriate 
anticommutation of vertex operators. We refer the reader to the Appendix of 
Ref.~\cite{Neupert11b} for more details on this procedure.

For bosons, each wire $j=1,\ldots,2N$ instead contains $M$ flavors of
\textit{nonchiral} scalar fields $\hat{\phi}^{\,}_{j,\alpha^{\,}_{1}}$ and
$\hat{\phi}^{\,}_{j,\alpha^{\,}_{2}}$, 
where $\alpha^{\,}_{1,2}=1,\dots,M$ label ``charge"
and ``spin" degrees of freedom, respectively.  
These fields obey the equal-time commutation relations
\begin{subequations}
\label{eq: def bosonic K}
\begin{align}
\[
\partial^{\,}_{z}\hat{\phi}^{\,}_{j,\alpha^{\,}_{1}}(z),
\hat{\phi}^{\,}_{\pri{j},\alpha^{\prime}_{1}}(\pri{z})
\]&=0 , 
\nonumber\\
\[
\partial^{\,}_{z}\hat{\phi}^{\,}_{j,\alpha^{\,}_{2}}(z),
\hat{\phi}^{\,}_{\pri{j},\alpha^{\prime}_{2}}(\pri{z})
\]&=0 ,
\label{eq: def bosonic K a}
\\
\[
\partial^{\,}_{z}\hat{\phi}^{\,}_{j,\alpha^{\,}_{1}}(z),
\hat{\phi}^{\,}_{\pri{j},\alpha^{\,}_{2}}(\pri{z})
\]&=
\mathrm{i}\,2\pi\, 
\delta^{\,}_{j \pri{j}}\, 
\delta^{\,}_{\alpha^{\,}_{1}\alpha^{\,}_{2}}\, 
\delta(z-\pri{z}), 
\nonumber
\end{align}
so that the $K$-matrix for bosons is
\begin{align}
K^{\,}_{\mathrm{b}}\:= 
\bigoplus^{M}_{\alpha = 1}\, 
\begin{pmatrix}
0&1
\\
1&0
\end{pmatrix}.
\label{eq: def bosonic K b}
\end{align}
We take the bosonic charge vector to be
\begin{align}
Q^{\,}_{\mathrm{b}}\:=
2\begin{pmatrix}
1 & 0& \dots & 1 & 0
\end{pmatrix}^{\T},
\label{eq: def bosonic K c}
\end{align}
\end{subequations}
in units where the electron charge $e$ is set to unity,
so that the fields $\hat{\phi}^{\,}_{j,\alpha^{\,}_{1}}$ carry a $U(1)$ electric charge,
while $\hat{\phi}^{\,}_{j,\alpha^{\,}_{2}}$ is neutral.  (Of course, one could define a
``spin vector" analogous to $Q$ that encodes the coupling to another
$U(1)$ gauge field for spin, but, for simplicity, we will work
exclusively with electric charges here.)

The fundamental excitations of a fermionic or bosonic wire can be
built out of the vertex operators
\begin{align}\label{fermi-bose vertex ops}
\hat{\psi}^{\dag}_{\mathrm{f,b}; j,\alpha}(z) \:= 
\exp
\(
-\mathrm{i}\, 
(K^{\,}_{\mathrm{f},b})^{\,}_{\alpha\pri{\alpha}}\, 
\hat{\phi}^{\,}_{j,\pri{\alpha}}(z)
\),
\end{align}
for any $j=1,\dots,2N$ and $\alpha=1,\dots,2M$,
where we have adopted the convention of summing over repeated indices.
Any local operator acting within a single wire can be built from these
vertex operators.  Similarly, operators spanning multiple wires can be
built by taking products of vertex operators from each constituent
wire.  The charges of the excitations created by these vertex
operators are measured by the charge operator
\begin{align}
\hat{Q}^{\,}_{j,\alpha}&\:= 
\frac{Q^{\,}_{\alpha}}{2\pi}\, 
\delta^{\,}_{\alpha\pri{\alpha}}
\int\limits^{L}_{0}
\mathrm{d}z\ 
\partial^{\,}_{z}\hat{\phi}^{\,}_{j,\pri{\alpha}}(z),
\end{align}
for any $j=1,\dots,2N$ and $\alpha=1,\dots,2M$,  
where $L$ is the length of a wire.
The normalization of the charge
operator is taken to be such that
\begin{align}
[
\hat{Q}^{\,}_{j,\alpha}, 
\hat{\psi}^{\dag}_{\mathrm{f,b};\pri{j},\pri{\alpha}}(z)
]&= 
Q^{\,}_\alpha\, 
\delta^{\,}_{j\pri{j}}\, 
\delta^{\,}_{\alpha\pri{\alpha}}\,
\hat{\psi}^{\dag}_{\mathrm{f,b};\pri{j},\pri{\alpha}}(z)
\end{align}
at equal times,
indicating that the vertex operator
$\hat{\psi}^{\dag}_{\mathrm{f,b};j,\alpha}$ 
carries the charge $Q^{\,}_\alpha$.

\subsection{Interwire couplings and criteria for producing gapped states 
of matter}
\label{subsec: 2D coupling discussion}

Given the two-dimensional array of decoupled and gapless quantum wires 
defined in
Sec.~\ref{subsec: decoupled wires d=2}, we would like to devise a systematic
way of introducing strong single-particle or many-body couplings
between adjacent wires in order to yield a variety of gapped
topologically-nontrivial three-dimensional phases of matter.  Our
strategy will be to extend the approach taken in
Ref.~\cite{Neupert14}, which considered one-dimensional chains of
wires, to two dimensions.  We begin by adding to the quadratic
Lagrangian $\hat{L}^{\,}_{0}$ defined in Eq.~\eqref{L0 d=2} a set of cosine
potentials
\begin{align}\label{Lint}
\hat{L}^{\,}_{\{\mathcal{T}\}}&:= 
\int\limits^{L}_{0}
\mathrm{d}z
\sum_{\mathcal{T}}
U^{\,}_{\mathcal{T}}(z)\, 
\cos
\(
\mathcal{T}^{\T}\,\mathcal{K}\,\hat{\Phi}(t,z)+\alpha^{\,}_{\mathcal{T}}(z)
\).
\end{align}
Here, the $4MN$-dimensional integer vectors $\mathcal{T}$ encode
tunneling processes between adjacent wires.  This interpretation
becomes transparent upon recognizing that, up to an overall phase,
\begin{align}\label{eq: tunneling}
e^{-\mathrm{i}\,\mathcal{T}^{\T}\,\mathcal{K}\,\hat{\Phi}(z)}= 
\prod_{j=1}^{2N}\prod_{\alpha=1}^{2M}
\[\hat{\psi}^{\dag}_{\mathrm{f,b};j,\alpha}(z)\]^{\mathcal{T}^{\,}_{j,\alpha}},
\end{align}
where $\hat{\psi}^{\dag}_{\mathrm{f,b};j,\alpha}$ are the vertex operators
defined in Eq.~(\ref{fermi-bose vertex ops}).  
[We follow
Ref.~\cite{Neupert14} in using the shorthand notation
$(\hat{\psi}^{\dag}_{\mathrm{f,b};j,\alpha})^{-1}\equiv
\hat{\psi}^{\phantom\dag}_{\mathrm{f,b};j,\alpha}$
and in employing an appropriate point-splitting prescription when
multiplying fermionic operators.]  For generic tunneling vectors
$\mathcal{T}$, Eq.~\eqref{eq: tunneling} describes a many-body or
correlated tunneling that amounts to an interaction term in the
Lagrangian  
\begin{equation} \label{def L as sum L0 and LT}
\hat{L}\:=\hat{L}^{\,}_{0}+\hat{L}^{\,}_{\{\mathcal{T}\}}.
\end{equation}
The real-valued functions
$U^{\,}_{\mathcal{T}}(z)\geq 0$ and $\alpha^{\,}_{\mathcal{T}}(z)$ in
Eq.~\eqref{Lint} encode the effects of disorder on the amplitude and
phase of these interwire couplings.

Distinct states of matter can be realized by restricting the sum over
tunneling vectors $\mathcal{T}$ in Eq.~\eqref{Lint} to ensure that the
interaction terms \eqref{eq: tunneling} satisfy certain symmetries.  For
all examples considered in this work, we will assume that either
charge or number-parity conservation holds.  The former is imposed by
demanding that
\begin{subequations}\label{charge or parity conservation}
\begin{align}\label{charge conservation}
\mathcal{Q}^{\T}\,\mathcal{T}=0\quad \forall\ \mathcal{T},
\end{align}
while the latter is imposed by relaxing the above requirement to
\begin{align}\label{parity conservation}
\mathcal{Q}^{\T}\,\mathcal{T}=0\ \text{mod 2}\quad \forall\ \mathcal{T}.
\end{align}
\end{subequations}
For a detailed discussion of how further symmetry requirements
constrain the tunneling vectors $\mathcal{T}$, see
Ref.~\cite{Neupert14}.

We are now prepared to discuss the strategy we employ to produce
gapped states of matter from the above construction.  We first recall
that the array of decoupled quantum wires consists of 
$4\,M\,N$ gapless degrees of freedom.  
As noted in Ref.~\cite{Haldane95}, and later employed in
Refs.~\cite{Neupert11b,Neupert14}, a single cosine term in the sum
in Eq.~\eqref{Lint} is capable of removing (i.e., gapping
out) at most two of these gapless degrees of freedom from the
low-energy sector of the theory.  This occurs in the limit
$U^{\,}_{\mathcal{T}}\to\infty$, where the argument of the cosine term
becomes pinned to its classical minimum.  Therefore, in principle it
takes only $2MN$ cosine terms to gap out all $4\,M\,N$
degrees of freedom in the bulk of the array of quantum wires 
when periodic boundary conditions are imposed.
Matters are complicated somewhat by the nontrivial commutation
relations \eqref{eq: K-matrix commutator}, which ensure that cosine terms
corresponding to distinct tunneling vectors $\mathcal{T}$ and $\mathcal
T^{\prime}$ do not commute in general.  Consequently, it is possible
that quantum fluctuations may lead to competition between the various
cosine terms that frustrates the optimization problem of
simultaneously minimizing all of these terms.  However, in Ref.~\cite{Haldane95},
Haldane observed that if the criterion
\begin{align}
\mathcal{T}^{\T}\,\mathcal{K}\,\mathcal{T}^{\prime}=0
\label{Haldane criterion}
\end{align}
holds, then the cosine terms associated with the tunneling vectors
$\mathcal{T}$ and $\mathcal{T}^{\prime}$ can be minimized independently,
and therefore do not compete with one another.
[Note that each tunneling vector $\mathcal{T}$
must also satisfy Eq.~\eqref{Haldane criterion}, i.e., we require
that $\mathcal{T}^{\T}\,\mathcal K\,\mathcal{T}=0$ 
for all $\mathcal{T}$.]  Therefore, if one
can find a ``Haldane set" $\mathbb{H}$ of $2MN$ linearly-independent
tunneling vectors, all of which satisfy Eq.~\eqref{Haldane criterion},
then it is possible to gap out all degrees of freedom in the 
array of quantum wires
by adding sufficiently strong interactions of the form
\eqref{Lint}. If such a set $\mathbb{H}$ is found, then it suffices to
restrict the sum in Eq.~\eqref{Lint} to $\mathcal{T}\in\mathbb{H}$, and
to posit that all couplings $U^{\,}_{\mathcal{T}}$ are sufficiently large in
magnitude to gap out all modes in the array of quantum wires.

\begin{figure}[t]
\begin{center}
\includegraphics[scale=0.5]{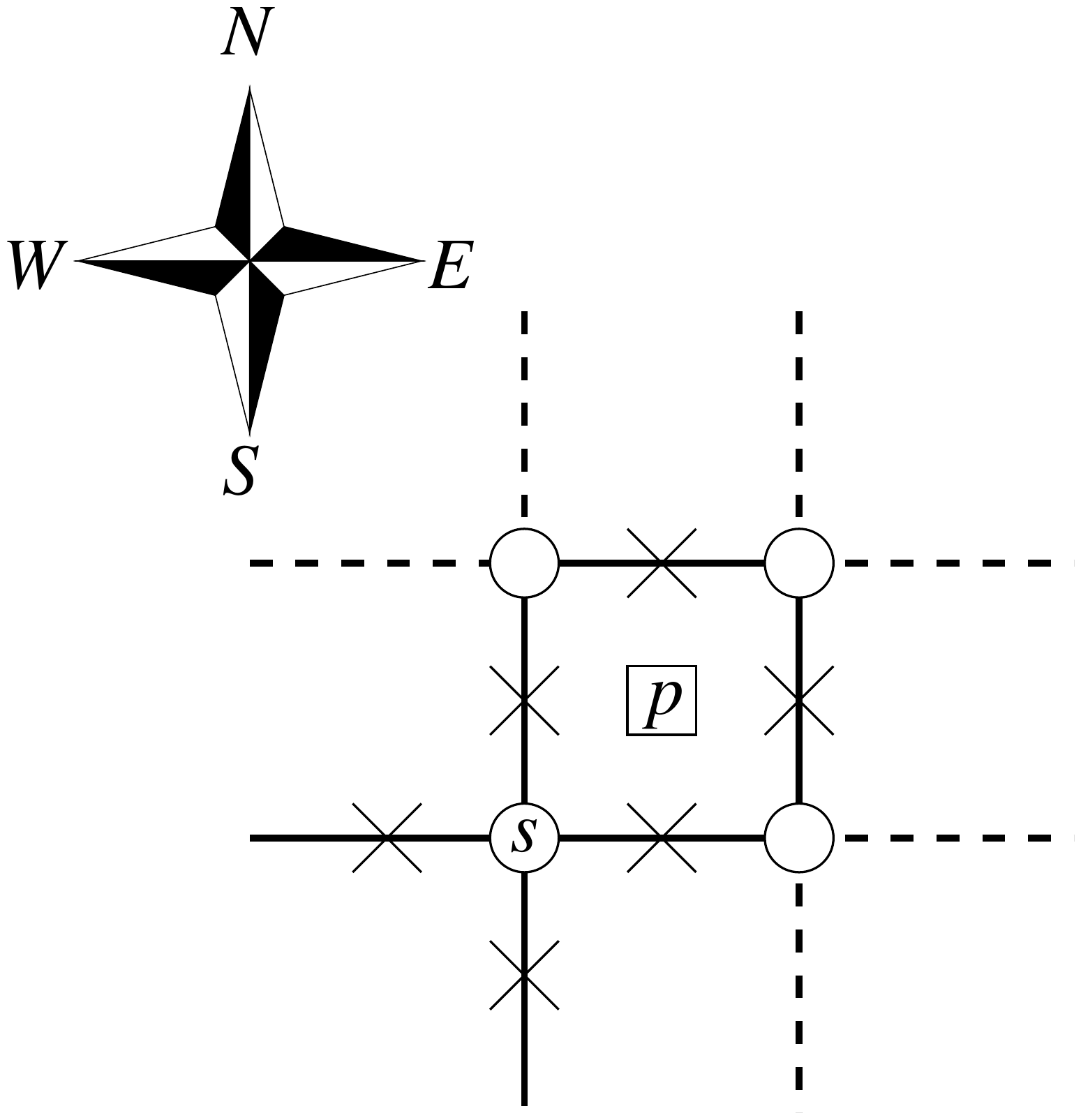}
\caption{ 
A single unit cell of the square array of wires, consisting of a
single star $s$ and plaquette $p$. The dashed nearest-neigbor links
belong to neighboring unit cells. The 
midpoint of each
link hosts a quantum wire, represented by the symbol $\times$,
aligned along the $z$-direction (out of the page). Any plaquette $p$ is
surrounded by four quantum wires located at the four cardinal points
$p^{\,}_{N}$, $p^{\,}_{W}$, $p^{\,}_{S}$, and $p^{\,}_{E}$,
repectively.  Similarly, any star $s$ is surrounded by four quantum
wires located at the four cardinal points $s^{\,}_{N}$,
$s^{\,}_{W}$, $s^{\,}_{S}$, and $s^{\,}_{E}$.} 
\label{Fig: square lattice with PBC}
\end{center}
\end{figure}

We now present a simple geometric prescription to aid in the
determination of the existence (or lack thereof) of a Haldane set
$\mathbb{H}$ for a two-dimensional array of quantum wires 
with some set of desired symmetries.  
This prescription capitalizes on the fact that we have
chosen all $2N$ quantum wires to lie on the links of a square lattice.
(In principle, this is not the only possible choice of lattice
geometry, but it provides a simple way of counting degrees of freedom
in any dimension, as we will see below and in Sec.~\ref{sec: Higher-dimensional wire constructions}.)
On a square lattice with $2N$ sites, there are $N$ ``stars" (centered
on the vertices of the lattice) and $N$ ``plaquettes" (centered on the
vertices of the \textit{dual lattice}), assuming that periodic boundary 
conditions are imposed as in Fig.\ \ref{Fig: square lattice with PBC}.  
If we associate the tunneling vectors
$\mathcal{T}^{\,}_{s}$ and $\mathcal{T}^{\,}_{p}$  
with each star $s$ and plaquette $p$, respectively,
then we have a set of $2N$ tunneling vectors.  
Since there are $4MN$ gapless degrees of freedom 
in the array of decoupled quantum wires, 
we can obtain the necessary number $2MN$ of tunneling vectors by expanding
this set to include $M$ ``flavors" of tunneling vectors 
$\mathcal{T}^{(\texttt{j})}_{s}$ and $\mathcal{T}^{(\texttt{j})}_{p}$ for each star
and plaquette, respectively.  We label these flavors using a teletype
index $\texttt{j}=1,\dots,M$.  Imposing the Haldane criterion
\eqref{Haldane criterion} on this set of tunneling vectors then yields
the set of equations
\begin{subequations}\label{Haldane 2D}
\begin{align}
&
\mathcal{T}^{(\texttt{j})\T}_{s}\ 
\mathcal{K}\ 
\mathcal{T}^{(\texttt{j}^{\prime})}_{s^{\prime}}=0 
\qquad \forall\ s,\pri s,\texttt{j},\texttt{j}^{\prime} ,
\label{star-star Haldane}
\\
&
\mathcal{T}^{(\texttt{j})\T}_{p}\ 
\mathcal{K}\ 
\mathcal{T}^{(\texttt{j}^{\prime})}_{p^{\prime}}=0 
\qquad \forall\ p,\pri{p},\texttt{j},\texttt{j}^{\prime} ,
\label{plaquette-plaquette Haldane}
\\
&
\mathcal{T}^{(\texttt{j})\T}_{s}\ 
\mathcal{K}\ 
\mathcal{T}^{(\texttt{j}^{\prime})}_{p}=0 
\qquad \forall\ s, p,\texttt{j},\texttt{j}^{\prime}.
\label{star-plaquette Haldane}
\end{align}
\end{subequations}
If the above equations are satisfied, then the set of $2MN$ tunneling
vectors is a Haldane set, and therefore capable of yielding a gapped
phase in the strong-coupling limit.

\begin{figure}[t]
\begin{center}
\begin{flushleft}
\hspace{.5cm} (a)
\end{flushleft}
\includegraphics[width=.4\textwidth]{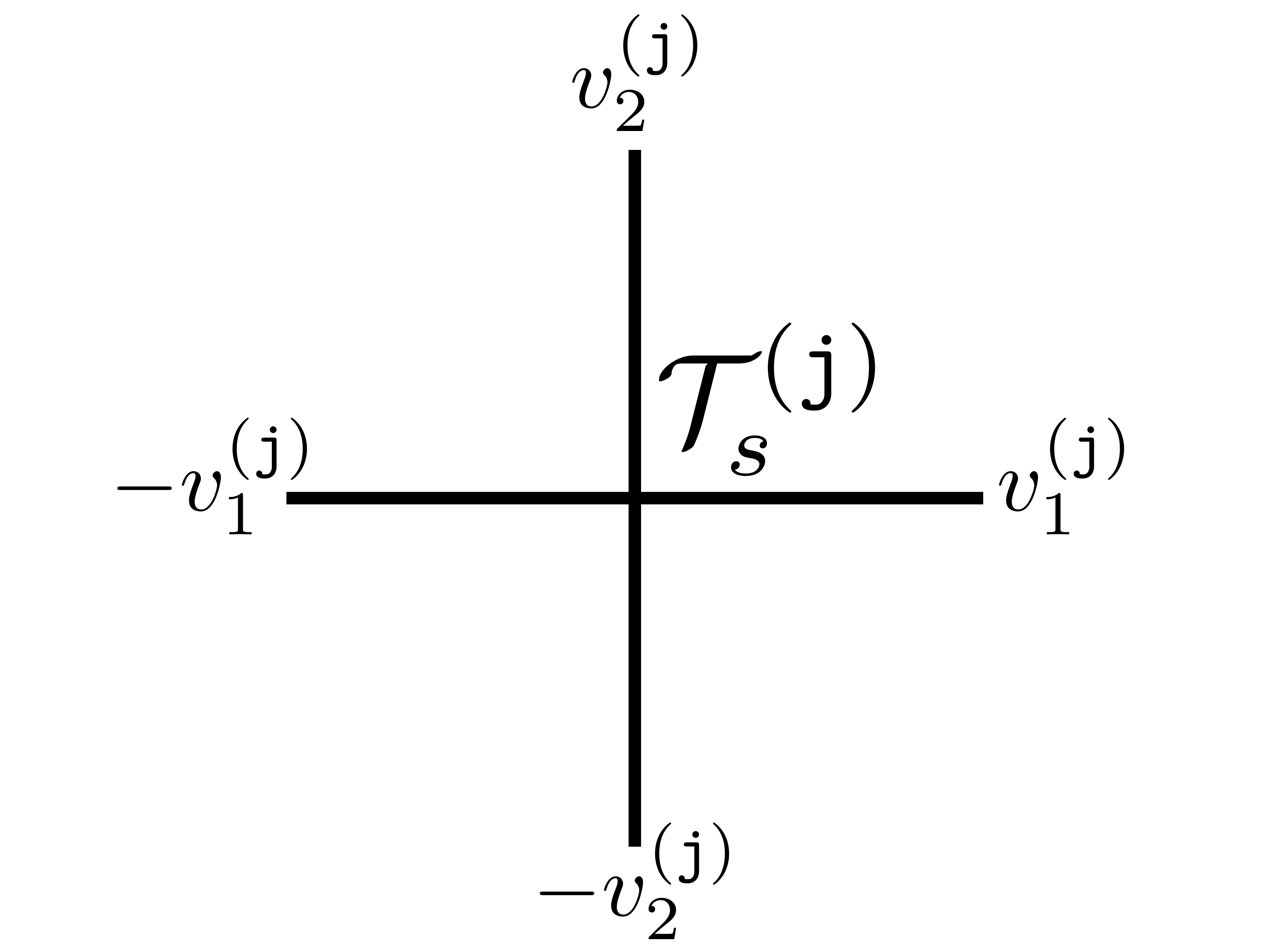}\\
\begin{flushleft}
\hspace{.5cm} (b)
\end{flushleft}
\includegraphics[width=.4\textwidth]{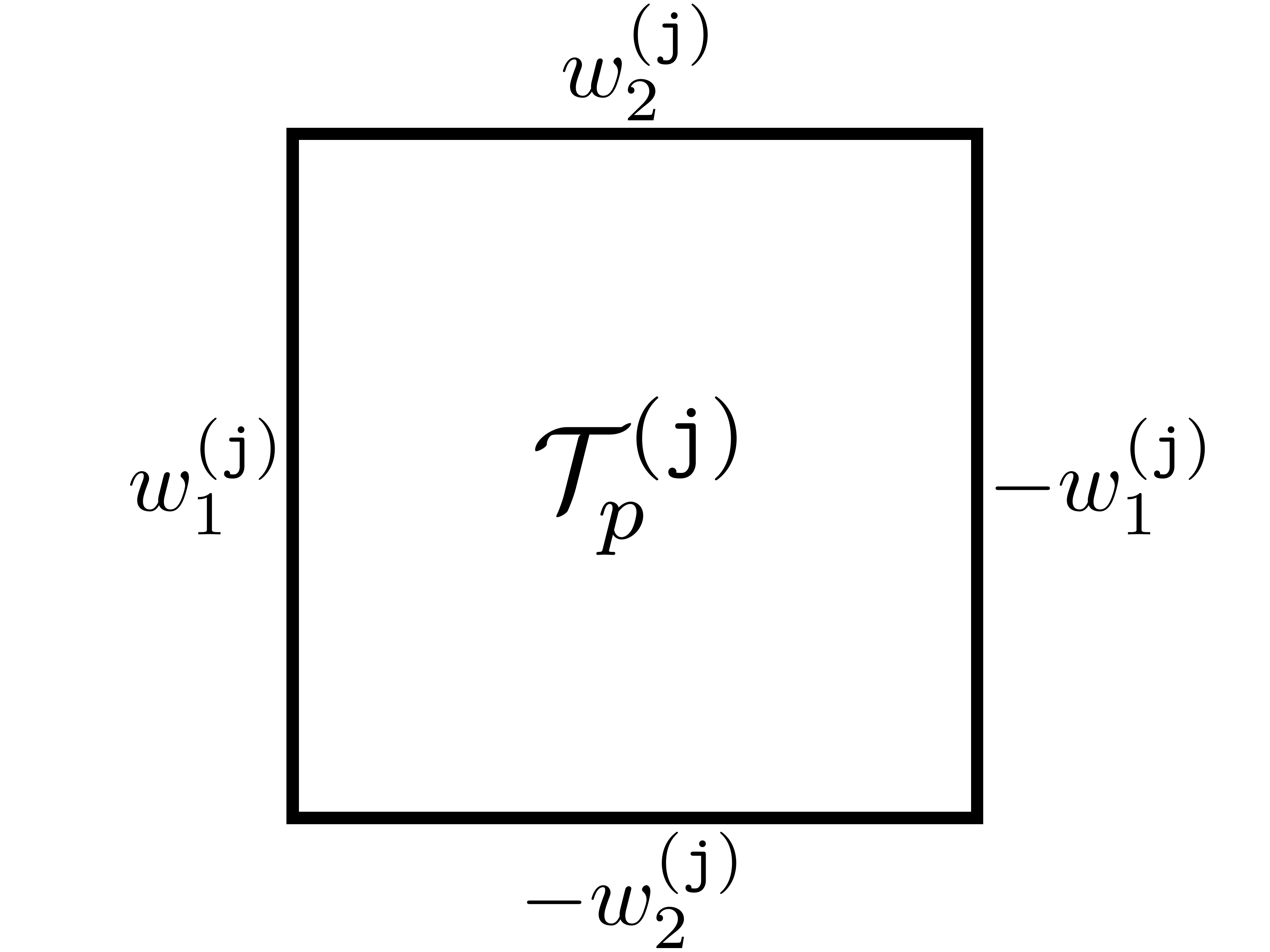}
\caption{
Pictorial representation of the tunneling vectors 
\eqref{2d T-vecs charge conserving}.  The $2M$-dimensional integer-valued
vectors $v^{(\texttt j)}_{1,2}$ and $w^{(\texttt j)}_{1,2}$ determine the
linear combinations of bosonic fields in each wire that enter the cosine
term associated with each star or plaquette, respectively.
\label{fig: generic tunneling vectors}
         }
\end{center}
\end{figure}

We now turn to the problem of building $2MN$ tunneling vectors
$\mathcal{T}^{(\texttt{j})}_{s}$ and $\mathcal{T}^{(\texttt{j})}_{p}$.
Enumerating all solutions to this problem for all matrices $\mathcal
K$ is beyond the scope of the present work.  However, we will present
below one way of constructing these tunneling vectors that builds in
the minimal symmetries of charge and/or parity conservation
[Eqs.~\eqref{charge or parity conservation}] and greatly reduces the
number of equations that must be solved [relative to
Eqs.~\eqref{Haldane 2D}, which contain an infinite number of linear
equations in the thermodynamic limit $N\to\infty$ if no
additional information is provided].  
In particular, if we desire charge
conservation [Eq.~\eqref{charge conservation}] to hold, we may define
the tunneling vectors by their nonvanishing components
\begin{subequations}\label{2d T-vecs charge conserving}
\begin{align}
\begin{split}
(\mathcal{T}^{(\texttt{j})}_{s})_{j,\alpha} &\:= 
v^{(\texttt{j})}_{1,\alpha}\, 
\left(
\delta^{\,}_{j,s^{\,}_{E}} 
- 
\delta^{\,}_{j,s^{\,}_{W}}
\right)
\\
&\qquad\qquad 
+ v^{(\texttt{j})}_{2,\alpha}\, 
\left(
\delta^{\,}_{j,s^{\,}_{N}} 
- 
\delta^{\,}_{j,s^{\,}_{S}}
\right),
\end{split}\\
\begin{split}
(\mathcal{T}^{(\texttt{j})}_{p})_{j,\alpha} &\:= 
w^{(\texttt{j})}_{1,\alpha}\, 
\left(
\delta^{\,}_{j,p^{\,}_{W}} 
- 
\delta^{\,}_{j,p^{\,}_{E}}
\right)
\\
&\qquad\qquad 
+ w^{(\texttt{j})}_{2,\alpha}\, 
\left(
\delta^{\,}_{j,p^{\,}_{N}} 
- 
\delta^{\,}_{j,p^{\,}_{S}} 
\right),
\end{split}
\end{align}
\end{subequations}
where we recall that $j=1,\dots,2N$ labels the quantum wires
and $\alpha=1,\dots,2M$ labels the degrees of freedom within a
wire. Here, $v^{(\texttt{j})}_{1}$, $v^{(\texttt{j})}_{2}$, $w^{(\texttt{j})}_{1}$,  
and $w^{(\texttt{j})}_{2}$ are arbitrary $2M$-dimensional integer vectors.
The Kronecker deltas in the tunneling vector $\mathcal{T}^{\,}_{s}$ 
ensure that its nonzero entries are defined within the quantum wires 
$s^{\,}_{N},\dots,s^{\,}_{W}$ to the north, $\dots$, west 
of the vertex on which star 
$s$ is centered.  
The Kronecker deltas in $\mathcal{T}^{\,}_{p}$ select
the quantum wires $p^{\,}_{N},\dots,p^{\,}_{W}$, 
which are defined similarly for the
plaquette $p$ (see Fig.~\ref{Fig: square lattice with PBC}).
With these definitions, one verifies that Eq.~\eqref{charge conservation}
holds independently of the form of $v^{(\texttt{j})}_{1,2}$, $w^{(\texttt{j})}_{1,2}$, 
and the charge-vector $Q$ for a single wire.  

Similarly, when we wish to
impose number-parity conservation [Eq.~\eqref{parity conservation}],
we may define for any $j=1,\dots,2N$ 
and any $\alpha=1,\dots,2M$
\begin{subequations}\label{2d T-vecs parity conserving}
\begin{align}
\begin{split}
(\mathcal{T}^{(\texttt{j})}_{s})_{j,\alpha} &\:= 
v^{(\texttt{j})}_{1,\alpha}\, 
\left(
\delta^{\,}_{j,s^{\,}_{E}} 
+ 
\delta^{\,}_{j,s^{\,}_{W}}
\right)
\\
&\qquad\qquad 
+ v^{(\texttt{j})}_{2,\alpha}\, 
\left(
\delta^{\,}_{j,s^{\,}_{N}} 
+\delta^{\,}_{j,s^{\,}_{S}} 
\right),
\end{split}
\\
\begin{split}
(\mathcal{T}^{(\texttt{j})}_{p})_{j,\alpha} &\:= 
w^{(\texttt{j})}_{1,\alpha}\, 
\left(
\delta^{\,}_{j,p^{\,}_{E}} 
+ 
\delta^{\,}_{j,p^{\,}_{W}}
\right)
\\
&\qquad\qquad 
+ w^{(\texttt{j})}_{2,\alpha}\, 
\left(
\delta^{\,}_{j,p^{\,}_{N}} 
+ 
\delta^{\,}_{j,p^{\,}_{S}}
\right),
\end{split}
\end{align}
\end{subequations}
and verify that Eq.\ (\ref{parity conservation}) holds independently of
the form of $v^{(\texttt{j})}_{1,2}$, $w^{(\texttt{j})}_{1,2}$, and $Q$.

Henceforth, we will focus on the charge-conserving tunneling vectors
defined in Eqs.\ (\ref{2d T-vecs charge conserving}), as all general
criteria discussed below have analogues for the parity-conserving
tunneling vectors defined in Eqs.\ (\ref{2d T-vecs parity conserving}).

The charge-conserving tunneling vectors defined in 
Eqs.\ (\ref{2d T-vecs charge conserving}) 
are expressed in a convenient pictorial form in
Fig.\ \ref{fig: generic tunneling vectors}.  
From this pictorial representation, it
is clear that any two distinct, adjacent stars (be they of the same
flavor or different flavors) 
share a single wire between them.  The
same statement holds for plaquettes.  However, adjacent stars and
plaquettes share two wires between them, regardless of the flavor.
Therefore, one can show that Eqs.\ (\ref{Haldane 2D}) are satisfied if
and only if
\begin{subequations}\label{eq: 2D criteria}
\begin{align}
&
v^{(\texttt{j})\T}_{\mu}\, K\, v^{(\texttt{j}^{\prime})}_{\mu} = 0 , \label{v criterion 2D}
\\
&
w^{(\texttt{j})\T}_{\mu}\, K\, w^{(\texttt{j}^{\prime})}_{\mu} = 0 , \label{w criterion 2D}
\\
&
v^{(\texttt{j})\T}_{1}\, K\, w^{(\texttt{j}^{\prime})}_{2}
-
v^{(\texttt{j})\T}_{2}\, K\, w^{(\texttt{j}^{\prime})}_{1} = 0, \label{star-plaq 2D}
\end{align}
\end{subequations}
for all $\texttt{j}$ and $\texttt{j}^{\prime}=1,\dots,M$ and $\mu=1,2$.
Equations\ \eqref{eq: 2D criteria} are fundamental to our construction, 
as each solution to these equations for a given dimension $2M$ of the matrix
$K$ may in principle describe a distinct gapped phase of matter.

Observe that Eqs.\ (\ref{eq: 2D criteria})
are symmetric under 
$1\leftrightarrow2$ and 
$\texttt{j}\leftrightarrow\texttt{j}^{\prime}$.  Therefore, these
criteria amount to a set of $5M(M+1)/2$ linear equations in $8M^{2}$
variables.  This is important for two reasons.  First, the number of
equations does not scale with the number $2N$ of quantum wires in the array.
This ensures that a single solution to these equations holds for any
system size when periodic boundary conditions are imposed. Second,
this set of equations is underconstrained for any $M$ (i.e., there are
always more variables than equations).  This means that for generic
matrices $K$ of fixed dimension $2M$, there is in principle more than
one solution to Eqs.~\eqref{eq: 2D criteria}.

We aim to construct gapped states of matter that have an isotropic low-energy
description. Consequently, it is natural to demand that the tunneling vectors
defined in Eqs.~\eqref{2d T-vecs charge conserving} and depicted in 
Fig.~\ref{fig: generic tunneling vectors} are independent of direction.
This can be achieved by imposing the additional constraints
\begin{subequations}\label{simplifying condition}
\begin{equation} 
v^{(\texttt{j})}_{1} = v^{(\texttt{j})}_{2} \=: v^{(\texttt{j})}
\end{equation}
and
\begin{equation} 
w^{(\texttt{j})}_{1} = w^{(\texttt{j})}_{2} \=: w^{(\texttt{j})}.
\end{equation}
\end{subequations}  
Note that Eq.~\eqref{star-plaq 2D} is solved
independently of the form of the $2M$-dimensional vectors 
$v^{(\texttt{j})}_{\mu}$ and $w^{(\texttt{j})}_{\mu}$ 
if Eqs.~\eqref{simplifying condition}
hold.  These constraints reduce the total number of variables
contained in the tunneling vectors 
$\mathcal{T}^{\,}_{s}$ and 
$\mathcal{T}^{\,}_{p}$ from $8M^{2}$ to $4M^{2}$, 
and the number of nontrivial
equations to $2M(M+1)/2$, i.e.,
\begin{subequations}\label{eq: simplest 2D criteria}
\begin{align}
&
v^{(\texttt{j})\T}\, K\, v^{(\texttt{j}^{\prime})}= 0 ,
\\
&
w^{(\texttt{j})\T}\, K\, w^{(\texttt{j}^{\prime})}= 0 ,
\end{align}
\end{subequations}
which are merely rewritings of Eqs.~\eqref{v criterion 2D} and
\eqref{w criterion 2D}.  With this, we have arrived at the simplest
incarnation of our construction.  We will henceforth assume
that Eqs.~\eqref{eq: simplest 2D criteria} hold for appropriate choices of
the $2M$, $2M$-dimensional vectors 
$v^{(\texttt{j})}$ 
and 
$w^{(\texttt{j})}$.  
However, note that Eqs.~\eqref{simplifying condition} 
are sufficient but not necessary in order to produce a state of matter 
that has an isotropic low-energy description.
We will therefore comment, as appropriate, 
on how our results below generalize to cases where 
$v^{(\texttt{j})}_{1}\neq 
v^{(\texttt{j})}_{2}$ 
and 
$w^{(\texttt{j})}_{1}\neq w^{(\texttt{j})}_{2}$.

\subsection{Fractionalization}
\label{subsec: Fractionalization in the coupled wire array}

\subsubsection{Change of basis}
\label{subsubsec: Change of basis}

In this section, we outline how to use two-dimensional arrays of coupled
quantum wires,
like those described in the previous two sections, to study phases of matter 
with fractionalized excitations.  To this end, let us assume that we have a
Haldane set $\mathbb{H}$ containing $2MN$ tunneling vectors 
$\mathcal{T}^{(\texttt{j})}_{s}$
and $\mathcal{T}^{(\texttt{j})}_{p}$ 
with $\texttt{j}=1,\ldots,M$
defined by Eqs.\
\eqref{2d T-vecs charge conserving} 
that satisfy \eqref{simplifying condition}
and the Haldane criterion
\eqref{eq: simplest 2D criteria}.  
With these assumptions, the two-dimensional array of coupled quantum wires 
acquires a gap in the strong-coupling limit, 
yielding a three-dimensional gapped state of matter.
 
As discussed in the previous section, the phase of matter obtained in
this way is a system of strongly-interacting fermions or bosons.
However, for the purposes of studying fractionalization, it is
convenient to work in a basis where the ``fundamental" constituents of
each wire are not fermions or bosons, but (possibly fractionalized)
quasiparticles.  This is achieved by making the change of basis

\begin{subequations}
\label{eq: def W transformation}
\begin{align}
&
\tilde{\hat{\Phi}}(z)\:=
\mathcal{W}^{-1}\,\hat{\Phi}(z),
\\
&
\tilde{\mathcal{V}}:= 
\mathcal{W}^{\T}\,\mathcal{V}\,\mathcal{W},
\\
&
\tilde{K}\:= 
W^{\T}\,K\, W,
\\
&
\tilde{Q}\:= 
W^{\T}\, Q,
\\
&
\tilde{\mathcal{T}}\:= 
\mathcal{W}^{-1}\,\mathcal{T},
\end{align}
where
\begin{align}
\mathcal{W} \:= 
\mathbbm{1}^{\,}_{2N}\otimes W,
\end{align}
for some invertible $2M\times2M$ integer-valued matrix $W$.
\end{subequations}
This change of variables has several virtues.
First,
$\tilde{K}$ remains symmetric and integer valued.
Second, $\tilde{Q}$ remains integer valued.
Third, this change of variables leaves
the quantity $\mathcal{T}^{\T}\, \mathcal{K}\, \hat{\Phi}(z)$, 
which enters the argument of the
cosine terms in Eq.~\eqref{Lint}, invariant, i.e.,
\begin{align}
\tilde{\mathcal{T}}^{\T}\,\tilde{\mathcal{K}}\,\tilde{\hat{\Phi}}(z)= 
\mathcal{T}^{\T}\, \mathcal{K}\, \hat{\Phi}(z).
\end{align}
Thus, the linear transformation \eqref{eq: def W transformation} does
not change the character of the interaction itself, although it alters
the tunneling vector $\mathcal{T}$ and the $4MN$-dimensional vector
$\hat{\Phi}$ of bosonic fields.  Furthermore, one verifies that the
linear transformation \eqref{eq: def W transformation} does not alter the
compatibility criteria \eqref{eq: 2D criteria} or the quantity
$\mathcal{Q}^{\T}\,\mathcal{T}$ that determines the presence or absence
of charge or number-parity conservation.

Given the possibility of performing a change of basis of the form
\eqref{eq: def W transformation},
we may now take a different approach.  
Instead of viewing the wire construction as a theory, with the Lagrangian
(\ref{def L as sum L0 and LT}),
of scalar fields obeying
the commutation relations \eqref{eq: K-matrix commutator} with a
$K$-matrix $K^{\,}_{\mathrm{f}}$ [Eq.~\eqref{eq: def fermionic K}] 
for fermions or $K^{\,}_{\mathrm{b}}$ 
[Eq.~\eqref{eq: def bosonic K}] for bosons, 
we may also view it as a theory, with the Lagrangian
\begin{equation} \label{def tilde L as sum tilde L0 and tilde LT}
\tilde{\hat{L}}\:=
\tilde{\hat{L}}^{\,}_{0}+\tilde{\hat{L}}^{\,}_{\{\tilde{\mathcal{T}}\}},
\end{equation}
of scalar fields obeying the new equal-time commutation relations
\begin{align}
\label{tilde commutator}
\[
\partial^{\,}_{z}\tilde{\hat{\phi}}^{\,}_{j,\alpha}(z),
\tilde{\hat{\phi}}^{\,}_{\pri{j},\pri{\alpha}}(\pri{z})
\]&=
\mathrm{i}\,2\pi\, 
\delta^{\,}_{j \pri{j}}\, 
\tilde{K}^{-1}_{\alpha\pri{\alpha}}\, 
\delta(z-\pri{z}),
\end{align}
for $j,j'=1,\dots,2N$ and $\alpha,\pri{\alpha}=1,\cdots,2M$,
which are neither fermionic nor bosonic in nature.  We allow 
$\tilde{K}$ to be any symmetric, invertible, $2M\times 2M$ integer matrix, as
long as it is related to $K^{\,}_{\mathrm{f}}$ or $K^{\,}_{\mathrm{b}}$ by a
transformation of the form \eqref{eq: def W transformation}.  Interactions
between wires that yield a gapped state of matter can be constructed
by following the procedures of the previous section.  The $2M\,N$
integer tunneling vectors $\tilde{\mathcal{T}}^{(\texttt{j})}_{s}$ and
$\tilde{\mathcal{T}}^{(\texttt{j})}_{p}$ obtained in this way form a
Haldane set $\tilde{\mathbb{H}}$ related to the Haldane set $\mathbb{H}$
by the transformation \eqref{eq: def W transformation}.  For reasons of
simplicity that will become clear momentarily, we will concern
ourselves in this paper primarily with the tunneling vectors
$\tilde{\mathcal{T}}^{(\texttt{j})}_{s}$ 
and
$\tilde{\mathcal{T}}^{(\texttt{j})}_{p}$ 
whose nonzero entries are equal to $\pm 1$.  
(Of course, nothing prevents us from also
considering cases where this does not hold.)  The counterparts
$\mathcal{T}^{(\texttt{j})}_{s}$ 
and
$\mathcal{T}^{(\texttt{j})}_{p}$ 
of these tunneling vectors under the
transformation \eqref{eq: def W transformation} generically have entries with
magnitude larger than 1.  This fact will be of importance to us now,
as we turn to the issue of compactification.

\subsubsection{Compactification, vertex operators, and fractional charges}
\label{subsubsec: Compactification, vertex operators, and fractional charges}

Although the transformation \eqref{eq: def W transformation} might
appear innocuous, there is a fundamental difference between the theory
with the Lagrangian $\tilde{\hat{L}}$ 
defined in Eq.\ (\ref{def tilde L as sum tilde L0 and tilde LT})
and the original fermionic or bosonic theory
with the Lagrangian $\hat{L}$
defined in Eq.\ (\ref{def L as sum L0 and LT})
when periodic boundary conditions are imposed in the $z$-direction (as
we have assumed from the outset).  In the latter theory, which is a
theory of interacting electrons or bosons treated within bosonization,
the traditional choice of compactification for the scalar fields
$\hat{\phi}^{\,}_{j,\alpha}(z)$ 
with $j=1,\ldots,2N$ and $\alpha=1,\cdots,2M$
is
\begin{align}
\hat{\phi}^{\,}_{j,\alpha}(z+L)&\equiv
\hat{\phi}^{\,}_{j,\alpha}(z)+2\pi\,n^{\,}_{\alpha},
\label{eq: traditional compactification}
\end{align}
for $n^{\,}_{\alpha}\in\mathbb{Z}^{2M}$.
This choice ensures the
single-valuedness of the fermionic or bosonic vertex operators
\eqref{fermi-bose vertex ops}
under $z\to z+L$, and, in turn, that of
the Lagrangian 
$\hat{L}=\hat{L}^{\,}_{0}+\hat{L}^{\,}_{\{\mathcal{T}\}}$, 
as one can re-write
$\hat{L}^{\,}_{\{\mathcal{T}\}}$ in terms of the correlated tunnelings
\eqref{eq: tunneling}, which reduce to products of these vertex operators.
However, depending on the tunneling vectors
$\mathcal{T}^{(\texttt{j})}_{s,p}$, there may be other, less
stringent, compactifications of these scalar fields that also render
the Lagrangian $\hat{L}$ single-valued under $z\to z+L$.  
The parsimonious course of action is to choose the ``minimal" compactification, 
i.e., the smallest
compactification radius that still maintains the single-valuedness of
$\hat{L}$ under $z\to z+L$.  

If the tunneling vectors 
$\mathcal{T}^{(\texttt{j})}_{s}$ 
and
$\mathcal{T}^{(\texttt{j})}_{p}$
correspond
to the tunneling vectors 
$\tilde{\mathcal{T}}^{(\texttt{j})}_{s}$ 
and
$\tilde{\mathcal{T}}^{(\texttt{j})}_{p}$ 
under
the transformation \eqref{eq: def W transformation} whose only nonzero
entries are equal to $\pm 1$, then there is a clear choice of minimal
compactification. This choice can be obtained as follows.  Working in the
tilde basis, we can rewrite the interactions using the relation
[analogous to \eqref{eq: tunneling}]
\begin{subequations}
\begin{align}
e^{-\mathrm{i}\, \tilde{\mathcal{T}}^{\T}\,\tilde{\mathcal{K}}\,\tilde{\hat{\Phi}}(z)}=
\prod_{j=1}^{2N}\prod_{\alpha=1}^{2M}
\[\tilde{\hat{\psi}}^{\dag}_{\mathrm{f,b};j,\alpha}(z)\]^{\tilde{\mathcal{T}}^{\,}_{j,\alpha}},
\label{eq: tilde tunneling}
\end{align}
thereby implicitly defining a new set of fermionic or bosonic vertex operators,
\begin{align}
\tilde{\hat{\psi}}^{\dag}_{\mathrm{f,b}; j,\alpha}(z) \:= 
\exp
\(
-\mathrm{i}\, 
\sum_{\pri{\alpha}=1}^{2M}
\tilde{K}^{\,}_{\alpha\pri{\alpha}}\, 
\tilde{\hat{\phi}}^{\,}_{j,\pri{\alpha}}(z)
\).
\label{tilde fermi-bose vertex ops}
\end{align}
\end{subequations}
The minimal compactification
is then obtained by demanding that this
new set of vertex operators be single valued
under $z\to z+L$.  For any $j=1,\ldots,2N$ and $\alpha=1,\cdots,2M$,
this is achieved by imposing the periodic boundary conditions
\begin{align}\label{eq: minimal compactification}
\tilde{\hat{\phi}}_{j,\alpha}(z+L) &\equiv 
\tilde{\hat{\phi}}^{\,}_{j,\alpha}(z)
+
2\pi\,\tilde{K}^{-1}_{\alpha\pri{\alpha}}\,n^{\,}_{\pri{\alpha}},
\end{align}
for $n^{\,}_{\alpha}\in\mathbb{Z}^{2M}$.  Here, there is an important
difference with respect to Eq.\
\eqref{eq: traditional compactification}. 
Because $\tilde{K}$ is an integer-valued matrix, 
$\tilde{K}^{-1}$ is generically a rational-valued matrix.
The field
$\tilde{\hat{\phi}}^{\,}_{j,\alpha}(z)$ is thus allowed to advance by
rational (rather than integer) multiples of $2\pi$ when the coordinate
$z$ is advanced through a full period $L$.  This crucial distinction
is what allows for the existence of fractionally-charged operators in
the coupled wire array, as we now demonstrate.

Fractional quantum numbers appear in the wire construction because the
compactification condition \eqref{eq: minimal compactification} allows for
the existence of ``quasiparticle" vertex operators
\begin{align}
\hat q^{\dag}_{j,\alpha}(z) \:= 
\exp
\(
-\mathrm{i}\, 
\tilde{\hat{\phi}}^{\,}_{j,\alpha}(z)
\)
\label{quasiparticle vertex ops}
\end{align}
for any $j=1,\ldots,2N$ and $\alpha=1,\cdots,2M$
that are multivalued under the operation $z\mapsto z+L$.  The fact
that these vertex operators generically carry fractional charges
can be seen by considering the transformed charge operator
\begin{align}\label{Q tilde}
\tilde{\hat{Q}}^{\,}_{j,\alpha}\:=
\frac{\tilde{Q}^{\,}_{\alpha}}{2\pi}\,
\sum_{\pri{\alpha}=1}^{2M}
\delta^{\,}_{\alpha\pri{\alpha}}
\int\limits^{L}_{0}\mathrm{d}z\ 
\partial^{\,}_{z}\, 
\tilde{\hat{\phi}}^{\,}_{j,\pri{\alpha}}(z)
\end{align}
for any $j=1,\ldots,2N$ and $\alpha=1,\cdots,2M$.
Its normalization is here chosen
such that the fermionic or bosonic vertex operators defined in 
Eq.\ \eqref{tilde fermi-bose vertex ops} 
have charge $\tilde{Q}^{\,}_{\alpha}$. Indeed,
for any $j,j'=1,\ldots,2N$ and $\alpha,\alpha'=1,\cdots,2M$, 
the equal-time commutator
\begin{align}
\[
\tilde{\hat{Q}}^{\,}_{j,\alpha}, 
\hat{q}^{\dag}_{\pri{j},\pri{\alpha}}(z)
\]= 
\tilde{Q}^{\,}_\alpha\, 
\delta^{\,}_{j\pri{j}}\, 
\tilde{K}^{-1}_{\alpha\pri{\alpha}}\, 
\hat{q}^{\dag}_{\pri{j},\pri{\alpha}}
\label{eq: q is eigenstate of charge Q}
\end{align}
indicates that, since $\tilde{K}^{-1}$ is generically a rational
matrix, the quasiparticle operator $\hat{q}^{\dag}_{\pri{j},\pri{\alpha}}$ 
generically has a rational charge.  In
particular, if $\tilde{Q}= Q$ under the transformation 
\eqref{eq: def W transformation}, and $\tilde{K}^{-1}$ has at least one rational
entry with magnitude smaller than 1, the operator 
$\hat{q}^{\dag}_{\pri{j},\pri{\alpha}}$ must then carry a fractional charge.

\begin{figure}[t]
\begin{center}
\includegraphics[width=.4\textwidth]{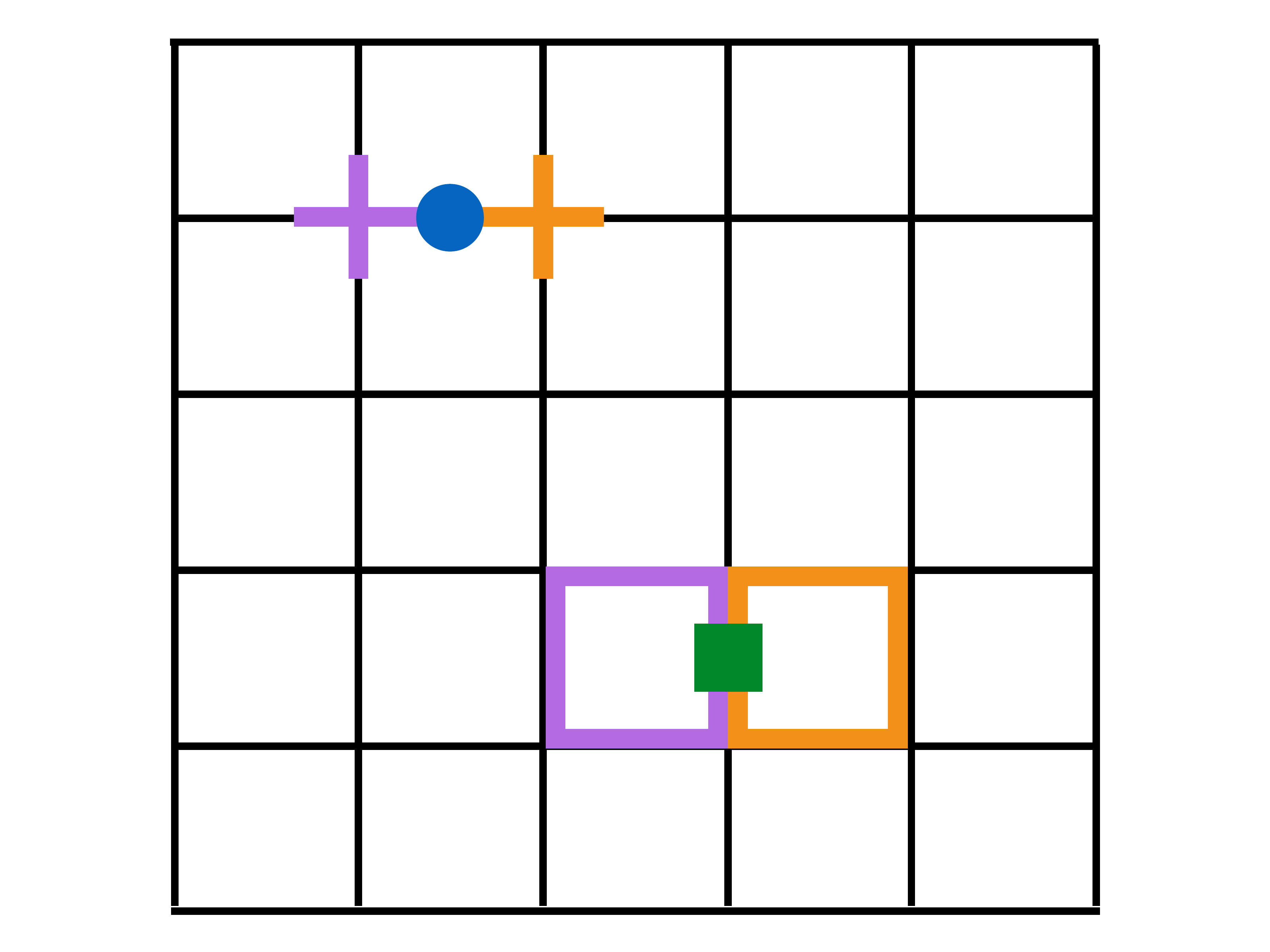}\\
\caption{(Color online) 
Pictorial representation of star and plaquette excitations
created by the operators
(\ref{eq: 2D defect hopping}).  
The filled blue circle represents an application of the
vertex operator \eqref{eq: 2D defect hopping a} in the corresponding wire,
while the purple and orange crosses represent defective stars hosting solitons
of opposite signs.  
Similarly, the filled green square represents an application of the vertex
operator \eqref{eq: 2D defect hopping b}, and the purple and orange squares
represent defective plaquettes hosting solitons of opposite signs.
\label{fig: star and plaquette excitations}
         }
\end{center}
\end{figure}

\begin{figure*}
\centering
(a)\includegraphics[width=0.3\textwidth]{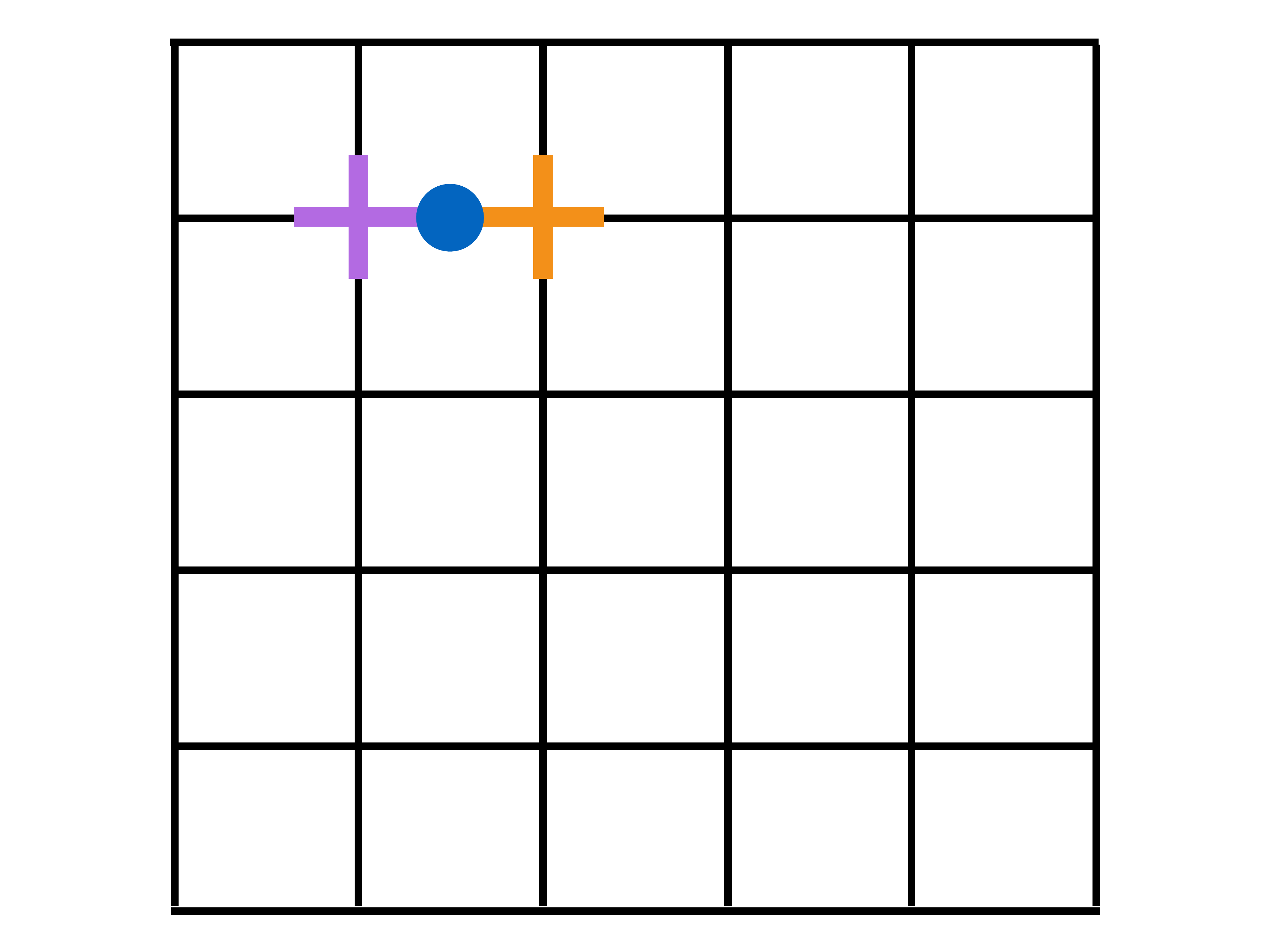}
(b)\includegraphics[width=0.3\textwidth]{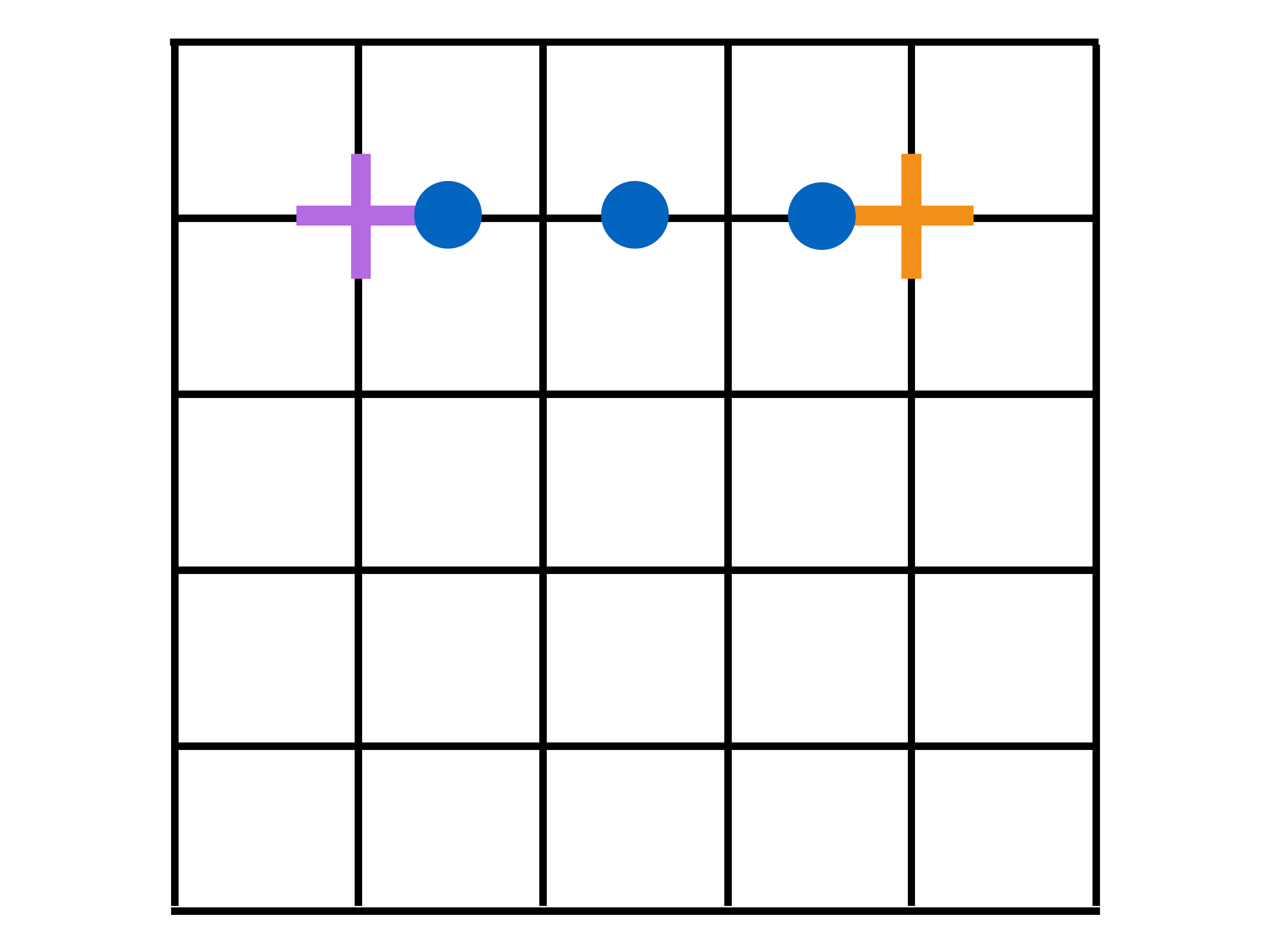}\\
(c)\includegraphics[width=0.3\textwidth]{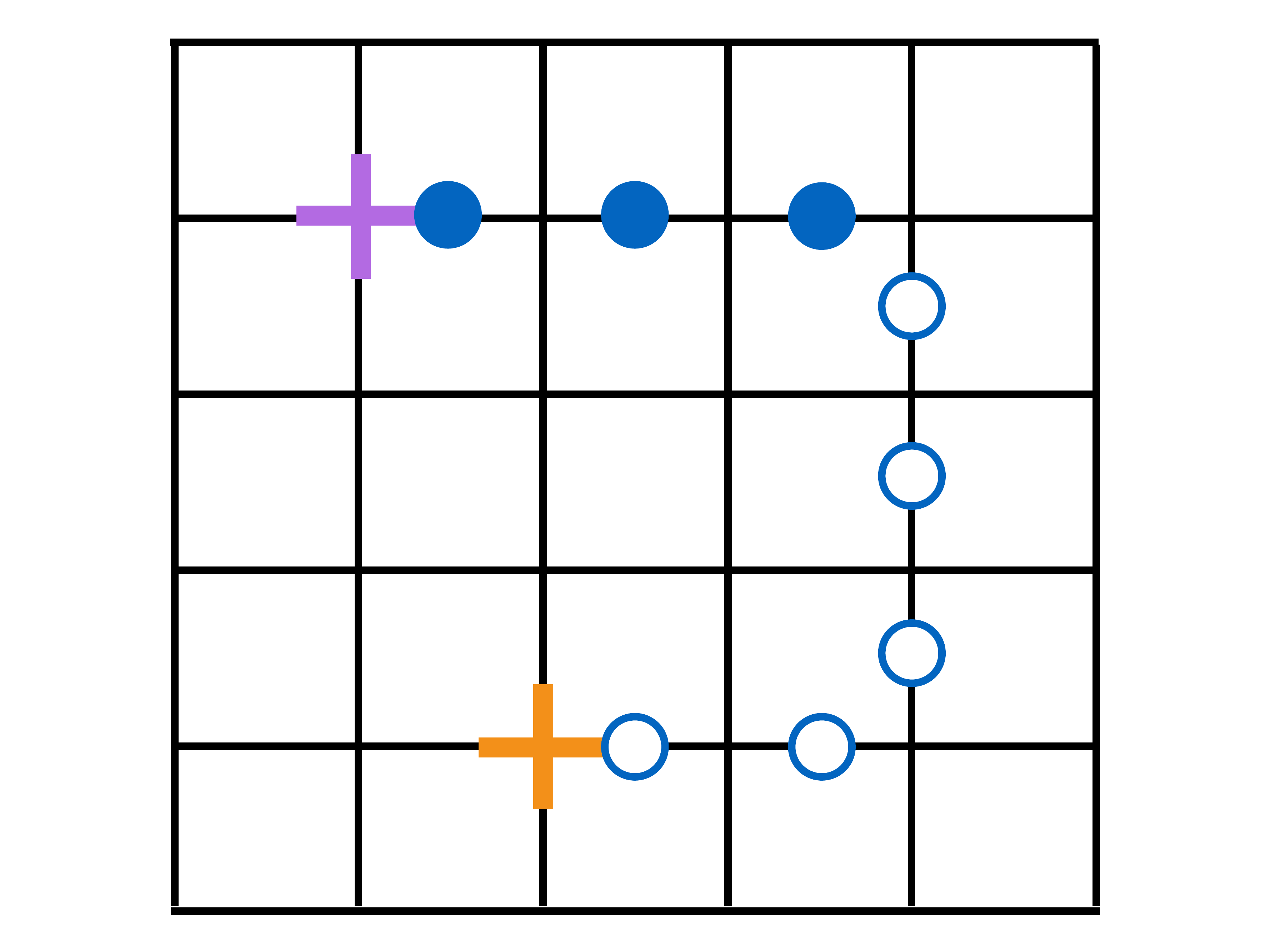}
(d)\includegraphics[width=0.3\textwidth]{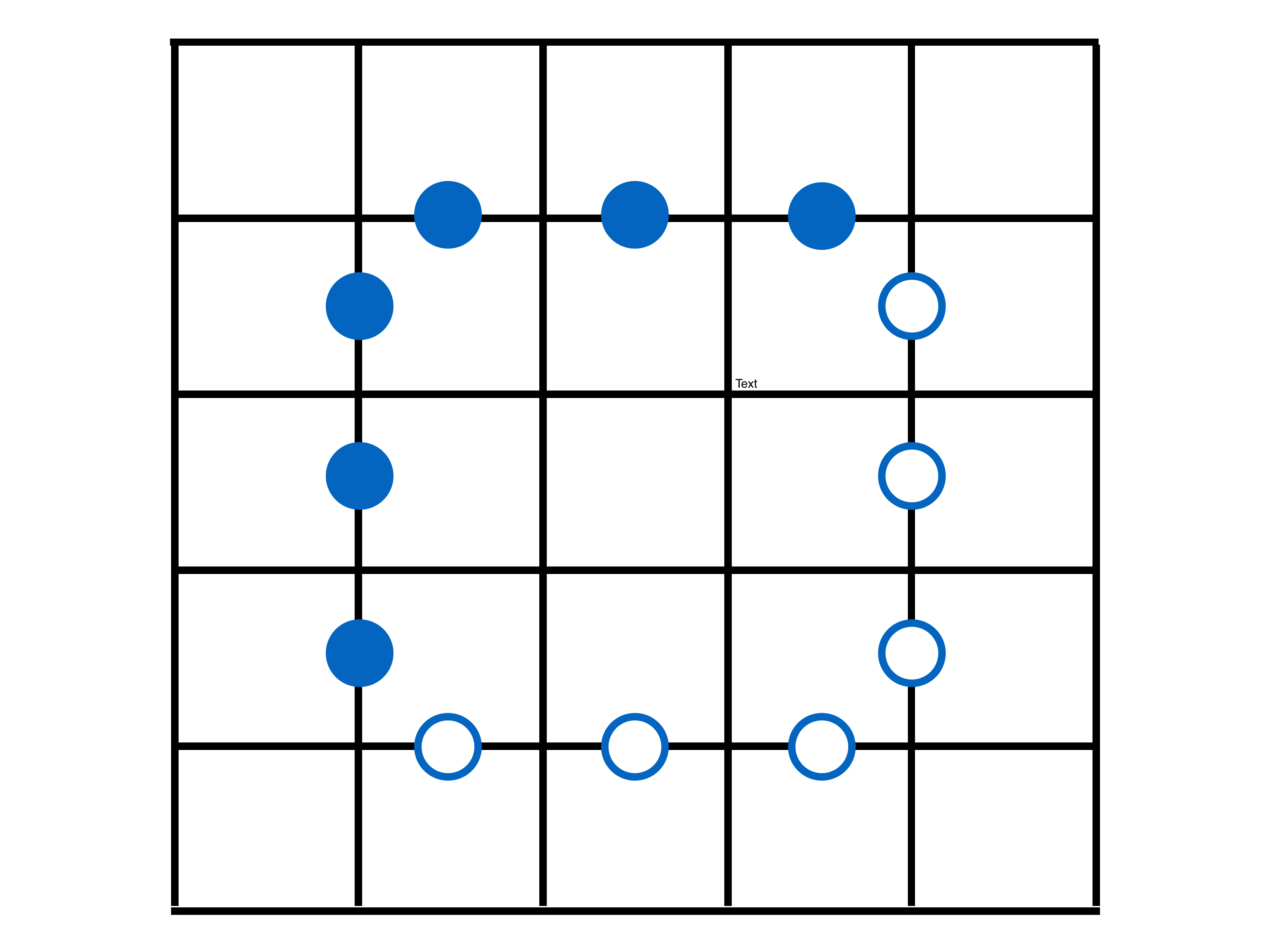}
\caption{(Color online) 
Deconfinement of star defects within a plane.
(a) A single application of the vertex operator \eqref{eq: 2D defect hopping a}  
(blue circle) creates a star defect (purple cross) and an anti-star defect (orange cross).
(b) Applying a string of vertex operators \eqref{eq: 2D defect hopping a} moves a star defect by
healing one while creating another. Consequently, it costs no extra
energy to separate the two star defects.  
(c) To turn a corner, 
it may be necessary to heal a defect with an application of
the inverse of the vertex operator \eqref{eq: 2D defect hopping a} in the
appropriate wire (white circle).  
(d) When two star defects meet, they annihilate one
another.  A completely analogous description of plaquette
defects also holds starting from the operators
Eq.\ (\ref{eq: 2D defect hopping b}).
        }
\label{fig: deconfined star defects}
\end{figure*}

\subsubsection{Pointlike and linelike excitations}
\label{subsubsec: Pointlike and linelike excitations}

We now outline the relationship between the quasiparticle vertex
operators defined in Eq.\ \eqref{quasiparticle vertex ops} and
(possibly fractionalized) excitations in the array of coupled quantum wires.  
In the strong-coupling limit
$U^{\,}_{\tilde{\mathcal{T}}}(z)\to\infty$, the compatibility criteria
\eqref{eq: 2D criteria} ensure that the quantity
\begin{align}
\tilde{\mathcal{T}}^{\T}\,
\tilde{\mathcal{K}}\,
\tilde{\hat{\Phi}}(z)
+
\alpha^{\,}_{\tilde{\mathcal{T}}}(z),
\end{align}
where 
$\tilde{\mathcal{T}}=
\tilde{\mathcal{T}}^{(\texttt{j})}_{s}$ or 
$\tilde{\mathcal{T}}^{(\texttt{j})}_{p}$,
is pinned to a classical minimum of the corresponding cosine potential
in $\hat{L}^{\,}_{\{\tilde{\mathcal{T}}\}}$.  Following
Refs.~\cite{Kane02,Teo14} and subsequent works, we identify
excitations in the coupled-wire theory with solitons that increment
the ``pinned field"
$\tilde{\mathcal{T}}\,\tilde{\mathcal{K}}\,\tilde{\hat{\Phi}}(z)$ by an
integer multiple of $2\pi$.  These excitations can therefore be viewed
as living on either the stars or the plaquettes 
of the square lattice, rather than within the wires themselves.

We now demonstrate that products of an appropriate number of
quasiparticle vertex operators of the form 
\eqref{quasiparticle vertex ops} 
can be used to move the soliton defects to adjacent stars and
plaquettes.  To see this, we write out the pinned fields
explicitly for all $MN$ tunneling vectors
$\tilde{\mathcal{T}}^{(\texttt{j})}_{s}$
with $\texttt{j}=1,\ldots,M$
defined on the stars $s=1,\ldots,N$,
\begin{widetext}
\begin{subequations}
\begin{align}
&
\tilde{\mathcal{T}}^{(\texttt{j})\mathsf{T}}_{s}\,
\tilde{\mathcal{K}}\,
\tilde{\hat{\Phi}}(z)=
\sum_{\alpha,\alpha'=1}^{2M}
\tilde{v}^{(\texttt{j})}_{\alpha}\, 
\tilde{K}^{\,}_{\alpha\pri{\alpha}}\,
\[
\tilde{\hat{\phi}}^{\,}_{s^{\,}_{E},\pri{\alpha}}(z)
-
\tilde{\hat{\phi}}^{\,}_{s^{\,}_{W},\pri{\alpha}}(z)
+
\tilde{\hat{\phi}}^{\,}_{s^{\,}_{N},\pri{\alpha}}(z)
-
\tilde{\hat{\phi}}^{\,}_{s^{\,}_{S},\pri{\alpha}}(z)
\],
\end{align}
and for all $MN$ tunneling vectors
$\tilde{\mathcal{T}}^{(\texttt{j})}_{p}$
with $\texttt{j}=1,\ldots,M$
defined on the plaquettes $p=1,\ldots,N$,
\begin{align}
&
\tilde{\mathcal{T}}^{(\texttt{j})\mathsf{T}}_{p}\,
\tilde{\mathcal{K}}\,
\tilde{\hat{\Phi}}(z)=
\sum_{\alpha,\alpha'=1}^{2M}
\tilde{w}^{(\texttt{j})}_{\alpha}\, 
\tilde{K}^{\,}_{\alpha\pri{\alpha}}
\[
\tilde{\hat{\phi}}_{p^{\,}_{W},\pri{\alpha}}(z)
-
\tilde{\hat{\phi}}_{p^{\,}_{E},\pri{\alpha}}(z)
+
\tilde{\hat{\phi}}^{\,}_{p^{\,}_{N},\pri{\alpha}}(z)
-
\tilde{\hat{\phi}}^{\,}_{p^{\,}_{S},\pri{\alpha}}(z)
\].
\end{align}
\end{subequations}
For any star $s=1,\ldots,N$ or plaquette $p=1,\ldots,N$ from the square lattice
and for any $\alpha,\beta=1,\ldots,2M$, 
observe that, by Eq.~\eqref{tilde commutator},
the equal-time commutators
\begin{align}
\left[
\sum_{\alpha'=1}^{2M}
\tilde{K}^{\,}_{\alpha\pri{\alpha}}\,
\tilde{\hat{\phi}}^{\,}_{s^{\,}_{C},\pri{\alpha}}(z),
\partial^{\,}_{\pri{z}}
\tilde{\hat{\phi}}^{\,}_{s^{\,}_{C},\beta}(\pri{z})
\right]
= 
\left[
\sum_{\alpha'=1}^{2M}
\tilde{K}^{\,}_{\alpha\pri{\alpha}}\,
\tilde{\hat{\phi}}^{\,}_{p^{\,}_{C},\pri{\alpha}}(z),
\partial^{\,}_{\pri{z}}
\tilde{\hat{\phi}}^{\,}_{p^{\,}_{C},\beta}(\pri{z})
\right]= 
-\mathrm{i}\,2\pi\,\delta^{\,}_{\alpha\beta}\,
\delta(z-\pri{z})
\label{eq: 2D conjugates to pinning fields}
\end{align}
hold.
Here, the uppercase Latin index $C=N,W,S,E$ labels the four cardinal directions.
Equation (\ref{eq: 2D conjugates to pinning fields})
indicates that the pair of fields 
$\partial^{\,}_{z}\tilde{\hat{\phi}}^{\,}_{s^{\,}_{C},\alpha}$ 
and 
$\partial^{\,}_{z}\tilde{\hat{\phi}}^{\,}_{p^{\,}_{C},\alpha}$ can be
viewed, up to  a multiplicative constant,
as canonical conjugates to the pair of fields 
$\sum_{\pri{\alpha}}\tilde{K}^{\,}_{\alpha\pri{\alpha}}\,
\tilde{\hat{\phi}}^{\,}_{s^{\,}_{C},\pri{\alpha}}$
and 
$\sum_{\pri{\alpha}}\tilde{K}^{\,}_{\alpha\pri{\alpha}}\,
\tilde{\hat{\phi}}^{\,}_{p^{\,}_{C},\pri{\alpha}}(z)$
that enter the pair of pinned fields
$\tilde{\mathcal{T}}^{(\texttt{j})\T}_{s}\,
\tilde{\mathcal{K}}\,\tilde{\hat{\Phi}}$
and
$\tilde{\mathcal{T}}^{(\texttt{j})\T}_{p}\,
\tilde{\mathcal{K}}\,\tilde{\hat{\Phi}}$,
respectively.
Interpreted this way, Eqs.\ 
\eqref{eq: 2D conjugates to pinning fields} 
suggest that, 
for any $\texttt{j}=1,\ldots,M$,
$s^{\,}_{C}=s^{\,}_{N},s^{\,}_{W},s^{\,}_{S},s^{\,}_{E}$,
and
$p^{\,}_{C}=p^{\,}_{N},p^{\,}_{W},p^{\,}_{S},p^{\,}_{E}$,
the operators
\begin{subequations}
\label{eq: 2D defect hopping}
\begin{align}\label{eq: 2D defect hopping a}
\hat{S}^{(\texttt{j})\dag}_{s^{\,}_{C}}(z)\:= 
\exp
\(
-\mathrm{i}\, 
\sum_{\alpha=1}^{2M}
\tilde{v}^{(\texttt{j})}_{\alpha}\, 
\tilde{\hat{\phi}}^{\,}_{s^{\,}_{C},\alpha}(z)
\)
\end{align}
and
\begin{align}\label{eq: 2D defect hopping b}
\hat{P}^{(\texttt{j})\dag}_{p^{\,}_{C}}(z)\:= 
\exp
\(
-\mathrm{i}\, 
\sum_{\alpha=1}^{2M}
\tilde{w}^{(\texttt{j})}_{\alpha}\, 
\tilde{\hat{\phi}}^{\,}_{p^{\,}_{C},\alpha}(z)
\)
\end{align}
\end{subequations}
act on the pinned fields as
[$|\tilde{v}^{(\texttt{j})}|^{2}$ denotes the magnitude squared
of the vector $\tilde{v}^{(\texttt{j})}\in\mathbb{Z}^{2M}$]
\begin{subequations}
\label{eq: 2D pinning soliton star}
\begin{align}
&
\hat{S}^{(\texttt{j})}_{s^{\,}_{N}}(\pri{z})\ 
\[
\tilde{\mathcal{T}}^{(\texttt{j})\mathsf{T}}_{s}\,
\tilde{\mathcal{K}}\,
\tilde{\hat{\Phi}}(z)
\]\ 
\hat{S}^{(\texttt{j})\dag}_{s^{\,}_{N}}(\pri{z})=
\tilde{\mathcal{T}}^{(\texttt{j})\mathsf{T}}_{s}\,
\tilde{\mathcal{K}}\,
\tilde{\hat{\Phi}}(z)
+
2\pi\, 
|\tilde{v}^{(\texttt{j})}|^{2}\,
\Theta(z-\pri{z})
+ 
\text{constant},
\\ 
&
\hat{S}^{(\texttt{j})}_{s^{\,}_{W}}(\pri{z})\ 
\[
\tilde{\mathcal{T}}^{(\texttt{j})\mathsf{T}}_{s}\,
\tilde{\mathcal{K}}\,
\tilde{\hat{\Phi}}(z)
\]\ 
\hat{S}^{(\texttt{j})\dag}_{s^{\,}_{W}}(\pri{z})=
\tilde{\mathcal{T}}^{(\texttt{j})\mathsf{T}}_{s}\,
\tilde{\mathcal{K}}\,
\tilde{\hat{\Phi}}(z)
-
2\pi\, 
|\tilde{v}^{(\texttt{j})}|^{2}\,
\Theta(z-\pri{z})
+ 
\text{constant},
\\ 
&
\hat{S}^{(\texttt{j})}_{s^{\,}_{S}}(\pri{z})\ 
\[
\tilde{\mathcal{T}}^{(\texttt{j})\mathsf{T}}_{s}\,
\tilde{\mathcal{K}}\,
\tilde{\hat{\Phi}}(z)
\]\ 
\hat{S}^{(\texttt{j})\dag}_{s^{\,}_{S}}(\pri{z})=
\tilde{\mathcal{T}}^{(\texttt{j})\mathsf{T}}_{s}\,
\tilde{\mathcal{K}}\,
\tilde{\hat{\Phi}}(z)
-
2\pi\, 
|\tilde{v}^{(\texttt{j})}|^{2}\,
\Theta(z-\pri{z})
+ 
\text{constant},
\\ 
&
\hat{S}^{(\texttt{j})}_{s^{\,}_{E}}(\pri{z})\ 
\[
\tilde{\mathcal{T}}^{(\texttt{j})\mathsf{T}}_{s}\,
\tilde{\mathcal{K}}\,
\tilde{\hat{\Phi}}(z)
\]\ 
\hat{S}^{(\texttt{j})\dag}_{s^{\,}_{E}}(\pri{z})=
\tilde{\mathcal{T}}^{(\texttt{j})\mathsf{T}}_{s}\,
\tilde{\mathcal{K}}\,
\tilde{\hat{\Phi}}(z)
+
2\pi\, 
|\tilde{v}^{(\texttt{j})}|^{2}\,
\Theta(z-\pri{z})
+ 
\text{constant},
\end{align}
\end{subequations}
and
[$|\tilde{w}^{(\texttt{j})}|^{2}$ denotes the magnitude squared
of the vector $\tilde{w}^{(\texttt{j})}\in\mathbb{Z}^{2M}$]
\begin{subequations}
\label{eq: 2D pinning soliton plaquette}
\begin{align}
&
\hat{P}^{(\texttt{j})}_{p^{\,}_{N}}(\pri{z})\ 
\[
\tilde{\mathcal{T}}^{(\texttt{j})\mathsf{T}}_{p}\,
\tilde{\mathcal{K}}\,
\tilde{\hat{\Phi}}(z)
\]\ 
\hat{P}^{(\texttt{j})\dag}_{p^{\,}_{N}}(\pri{z})=
\tilde{\mathcal{T}}^{(\texttt{j})\mathsf{T}}_{p}\,
\tilde{\mathcal{K}}\,
\tilde{\hat{\Phi}}(z)
+
2\pi\, 
|\tilde{w}^{(\texttt{j})}|^{2}\,
\Theta(z-\pri{z})
+ 
\text{constant},
\\ 
&
\hat{P}^{(\texttt{j})}_{p^{\,}_{W}}(\pri{z})\ 
\[
\tilde{\mathcal{T}}^{(\texttt{j})\mathsf{T}}_{p}\,
\tilde{\mathcal{K}}\,
\tilde{\hat{\Phi}}(z)
\]\ 
\hat{P}^{(\texttt{j})\dag}_{p^{\,}_{W}}(\pri{z})=
\tilde{\mathcal{T}}^{(\texttt{j})\mathsf{T}}_{p}\,
\tilde{\mathcal{K}}\,
\tilde{\hat{\Phi}}(z)
+
2\pi\, 
|\tilde{w}^{(\texttt{j})}|^{2}\,
\Theta(z-\pri{z})
+ 
\text{constant},
\\ 
&
\hat{P}^{(\texttt{j})}_{p^{\,}_{S}}(\pri{z})\ 
\[
\tilde{\mathcal{T}}^{(\texttt{j})\mathsf{T}}_{p}\,
\tilde{\mathcal{K}}\,
\tilde{\hat{\Phi}}(z)
\]\ 
\hat{P}^{(\texttt{j})\dag}_{p^{\,}_{S}}(\pri{z})=
\tilde{\mathcal{T}}^{(\texttt{j})\mathsf{T}}_{p}\,
\tilde{\mathcal{K}}\,
\tilde{\hat{\Phi}}(z)
-
2\pi\, 
|\tilde{w}^{(\texttt{j})}|^{2}\,
\Theta(z-\pri{z})
+ 
\text{constant},
\\ 
&
\hat{P}^{(\texttt{j})}_{p^{\,}_{E}}(\pri{z})\ 
\[
\tilde{\mathcal{T}}^{(\texttt{j})\mathsf{T}}_{p}\,
\tilde{\mathcal{K}}\,
\tilde{\hat{\Phi}}(z)
\]\ 
\hat{P}^{(\texttt{j})\dag}_{p^{\,}_{E}}(\pri{z})=
\tilde{\mathcal{T}}^{(\texttt{j})\mathsf{T}}_{p}\,
\tilde{\mathcal{K}}\,
\tilde{\hat{\Phi}}(z)
-
2\pi\, 
|\tilde{w}^{(\texttt{j})}|^{2}\,
\Theta(z-\pri{z})
+ 
\text{constant},
\end{align}
\end{subequations}
\end{widetext}
respectively.
To verify
Eqs.\ \eqref{eq: 2D pinning soliton star},
one integrates both sides of
the equalities entering
Eq.\ \eqref{eq: 2D conjugates to pinning fields} over the variable
$\pri{z}$ and uses the identity
\begin{equation}
\left(\frac{\mathrm{d}\Theta}{\mathrm{d}z}\right)(z)=
\delta(z),
\end{equation}  
where $\Theta(z)$ is the Heaviside step function.
If the underlying quantum wires in the theory are fermionic,
the arbitrary integration constants above are identified with Klein factors
that are necessary in order to ensure the anticommutation of fermionic vertex
operators in different wires.  If the underlying wires are bosonic, however,
the integration constants can be set to zero.

Evidently, the operators $\hat{S}^{(\texttt{j})\dag}_{s^{\,}_{C}}$ and
$\hat{P}^{(\texttt{j})\dag}_{p^{\,}_{C}}$ defined in Eqs.\
\eqref{eq: 2D defect hopping} 
create $2\pi$ solitons in the pinned fields
$\tilde{\mathcal{T}}^{(\texttt{j})\T}_{s}\,\tilde{\mathcal{K}}\,\tilde{\hat{\Phi}}$
and
$\tilde{\mathcal{T}}^{(\texttt{j})\T}_{p}\,\tilde{\mathcal{K}}\,\tilde{\hat{\Phi}}$,
respectively.  However, the link $s^{\,}_{C}$ on star $s$ is shared with the
star $\pri{s}$ adjacent to $s$ along the cardinal direction
$C=N,W,S,E$, 
and, likewise,
the link $p^{\,}_{C}$ on plaquette $p$ is shared with the plaquette $\pri{p}$ 
adjacent to $p$ along the cardinal direction $C=N,W,S,E$.
Therefore,
\begin{widetext}
\begin{subequations}
\label{eq: 2D pinning soliton star prime}
\begin{align}
&
\hat{S}^{(\texttt{j})}_{s^{\,}_{N}}(\pri{z})\ 
\[
\tilde{\mathcal{T}}^{(\texttt{j})\mathsf{T}}_{\pri{s}}\,
\tilde{\mathcal{K}}\,
\tilde{\hat{\Phi}}(z)
\]\ 
\hat{S}^{(\texttt{j})\dag}_{s^{\,}_{N}}(\pri{z})=
\tilde{\mathcal{T}}^{(\texttt{j})\mathsf{T}}_{\pri{s}}\,
\tilde{\mathcal{K}}\,
\tilde{\hat{\Phi}}(z)
-
2\pi\, 
|\tilde{v}^{(\texttt{j})}|^{2}\,
\Theta(z-\pri{z})
+ 
\text{constant},
\\ 
&
\hat{S}^{(\texttt{j})}_{s^{\,}_{W}}(\pri{z})\ 
\[
\tilde{\mathcal{T}}^{(\texttt{j})\mathsf{T}}_{\pri{s}}\,
\tilde{\mathcal{K}}\,
\tilde{\hat{\Phi}}(z)
\]\ 
\hat{S}^{(\texttt{j})\dag}_{s^{\,}_{W}}(\pri{z})=
\tilde{\mathcal{T}}^{(\texttt{j})\mathsf{T}}_{\pri{s}}\,
\tilde{\mathcal{K}}\,
\tilde{\hat{\Phi}}(z)
+
2\pi\, 
|\tilde{v}^{(\texttt{j})}|^{2}\,
\Theta(z-\pri{z})
+ 
\text{constant},
\\ 
&
\hat{S}^{(\texttt{j})}_{s^{\,}_{S}}(\pri{z})\ 
\[
\tilde{\mathcal{T}}^{(\texttt{j})\mathsf{T}}_{\pri{s}}\,
\tilde{\mathcal{K}}\,
\tilde{\hat{\Phi}}(z)
\]\ 
\hat{S}^{(\texttt{j})\dag}_{s^{\,}_{S}}(\pri{z})=
\tilde{\mathcal{T}}^{(\texttt{j})\mathsf{T}}_{\pri{s}}\,
\tilde{\mathcal{K}}\,
\tilde{\hat{\Phi}}(z)
+
2\pi\, 
|\tilde{v}^{(\texttt{j})}|^{2}\,
\Theta(z-\pri{z})
+ 
\text{constant},
\\ 
&
\hat{S}^{(\texttt{j})}_{s^{\,}_{E}}(\pri{z})\ 
\[
\tilde{\mathcal{T}}^{(\texttt{j})\mathsf{T}}_{\pri{s}}\,
\tilde{\mathcal{K}}\,
\tilde{\hat{\Phi}}(z)
\]\ 
\hat{S}^{(\texttt{j})\dag}_{s^{\,}_{E}}(\pri{z})=
\tilde{\mathcal{T}}^{(\texttt{j})\mathsf{T}}_{\pri{s}}\,
\tilde{\mathcal{K}}\,
\tilde{\hat{\Phi}}(z)
-
2\pi\, 
|\tilde{v}^{(\texttt{j})}|^{2}\,
\Theta(z-\pri{z})
+ 
\text{constant},
\end{align}
\end{subequations}
and
\begin{subequations}
\label{eq: 2D pinning soliton plaquette prime}
\begin{align}
&
\hat{P}^{(\texttt{j})}_{p^{\,}_{N}}(\pri{z})\ 
\[
\tilde{\mathcal{T}}^{(\texttt{j})\mathsf{T}}_{\pri{p}}\,
\tilde{\mathcal{K}}\,
\tilde{\hat{\Phi}}(z)
\]\ 
\hat{P}^{(\texttt{j})\dag}_{p^{\,}_{N}}(\pri{z})=
\tilde{\mathcal{T}}^{(\texttt{j})\mathsf{T}}_{\pri{p}}\,
\tilde{\mathcal{K}}\,
\tilde{\hat{\Phi}}(z)
-
2\pi\, 
|\tilde{w}^{(\texttt{j})}|^{2}\,
\Theta(z-\pri{z})
+ 
\text{constant},
\\ 
&
\hat{P}^{(\texttt{j})}_{p^{\,}_{W}}(\pri{z})\ 
\[
\tilde{\mathcal{T}}^{(\texttt{j})\mathsf{T}}_{\pri{p}}\,
\tilde{\mathcal{K}}\,
\tilde{\hat{\Phi}}(z)
\]\ 
\hat{P}^{(\texttt{j})\dag}_{p^{\,}_{W}}(\pri{z})=
\tilde{\mathcal{T}}^{(\texttt{j})\mathsf{T}}_{\pri{p}}\,
\tilde{\mathcal{K}}\,
\tilde{\hat{\Phi}}(z)
-
2\pi\, 
|\tilde{w}^{(\texttt{j})}|^{2}\,
\Theta(z-\pri{z})
+ 
\text{constant},
\\ 
&
\hat{P}^{(\texttt{j})}_{p^{\,}_{S}}(\pri{z})\ 
\[
\tilde{\mathcal{T}}^{(\texttt{j})\mathsf{T}}_{\pri{p}}\,
\tilde{\mathcal{K}}\,
\tilde{\hat{\Phi}}(z)
\]\ 
\hat{P}^{(\texttt{j})\dag}_{p^{\,}_{S}}(\pri{z})=
\tilde{\mathcal{T}}^{(\texttt{j})\mathsf{T}}_{\pri{p}}\,
\tilde{\mathcal{K}}\,
\tilde{\hat{\Phi}}(z)
+
2\pi\, 
|\tilde{w}^{(\texttt{j})}|^{2}\,
\Theta(z-\pri{z})
+ 
\text{constant},
\\ 
&
\hat{P}^{(\texttt{j})}_{p^{\,}_{E}}(\pri{z})\ 
\[
\tilde{\mathcal{T}}^{(\texttt{j})\mathsf{T}}_{\pri{p}}\,
\tilde{\mathcal{K}}\,
\tilde{\hat{\Phi}}(z)
\]\ 
\hat{P}^{(\texttt{j})\dag}_{p^{\,}_{E}}(\pri{z})=
\tilde{\mathcal{T}}^{(\texttt{j})\mathsf{T}}_{\pri{p}}\,
\tilde{\mathcal{K}}\,
\tilde{\hat{\Phi}}(z)
+
2\pi\, 
|\tilde{w}^{(\texttt{j})}|^{2}\,
\Theta(z-\pri{z})
+ 
\text{constant},
\end{align}
\end{subequations}
\end{widetext}
respectively.
Note the sign difference with respect to 
Eqs.\ 
\eqref{eq: 2D pinning soliton star} 
and 
\eqref{eq: 2D pinning soliton plaquette}.
This difference
also stems from Fig.\ \ref{fig: generic tunneling vectors}.  
Consequently, we interpret the operators
$\hat{S}^{(\texttt{j})\dag}_{s^{\,}_{C}}$ 
and
$\hat{P}^{(\texttt{j})\dag}_{p^{\,}_{C}}$ as creating a 
soliton-antisoliton pair straddling the links 
$s^{\,}_{C}=s^{\,}_{N},s^{\,}_{W},s^{\,}_{S},s^{\,}_{E}$
and
$p^{\,}_{C}=p^{\,}_{N},p^{\,}_{W},p^{\,}_{S},p^{\,}_{E}$, 
respectively 
(see Fig.\ \ref{fig: star and plaquette excitations}).
By taking the derivative with respect to $z$ of
Eqs.\
(\ref{eq: 2D pinning soliton star}),
(\ref{eq: 2D pinning soliton plaquette}),
(\ref{eq: 2D pinning soliton star prime}),
and
(\ref{eq: 2D pinning soliton plaquette prime}),
we can interpret the operators
$\hat{S}^{(\texttt{j})\dag}_{s^{\,}_{C}}$ 
and
$\hat{P}^{(\texttt{j})\dag}_{p^{\,}_{C}}$ as creating a
dipole in the soliton density across the links 
$s^{\,}_{C}=s^{\,}_{N},s^{\,}_{W},s^{\,}_{S},s^{\,}_{E}$
and
$p^{\,}_{C}=p^{\,}_{N},p^{\,}_{W},p^{\,}_{S},p^{\,}_{E}$, 
respectively. Correspondingly, the annihilation vertex operators
$\hat{S}^{(\texttt{j})}_{s^{\,}_{C}}$ 
and
$\hat{P}^{(\texttt{j})}_{p^{\,}_{C}}$
reverse the orientations of these dipoles in the soliton density.

The defect and antidefect created by applying one of the operators
$\hat{S}^{(\texttt{j})\dag}_{s^{\,}_{C}}$ and
$\hat{P}^{(\texttt{j})\dag}_{p^{\,}_{C}}$ can be propagated away from one
another in the $x$-$y$ plane by subsequent applications of the same
operators on adjacent links, each of which ``heal" one defect while
creating another.  An example of such a process is shown in
Fig.\ \ref{fig: deconfined star defects}.
In the strong-coupling limit in which we work, this process does not generate
any additional excitations, indicating that star and plaquette defects
are deconfined in the $x$-$y$ plane.  Furthermore, one can show that,
in the same strong-coupling limit, these defects are also deconfined
in the $z$-direction (see Appendix 
\ref{appsec: Deconfinement of defects along the direction of a wire}
for more details).  Consequently,
we conclude that the wire construction supports deconfined pointlike
excitations, namely the star and plaquette defects.  When these
defects are separated from one another, there is a ``string" of
vertex operators connecting them.  These strings are a crucial ingredient for
determining the topological degeneracy, as we will see in the next
section.

\begin{figure*}
\centering
(a)\includegraphics[width=0.4\textwidth]{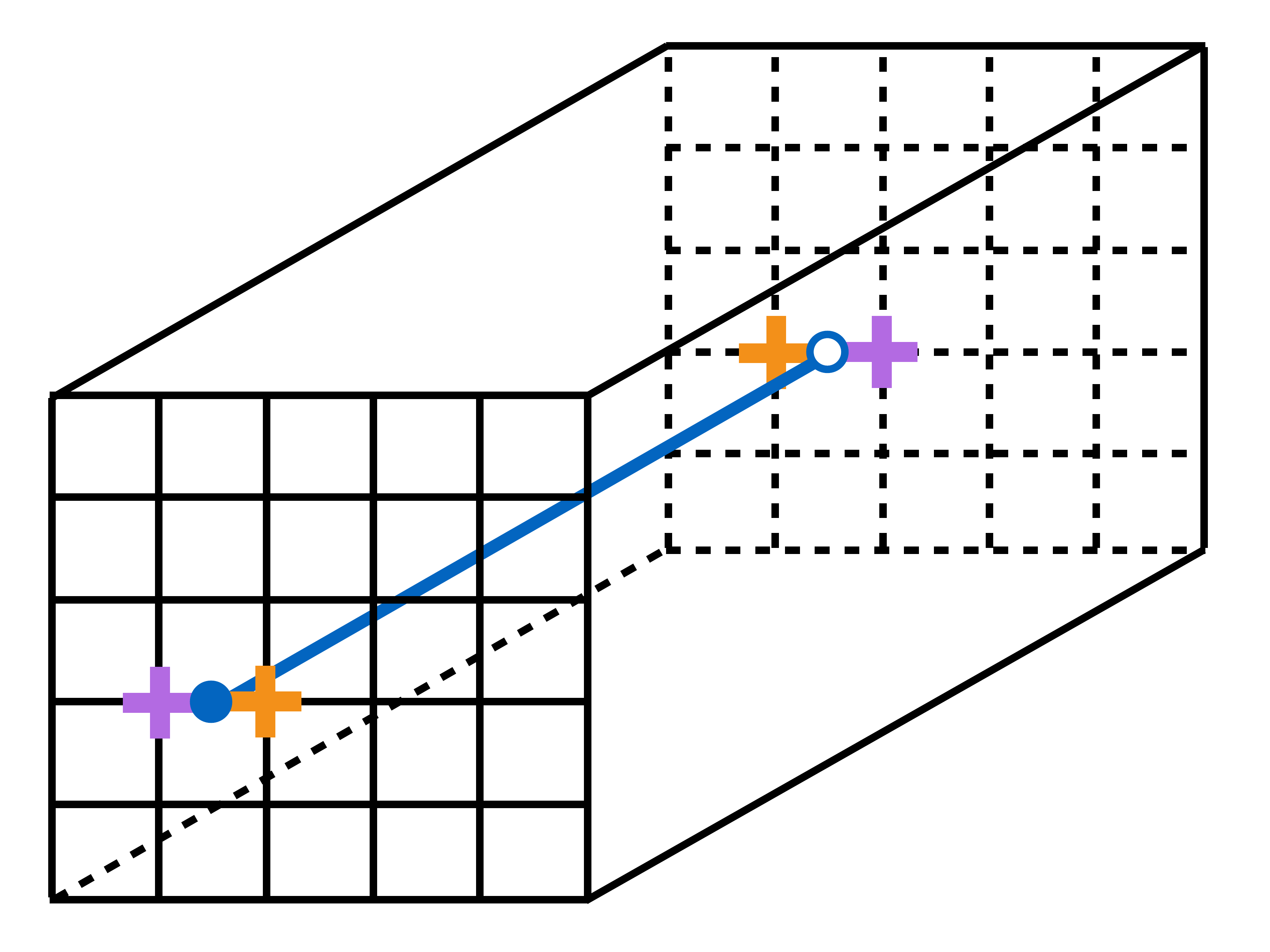}
(b)\includegraphics[width=0.4\textwidth]{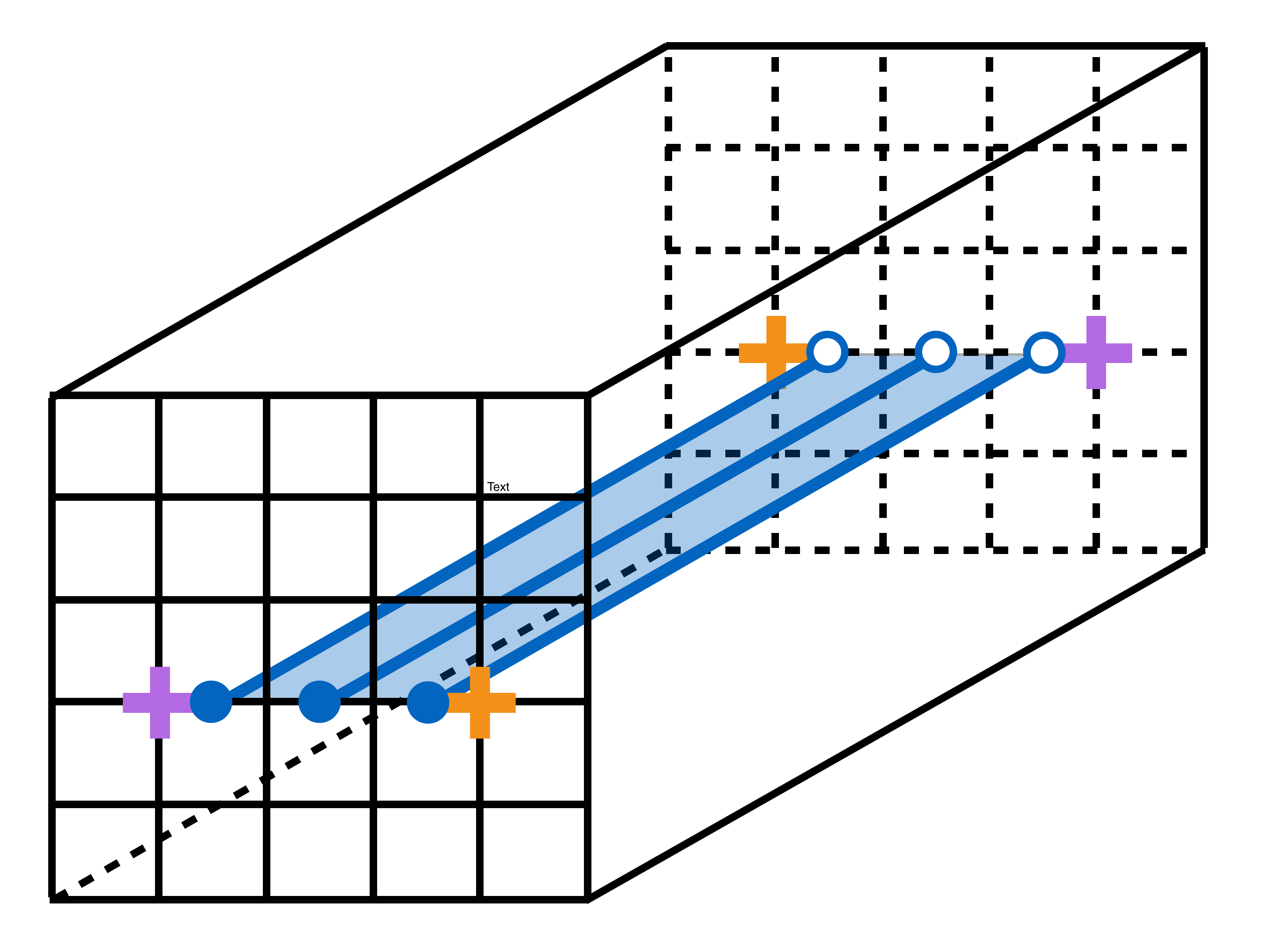}
(c)\includegraphics[width=0.4\textwidth]{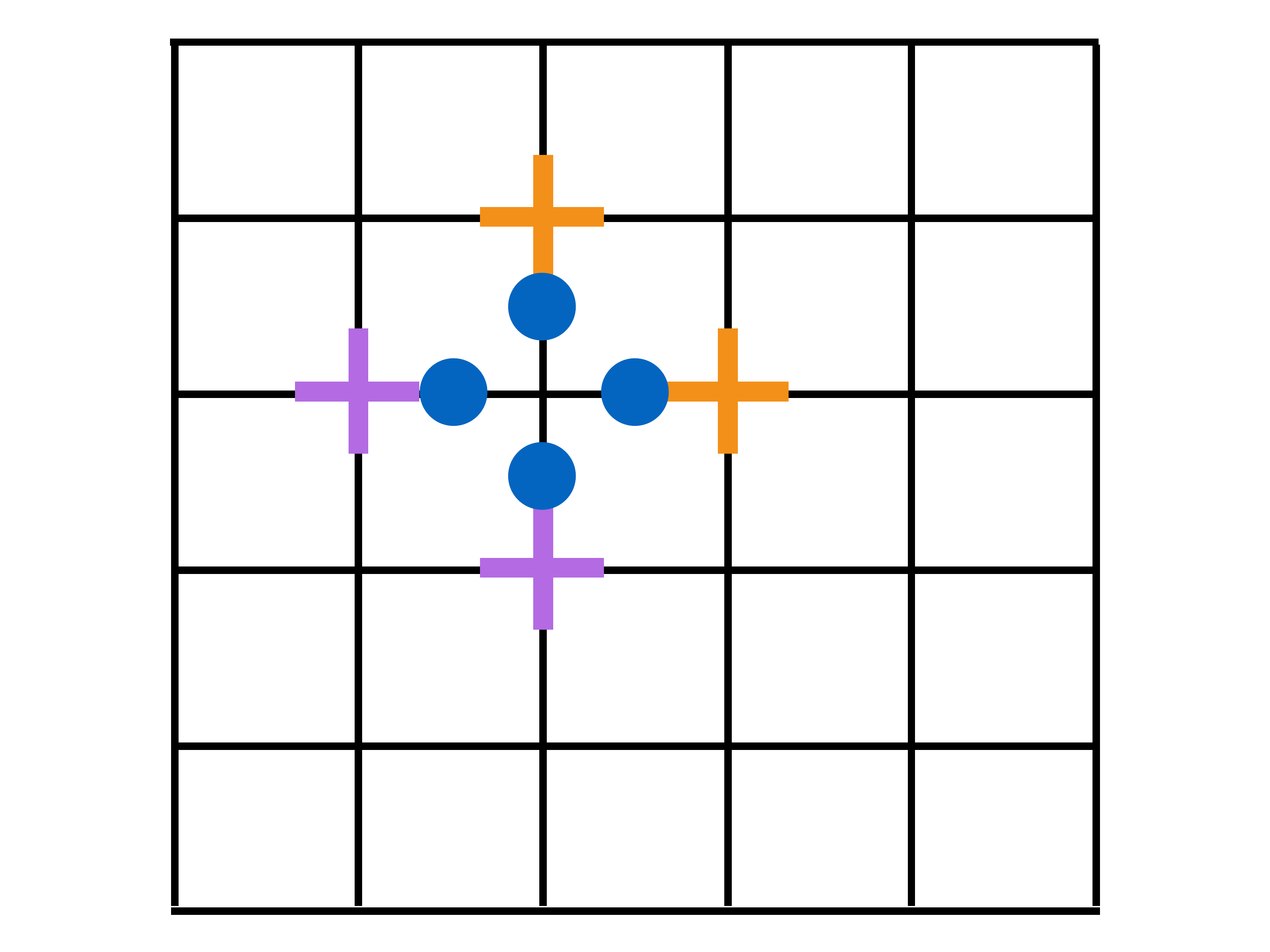}
(d)\includegraphics[width=0.4\textwidth]{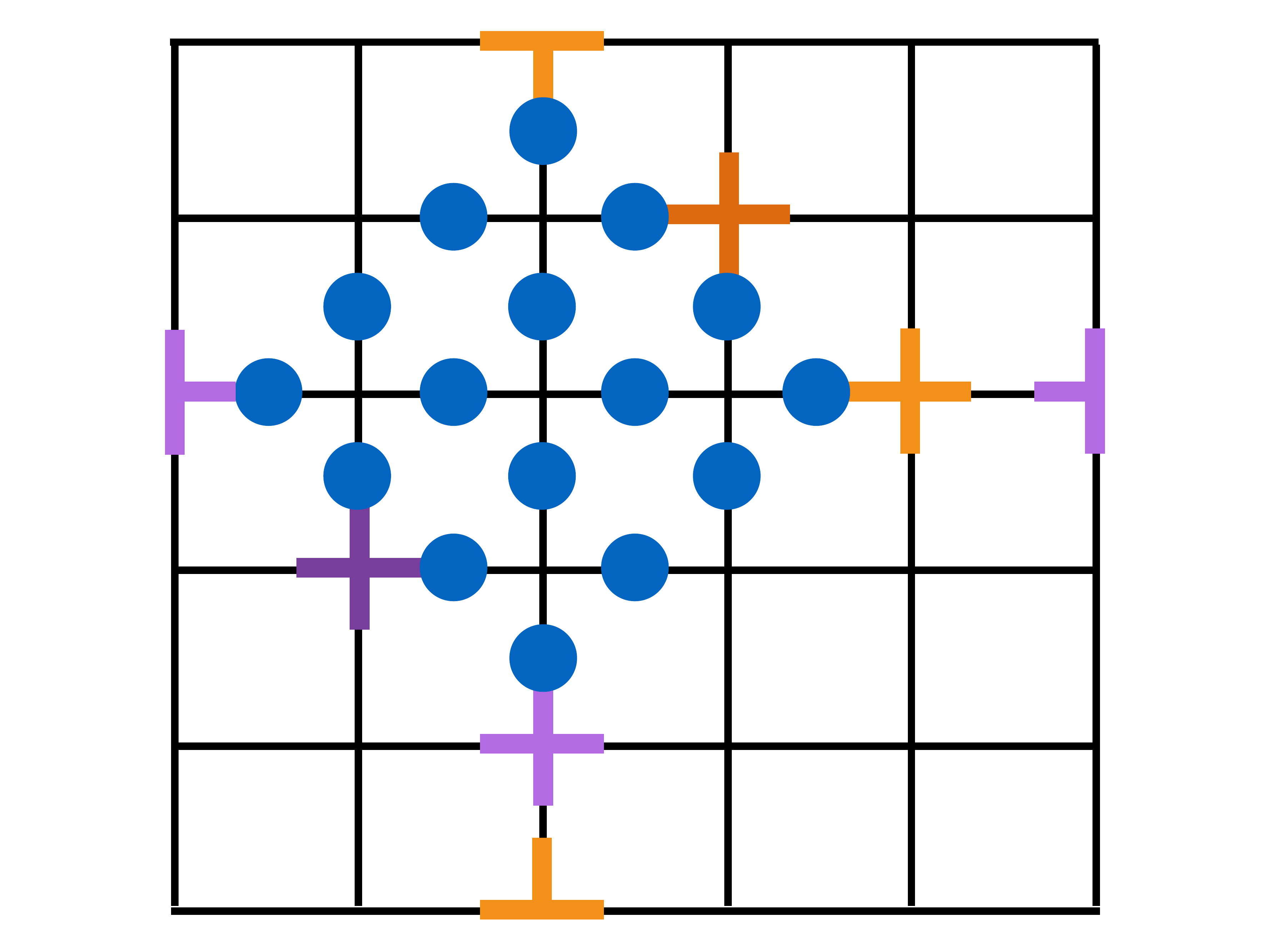}
\caption{(color online) 
(a) A pair of linelike defects along the $z$ direction created by the operators
(\ref{eq: def 2D z-string defs}).
(b) Propagating one linelike defect away from the other using the operators
(\ref{eq: def 2D z-string defs})
creates a membrane in the $x$-$z$ plane.
(c) Growing a star membrane in the $x$-$y$ plane
from the product of the operators
(\ref{eq: 2D defect hopping a}).
(d) Membrane in the $x$-$y$ plane with defect lines on its boundaries.  
Darker purple or orange crosses indicate defective stars with a 
``double-strength" soliton, where two
kinks of the same sign coexist in the same star.
        }  
\label{fig: xy line and membrane defects}
\end{figure*}

The wire construction also supports deconfined \textit{linelike}
excitations. For any $\texttt{j}=1,\ldots,M$,
pairs of linelike defects connecting the points $z^{\,}_{1}$
and $z^{\,}_{2}$ in a wire labelled by the link
$s^{\,}_{C}=s^{\,}_{N},s^{\,}_{W},s^{\,}_{S},s^{\,}_{E}$
or
$p^{\,}_{C}=p^{\,}_{N},p^{\,}_{W},p^{\,}_{S},p^{\,}_{E}$ 
can be created by the bi-local operators 
\begin{subequations}
\label{eq: def 2D z-string defs}
\begin{align}
\begin{split}
\hat{S}^{(\texttt{j})\dag}_{s^{\,}_{C}}(z^{\,}_{1},z^{\,}_{2})&\:=
\hat{S}^{(\texttt{j})\dag}_{s^{\,}_{C}}(z^{\,}_{2})\,
\hat{S}^{(\texttt{j})\phantom\dag}_{s^{\,}_{C}}(z^{\,}_{1})
\\
&=
\exp
\bigg(
-\mathrm{i}\, 
\sum_{\alpha=1}^{2M}
\tilde{v}^{(\texttt{j})}_{\alpha}
\int\limits^{z^{\,}_{2}}_{z^{\,}_{1}}\mathrm{d}z\ 
\partial^{\,}_{z}\tilde{\hat{\phi}}^{\,}_{s^{\,}_{C},\alpha}(z)
\bigg),
\end{split}
\label{eq: def 2D z-string defs a}
\\
\begin{split}
\hat{P}^{(\texttt{j})\dag}_{p^{\,}_{C}}(z^{\,}_{1},z^{\,}_{2})&\:=
\hat{P}^{(\texttt{j})\dag}_{p^{\,}_{C}}(z^{\,}_{2})\,
\hat{P}^{(\texttt{j})\phantom\dag}_{p^{\,}_{C}}(z^{\,}_{1})\\
&=
\exp
\bigg(
-\mathrm{i}\, 
\sum_{\alpha=1}^{2M}
\tilde{w}^{(\texttt{j})}_{\alpha}
\int\limits^{z^{\,}_{2}}_{z^{\,}_{1}}\mathrm{d}z\ 
\partial^{\,}_{z}\tilde{\hat{\phi}}^{\,}_{p^{\,}_{C},\alpha}(z)
\bigg),
\end{split}
\label{eq: def 2D z-string defs b}
\end{align}
\end{subequations}
a pictorial example of which is depicted in Fig.\
\ref{fig: xy line and membrane defects}(a).  
Similarly to the propagation of star and plaquette
defects outlined in the previous paragraph, applying a string of the
above operators creates a \textit{membrane} with linelike defects at
its boundaries in the $z$-$x$ and $y$-$z$ planes
as is illustrated Fig.\ \ref{fig: xy line and membrane defects}(b).

The membranes created by applying strings of the operators defined in
Eqs.\ \eqref{eq: def 2D z-string defs} necessarily extend in the
$x$-$z$ or $y$-$z$ planes.  Membranes extending in the $x$-$y$ plane
can also be created by applying the operators defined in
Eqs.~\eqref{eq: 2D defect hopping} over a membrane as opposed to a
string, as in Fig.\ \ref{fig: xy line and membrane defects}(c-d).  
As with the $x$-$z$ and $y$-$z$ membranes, the boundary of an $x$-$y$ membrane
supports linelike defects.

It is important to note that the strings and membranes connecting
pairs of pointlike and linelike defects, respectively, may fluctuate
in all directions. The origin of these fluctuations lies in the
existence of a discrete gauge symmetry that can be formulated
explicitly in the strong coupling limit
$|U^{\,}_{\tilde{\mathcal{T}}}|\to\infty$.  
(We elaborate on the physical meaning of this limit in the next section.)  
We carry out this formulation in Appendix
\ref{appsec: Discrete gauge symmetry and ground state in the limit}.  
Strings and membranes that fluctuate in this way 
are familiar from the toric code and other
string-net models.

\subsubsection{Energetics of pointlike and linelike defects}
\label{subsubsec: Energetics of pointlike and linelike defects}

At this stage, a brief comment is in order regarding the energetics of
the pointlike and linelike defects defined in 
Sec.~\ref{subsubsec: Pointlike and linelike excitations}.  
It is misleading to compute
the energy cost of such a defect in the strong-coupling limit
$|U^{\,}_{\tilde{\mathcal{T}}}|\to\infty$, as in this limit, the
perfectly sharp solitons created by the operators
$\hat{S}^{(\texttt{j})\dag}_{s^{\,}_{C}}$ and
$\hat{P}^{(\texttt{j})\dag}_{p^{\,}_{C}}$ 
[recall Eqs.\
\eqref{eq: 2D pinning soliton star} 
and
\eqref{eq: 2D pinning soliton plaquette}] 
cost no energy from the point of view of the
cosine terms \eqref{Lint}. This is simply because these solitons
increment the argument of a cosine term abruptly at some $z$ by an
integer multiple of $2\pi$, which amounts to a discontinuous jump
between exact minima of the cosine potential. However, the presence
of an infinitesimal kinetic term of the form \eqref{L0 d=2 a} gives a
finite stiffness $\kappa$ to the pinned field.  In this case, the
optimal soliton profile is no longer the perfectly sharp one generated
by the operators 
$\hat{S}^{(\texttt{j})\dag}_{s^{\,}_{C}}$ 
and
$\hat{P}^{(\texttt{j})\dag}_{p^{\,}_{C}}$, but a slightly deformed one
where the interpolation between minima of the cosine potential is
smeared over a length scale $\xi$.  

Suppose that this optimal soliton profile is known.  
Then, it is possible to redefine the operators 
$\hat{S}^{(\texttt{j})\dag}_{s^{\,}_{C}}$ and $\hat{P}^{(\texttt{j})\dag}_{p^{\,}_{C}}$
in such a way that they act on the pinned fields as in 
Eqs.\ 
\eqref{eq: 2D pinning soliton star} 
and
\eqref{eq: 2D pinning soliton plaquette},
but now with the
perfectly sharp soliton profile replaced by the optimal one.  
[Note that this redefinition can be done without
altering the fundamental commutation relations \eqref{tilde commutator} 
on length scales longer than $\xi$.]  
The energy cost of such an optimal soliton is composed of two contributions:
one from the cosine potential (assumed to be large) and one from the stiffness 
(assumed to be small but finite).

Once the finite energy cost of a single soliton has been determined, 
it is readily seen that the stringlike and membranelike operators defined in 
Sec.~\ref{subsubsec: Pointlike and linelike excitations}
cannot dissociate into smaller pointlike or linelike operators 
without an energy cost that is extensive in the number
of vertex operators used to build the string or membrane.
For example, if one tries to pull apart the string of vertex operators shown in 
Fig.~\ref{fig: deconfined star defects}(b) so that
all operators in the string are disconnected, one necessarily increases 
the energy by an amount proportional to the
number of vertex operators in the chain.  This is because each application
of a vertex operator in the latter 
scenario costs energy due to \textit{two}
cosine terms (in addition to the stiffness).  
In contrast, when the vertex operators form a string, the only energy cost due to the
cosine terms occurs at the two ends of the string.

Finally, one might be concerned that the energetic effects discussed 
above could lead to confinement of star and plaquette defects.
Indeed, if the stiffness $\kappa$ is finite, then strings of vertex operators
like the ones depicted in Fig.~\ref{fig: deconfined star defects}
necessarily cost an energy proportional to their length.  In fact, there is a
direct parallel here with the confinement-deconfinement
transition in the toric code~\cite{Kitaev03}.  In that case, two
star defects (say) are connected by a string of flipped spins.  Thus,
there is a measurable trail of magnetization that connects the two defects.  
However, in the absence of an external
magnetic field, there is no energy cost associated with such a string.
The presence of a sufficiently large external field leads to
confinement, but, below a critical field strength, entropic effects are
sufficient to deconfine the defects.  In our system, the role of the
external magnetic field is played by the kinetic term, which is the origin
of the stiffness $\kappa$.

Thus, provided that the stiffness 
$\kappa\ll|U^{\,}_{\tilde{\mathcal{T}}}|$, 
we expect that defects in our model are deconfined because entropic effects
favor deconfinement, as is the case in the toric code. 
When $\kappa$ reaches some critical value, however, the defects become confined.

\subsubsection{Statistics of pointlike and linelike defects}
\label{subsubsec: Statistics of point, line, and membrane defects}

We have enumerated the pointlike and linelike excitations 
for a class of two-dimensional arrays of coupled quantum wires
by showing how to use vertex operators to
build open stringlike and open 
membranelike operators supporting these defects
on their boundaries.  The statistics of these excitations are
readily accessible within the wire formalism, as we now explain.
  
In principle, there are several types of statistics to
consider.  The first type, that of different types of pointlike
excitations, must be trivial in three dimensions by homotopy
arguments.~\cite{Leinaas77} (Essentially, such arguments hinge on
the fact that any loop that one particle makes around another in three
dimensions can be deformed to a point without passing through the
other particle.)  

The second type, that of pointlike and linelike
excitations, can be nontrivial in three dimensions, and will be
computed below for the class of models defined here.  

The third type
of statistics, that between linelike excitations, can also be
nontrivial in three dimensions, but can be shown to be trivial in the
present class of models.

Let us first examine the mutual statistics between
pointlike and linelike defects.  Using the identity
\begin{align}
e^{\hat A}\, 
e^{\hat B} = 
e^{\hat B}\, 
e^{\hat A}\ 
e^{[\hat A,\hat B]}
\end{align}
which follows from the Baker-Campbell-Hausdorff lemma whenever 
$[\hat A,\hat B]$ is a $c$-number, one can show from 
Eq.~\eqref{tilde commutator} that,
for any $\texttt{j},\texttt{j}'=1,\ldots,N$
and for any
$s^{\,}_{C},s^{\,}_{C'}=s^{\,}_{N},s^{\,}_{W},s^{\,}_{S},s^{\,}_{E}$
or
$p^{\,}_{C},p^{\,}_{C'}=p^{\,}_{N},p^{\,}_{W},p^{\,}_{S},p^{\,}_{E}$,
\begin{subequations}\label{string-membrane algebra}
\begin{align}
\begin{split}
&\hat{S}^{(\texttt{j})\dag}_{s^{\,}_{C}}(z)\, 
\hat{P}^{(\texttt{j}^{\prime})\dag}_{p^{\,}_{\pri{C}}}(\pri{z}_{1},\pri{z}_{2})
\\
&\qquad =
\hat{P}^{(\texttt{j}^{\prime})\dag}_{p^{\,}_{\pri{C}}}(\pri{z}_{1},\pri{z}_{2})\, 
\hat{S}^{(\texttt{j})\dag}_{s^{\,}_{C}}(z)\, 
e^{
+\mathrm{i}\, 
2\pi\, 
\tilde{v}^{(\texttt{j})\T}
\,\tilde{K}^{-1}\,
\tilde{w}^{(\texttt{j}^{\prime})}\, 
\delta^{\,}_{s^{\,}_{C},p^{\,}_{\pri{C}}}
  },
\end{split}\\
\begin{split}
&\hat{S}^{(\texttt{j})\dag}_{s^{\,}_{C}}(\pri{z}_{1},\pri{z}_{2})\, 
\hat{P}^{(\texttt{j}^{\prime})\dag}_{p^{\,}_{\pri{C}}}(z)
\\
&\qquad =
\hat{P}^{(\texttt{j}^{\prime})\dag}_{p^{\,}_{\pri{C}}}(z)\, 
\hat{S}^{(\texttt{j})\dag}_{s^{\,}_{C}}(\pri{z}_{1},\pri{z}_{2})\, 
e^{
-\mathrm{i}\, 
2\pi\, 
\tilde{v}^{(\texttt{j})\T}\,
\tilde{K}^{-1}\,
\tilde{w}^{(\texttt{j}^{\prime})}\, 
\delta^{\,}_{s^{\,}_{C},p^{\,}_{\pri{C}}}
  },
\end{split}
\end{align}
\end{subequations}
whenever $z^{\prime}_{1}<z<z^{\prime}_{2}$.

\begin{figure}[t]
\begin{center}
\includegraphics[width=.45\textwidth]{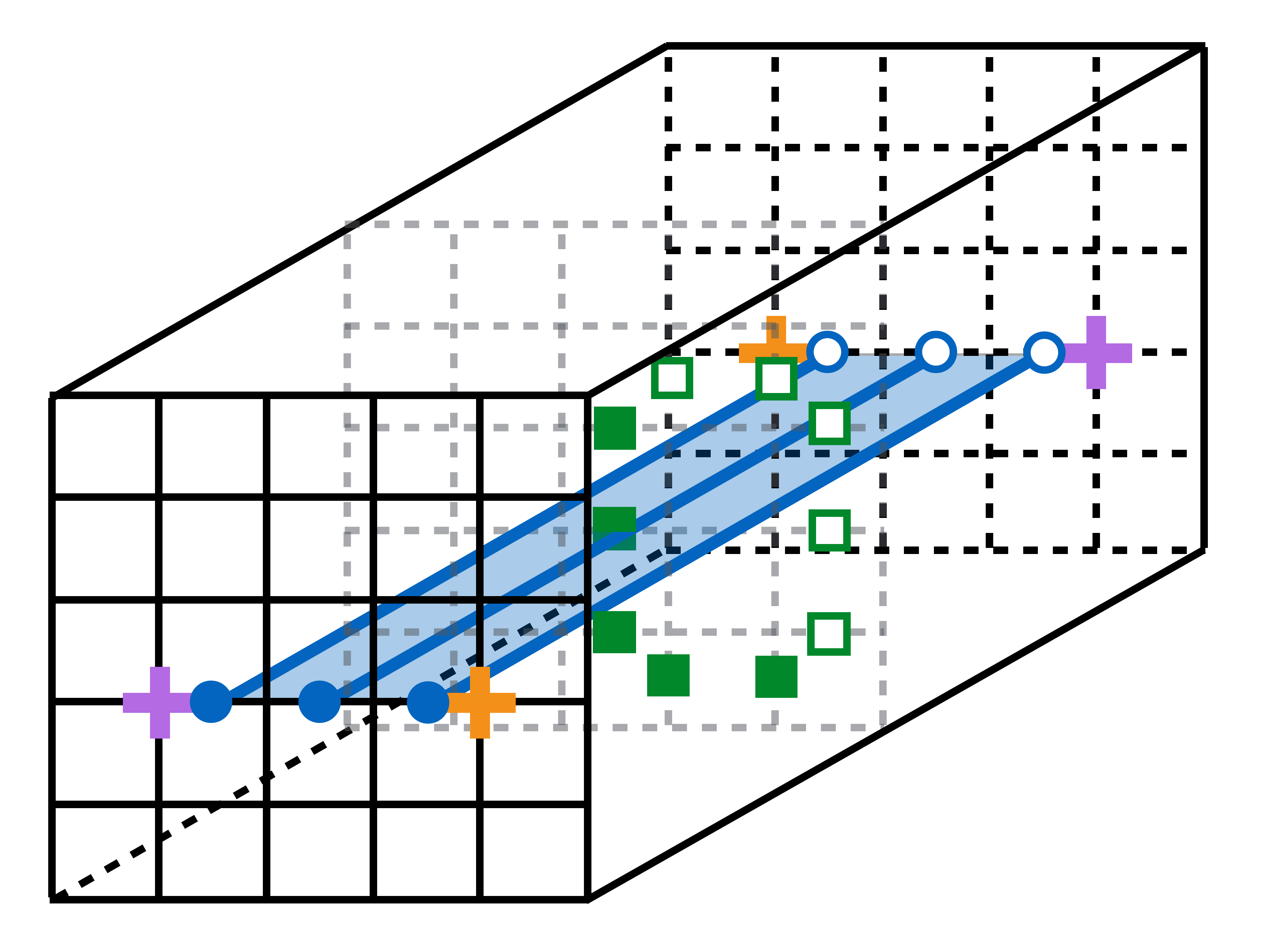}\\
\caption{
(Color online)
Pictorial representation for braiding a pointlike plaquette defect
[defined in Eq.\ (\ref{eq: 2D defect hopping b})] 
around a linelike star defect along the $z$-direction [defined in 
Eq.\ (\ref{eq: def 2D z-string defs a})].
The green colored loop represents the worldline during adiabatic
braiding of one end of an open plaquette string, say the plaquette
counterpart to the open string depicted in
Fig.\ \ref{fig: deconfined star defects}(c).
By choice, the world line encloses the rightmost boundary of the
star membrane in
Fig.\ \ref{fig: xy line and membrane defects}(b).
\label{fig: braiding of point around line defects}
         }
\end{center}
\end{figure}

With these relations in hand, one can readily compute the algebra of
the membrane and string operators that are used to create and
propagate pointlike and linelike defect-antidefect pairs.
As discussed in Ref.~\cite{Lin15},
this algebra determines the phase obtained by winding a pointlike
excitation around a linelike excitation.  The computation of this
phase is cumbersome to write down, but nevertheless quite
straightforward---a convenient way to see this comes from the
pictorial representation of such a braiding process
(see~Fig.~\ref{fig: braiding of point around line defects} for
an example).  From this pictorial representation, one sees immediately
that the membrane and string operators associated with stars commute
with one another (and likewise for plaquettes), as they never
intersect in a wire.  However, membranes associated with star defects
and strings associated with plaquette defects (and vice versa) always
intersect with one another during a braiding process.  The total phase
arising from commuting one operator past the other can then be read
off from the picture using Eqs.~\eqref{string-membrane algebra}.  We
find that the statistical phase obtained by braiding a pointlike
plaquette defect around a linelike star defect is given by
\begin{align}\label{eq: chosen case for statistical angle between linelike}
\frac{\theta^{\,}_{\texttt{j}\texttt{j}^{\prime}}}{2\pi}=
\tilde{v}^{(\texttt{j})\T}\,
\tilde{K}^{-1}\,
\tilde{w}^{(\texttt{j}^{\prime})}=
\tilde{w}^{(\texttt{j}^{\prime})\T}\,
\tilde{K}^{-1}\tilde{v}^{(\texttt{j})}
\end{align}
for any $\texttt{j},\texttt{j}'=1,\ldots,M$,
where the second equality follows from the fact that $\tilde{K}^{-1}$
is a symmetric matrix.

At this point, we remark that, although the construction of operators
undertaken in this section and in the previous section has assumed
that 
$\tilde{v}^{(\texttt{j})}_{1}=\tilde{v}^{(\texttt{j})}_{2}=\tilde{v}^{(\texttt{j})}$ 
and 
$\tilde{w}^{(\texttt{j})}_{1}=\tilde{w}^{(\texttt{j})}_{2}=\tilde{w}^{(\texttt{j})}$, 
this construction proceeds with only minor modifications in the more general case
$\tilde{v}^{(\texttt{j})}_{1}\neq \tilde{v}^{(\texttt{j})}_{2}$ 
and
$\tilde{w}^{(\texttt{j})}_{1}\neq \tilde{w}^{(\texttt{j})}_{2}$.
However, in the latter case, one finds that the statistical angle must obey
\begin{align}\label{eq: general case for statistical angle between linelike}
\frac{\theta_{\texttt{j}\texttt{j}^{\prime}}}{2\pi}=
\tilde{v}^{(\texttt{j})\T}_{1}\,
\tilde{K}^{-1}\,
\tilde{w}^{(\texttt{j}^{\prime})}_{2}\overset{!}{=}
\tilde{v}^{(\texttt{j})\T}_{2}\,
\tilde{K}^{-1}\,\tilde{w}^{(\texttt{j}^{\prime})}_{1}
\end{align}
for any $\texttt{j},\texttt{j}'=1,\ldots,M$.
The second equality in 
Eq.\ (\ref{eq: general case for statistical angle between linelike})
must be imposed as a consistency condition.  Otherwise, the statistical
angle $\theta^{\,}_{\texttt{j}\texttt{j}^{\prime}}$ would depend on whether
the string and membrane operators used to compute the statistics
intersected on vertical or horizontal bonds.  
(See, e.g., Fig.~\ref{fig: braiding of point around line defects},
where the string and membrane intersect on a horizontal bond.)  Thus, we must
demand that Eq.~\eqref{eq: general case for statistical angle between linelike}
holds, as otherwise the low-energy description of the theory 
would be anisotropic.

\begin{figure}[t]
\begin{center}
\includegraphics[width=.45\textwidth]{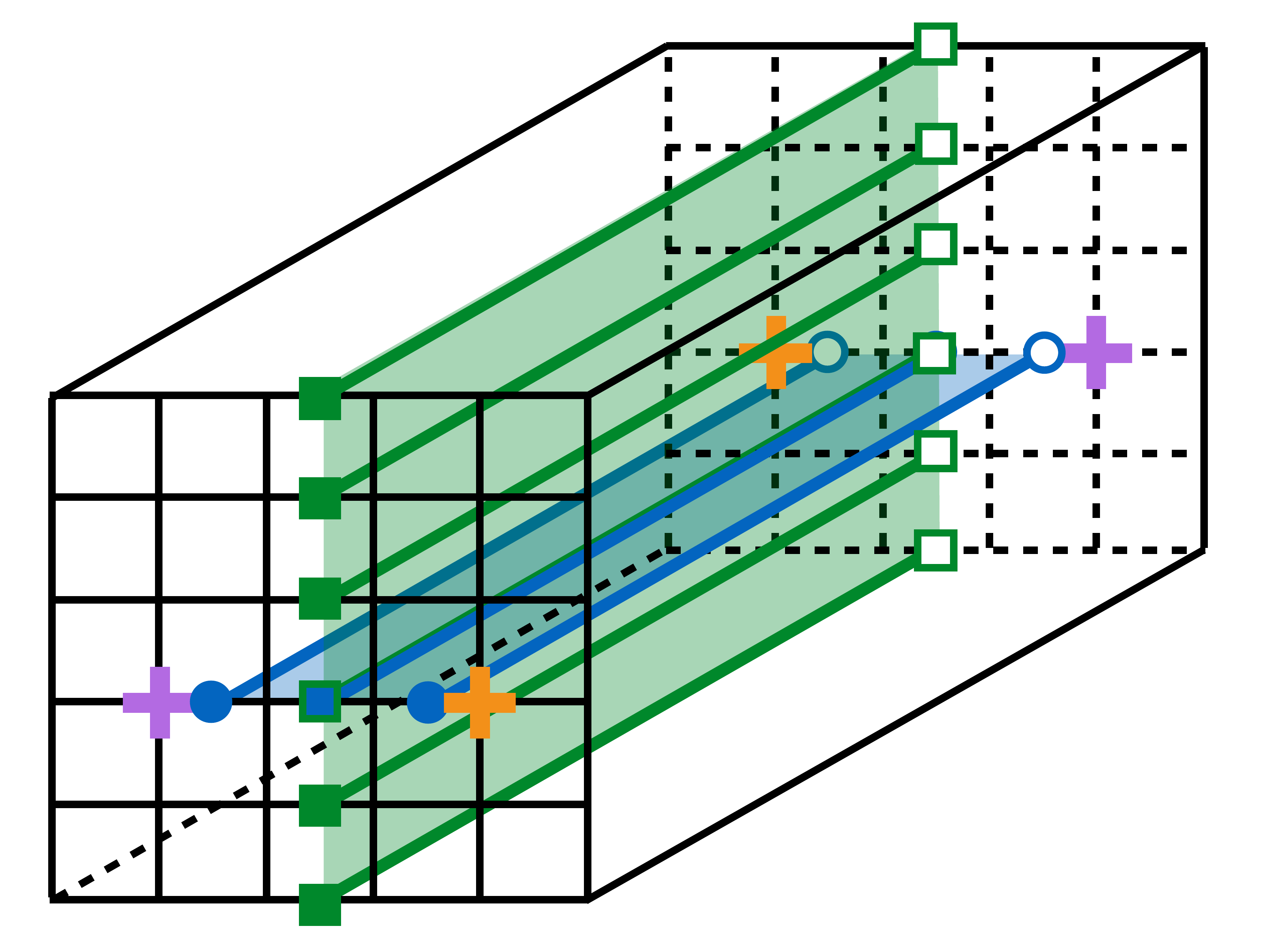}\\
\caption{
(Color online)
Pictorial representation of braiding a star line and a
plaquette line.  Because periodic boundary conditions are
imposed in all directions and because
we choose to represent the worldsheet induced by the adiabatic evolution of
plaquette linelike defects by the green membrane 
that extends across the width of the lattice
in the $y$ direction, this world sheet forms a
cylinder with its symmetry axis parallel to the $z$ direction.
By choice, this cylinder encircles the star-type line that defines
the rightmost boundary of the blue membrane. 
The algebra encoded by Eq.\ \eqref{line-line statistics}
implies that such a braiding yields no overall statistical phase.
\label{fig: star and plaquette excitations bis}
         }
\end{center}
\end{figure}

We close this section by outlining the reason why the line-line
statistics in this class of models is trivial.  The statistical phase
describing the line-line statistics is computed using membrane
surfaces arranged as in Fig.~\ref{fig: star and plaquette excitations bis}.  
From this, it is clear that the
relevant operator product to consider is of the form
\begin{widetext}
\begin{align}\label{line-line statistics}
&
\hat{S}^{(\texttt{j})\dag}_{s^{\,}_{C}}(z^{\,}_{1},z^{\,}_{2})\, 
\hat{P}^{(\texttt{j}^{\prime})\dag}_{p^{\,}_{\pri{C}}}(\pri{z}_{1},\pri{z}_{2})=
\hat{S}^{(\texttt{j})\dag}_{s^{\,}_{C}}(z^{\,}_{1},z^{\,}_{2})\, 
\hat{P}^{(\texttt{j}^{\prime})\dag}_{p^{\,}_{\pri{C}}}(\pri{z}_{1},\pri{z}_{2})
\, 
\exp\!
\left(\!
-\mathrm{i}\! 
\sum_{\alpha,\beta=1}^{2M}
\tilde{v}^{(\texttt{j})}_{\alpha}
\tilde{w}^{(\texttt{j}^{\prime})}_{\beta}
\int\limits^{z^{\,}_{2}}_{z^{\,}_{1}}\!\mathrm{d}z
\int\limits^{z^{\prime}_{2}}_{z^{\prime}_{1}}\!\mathrm{d}\pri{z}
\left[
\partial^{\,}_{z}\tilde{\hat{\phi}}^{\,}_{s^{\,}_{C},\alpha}(z),
\partial^{\,}_{\pri{z}}\tilde{\hat{\phi}}^{\,}_{p^{\,}_{\pri{C}},\beta}(\pri{z})
\right]
\right)
\end{align}
\end{widetext}
for any 
$\texttt{j}=1,\ldots,M$,
$s^{\,}_{C}=s^{\,}_{N},s^{\,}_{W},s^{\,}_{S},s^{\,}_{E}$,
and
$p^{\,}_{C'}=p^{\,}_{N},p^{\,}_{W},p^{\,}_{S},p^{\,}_{E}$,
where it is assumed that the interval $[z^{\,}_{1},z^{\,}_{2}]\subset
[z^{\prime}_{1},z^{\prime}_{2}]$ or vice versa.  However, by
differentiating Eq.\ (\ref{tilde commutator}) with respect to $\pri{z}$,
one sees that the commutator in the exponential is proportional to the
\textit{derivative} of a delta function.  Integrated over both $z$ and
$\pri{z}$, this yields zero for the statistical angle between two
lines.  In a similar manner, one can show that the \textit{three-line}
statistics (c.f.~Ref.~\cite{Lin15}) is trivial in this class of
models.

This discussion of the
excitations of the coupled-wire construction provides sufficient
information to determine the minimal topological ground-state degeneracy of the theory,
as we now show.

\subsubsection{Topological ground-state degeneracy}
\label{subsubsec: Topological ground state degeneracy}

The theory defined 
in Eq.\ 
(\ref{def tilde L as sum tilde L0 and tilde LT})
by the Lagrangian
$\tilde{\hat{L}}=
\tilde{\hat{L}}^{\,}_{0}+\tilde{\hat{L}}^{\,}_{\{\tilde{\mathcal{T}}\}}$
generically exhibits a ground-state degeneracy when defined on the
three-torus obtained by imposing periodic boundary conditions in the
$x$-, $y$-, and $z$-directions. We present an
argument as to why this is the case.  

First, recall that, when
periodic boundary conditions are imposed in the $x$- and
$y$-directions, for a square lattice with $2N$ links, there are $4\,M\,N$
degrees of freedom ($2\,M$ per wire).  There are also $M\,N$ star and $M\,N$
plaquette terms entering the interaction
$\tilde{\hat{H}}^{\,}_{\mathrm{int}}\equiv-
\tilde{\hat{L}}^{\,}_{\{\tilde{\mathcal{T}}\}}$
defined in Eq.\ (\ref{def tilde L as sum tilde L0 and tilde LT}), 
each of which gaps
out two of these degrees of freedom.  The number of star and plaquette
operators in $\tilde{\hat{L}}^{\,}_{\{\tilde{\mathcal{T}}\}}$ is
therefore sufficient to gap out the bulk of the wire array, as
mentioned above.

It is shown in Appendix 
\ref{appsec: Discrete gauge symmetry and ground state in the limit}
for $M=1$ with the choice made in Sec.\ \ref{subsec: Zm example}
for the integer-valued vectors $\tilde{v}$ and $\tilde{w}$ that
there are $4\,N$ local vertex operators,
namely two per star $s$ and two per plaquette $p$, that commute
with the interaction $\tilde{\hat{H}}^{\,}_{\mathrm{int}}\equiv-
\tilde{\hat{L}}^{\,}_{\{\tilde{\mathcal{T}}\}}$.
The proof in Appendix 
\ref{appsec: Discrete gauge symmetry and ground state in the limit}
readily generalizes to arbitrary $M=1,2,\dots$ and
the integer-valued vectors 
$\tilde{v}^{(\texttt{j})}$ and $\tilde{w}^{(\texttt{j})}$
with $\texttt{j}=1,\dots,M$
entering Fig.\ \ref{fig: generic tunneling vectors}.
This local gauge symmetry implies that all the cosines
entering the interaction $\tilde{\hat{H}}^{\,}_{\mathrm{int}}\equiv-
\tilde{\hat{L}}^{\,}_{\{\tilde{\mathcal{T}}\}}$
commute pairwise. 
This local gauge symmetry also implies
that deconfining the pointlike and stringlike defects
costs no energy in the strong coupling limit
where the full Hamiltonian $\tilde{\hat{H}}$ defined by Eq.\ 
(\ref{def tilde L as sum tilde L0 and tilde LT})
reduces to 
$\tilde{\hat{H}}^{\,}_{\mathrm{int}}\equiv-
\tilde{\hat{L}}^{\,}_{\{\tilde{\mathcal{T}}\}}$.
Hence, the strong coupling limit
$\tilde{\hat{H}}=\tilde{\hat{H}}^{\,}_{\mathrm{int}}$
is very singular since the gap induced by the cosines
from the Haldane set $\tilde{\mathbb{H}}$ collapses.
(As argued in Sec.~\ref{subsubsec: Energetics of pointlike and linelike defects}, 
including an infinitesimal kinetic term rectifies this singularity and yields a finite
energy cost for the creation of star and plaquette defects.)

Now, for any given coordinate 
$0\leq z<L$ along a wire,
there are $2\,M$ global constraints 
obeyed by the generators of this local gauge symmetry.
Indeed,
\begin{subequations}
\label{2D constraints}
\begin{align}
\prod_{p\in\mathcal{P}}
e^{
\mathrm{i}\,
\tilde{\mathcal{T}}^{(\texttt{j})\T}_{p}\,
\tilde{\mathcal{K}}\,
\tilde{\hat{\Phi}}(z)
  }&=1
\end{align}
and
\begin{align}
\prod_{s\in\mathcal{S}}
e^{
\mathrm{i}\, 
\tilde{\mathcal{T}}^{(\texttt{j})\T}_{s}\,
\tilde{\mathcal{K}}\,
\tilde{\hat{\Phi}}(z)}=1,
\end{align}
\end{subequations}
hold for all $\texttt{j}=1,\dots M$ and all $z\in[0,L)$.
Here, $\mathcal{P}$ and $\mathcal{S}$ 
are the sets of all plaquettes and stars
in the square lattice, respectively.
These constraints result from the fact that 
\begin{equation}
\sum_{p\in\mathcal{P}}
\tilde{\mathcal{T}}^{(\texttt{j})}_{p}=
\sum_{s\in\mathcal{S}}
\tilde{\mathcal{T}}^{(\texttt{j})}_{s}=0
\end{equation}
for each $\texttt{j}=1,\dots M$.

The $2\,M$ constraints (\ref{2D constraints})
are inherently nonlocal. Removing any
of the $\tilde{\mathcal{T}}^{(\texttt{j})}_{s}$
or 
$\tilde{\mathcal{T}}^{(\texttt{j})}_{p}$ 
for fixed $\texttt{j}$ from the set
$\tilde{\mathbb{H}}$ invalidates the
constraints (\ref{2D constraints}).  
If the number of independent commuting operators 
that commute with 
$\tilde{\hat{H}}^{\,}_{\mathrm{int}}\equiv-
\tilde{\hat{L}}^{\,}_{\{\tilde{\mathcal{T}}\}}$
defined in Eq.\ 
(\ref{def tilde L as sum tilde L0 and tilde LT})
is to match the number of degrees of freedom,
the constraints (\ref{2D constraints}) necessitate the existence
of additional nonlocal operators
that commute with $\tilde{\hat{H}}^{\,}_{\mathrm{int}}$.
It turns out that such operators exist and that 
the ground-state degeneracy is 
related to the representation of the algebra of these nonlocal operators
that has the smallest dimensionality.
We will now enumerate these operators, compute their algebra, and
deduce from this algebra the ground state degeneracy of the wire
construction.  One important result will be
that there is a \textit{unique} (i.e., non-degenerate) ground state if 
$|\det\,\tilde{K}|=1$, a result that is familiar from Abelian Chern-Simons
theories in (2+1) dimensions~\cite{Wen89b,Wen90,Wen91a,Wen92} and reappears in
the present (3+1)-dimensional context.

We are going to define two types of non-local operators
out of the local operators (\ref{eq: 2D defect hopping})
and the bi-local operators (\ref{eq: def 2D z-string defs}).

\begin{figure}[t]
\begin{center}
\includegraphics[width=.45\textwidth]{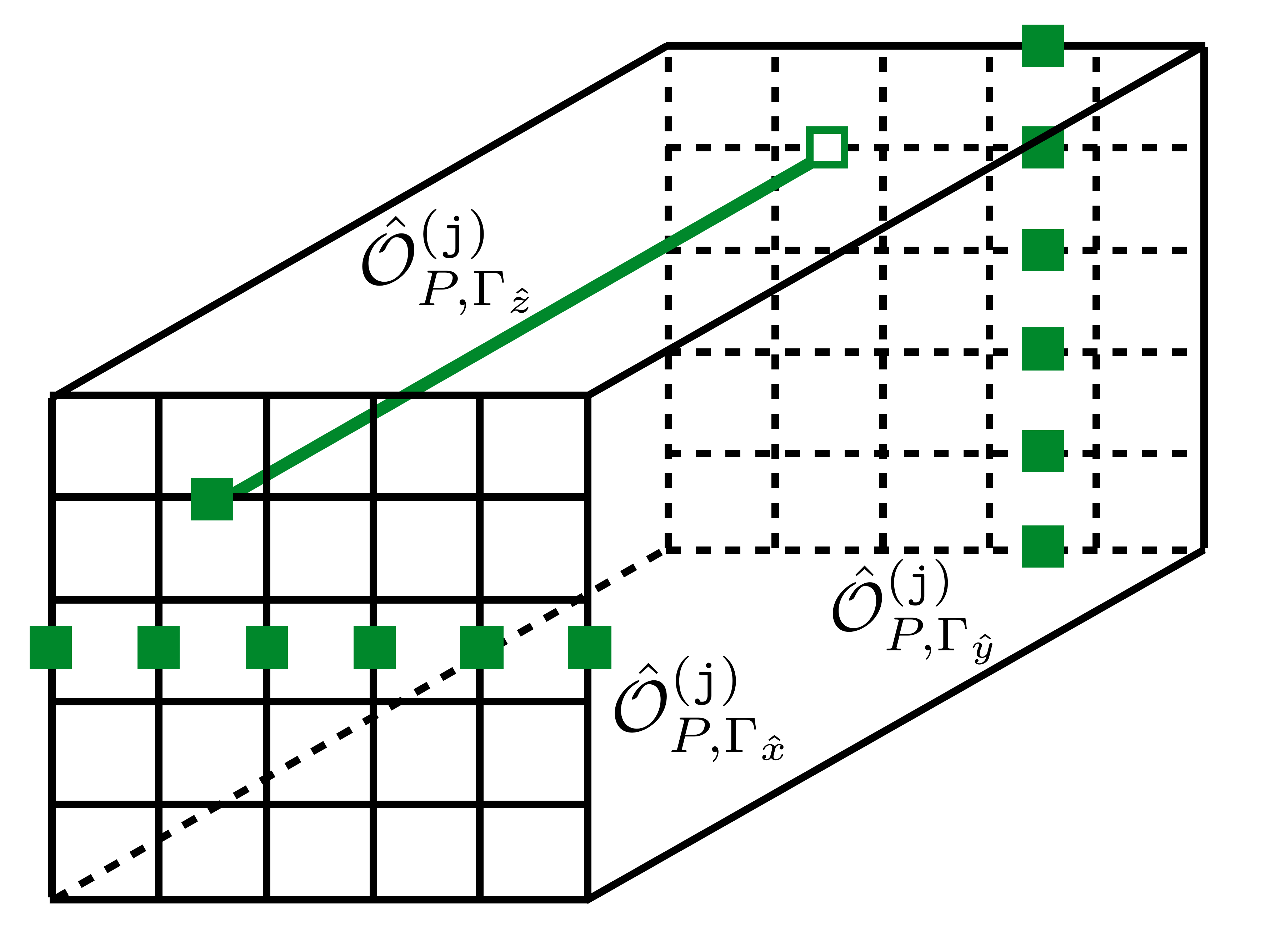}\\
\caption{(Color online)
Pictorial representation of the plaquette-type string
operators defined in Eqs.\
\eqref{eq: def Wilson loops}.  
The star-type string operators are defined similarly.
\label{fig: plaquette-type string operators}
         }
\end{center}
\end{figure}

First, for any $\texttt{j}=1,\cdots, M$, any cardinal directions
$C,C^\prime=N,W,S,E$, and any $0\leq z<L$,
we define the non-local \textit{string operators} 
\begin{subequations} 
\label{eq: def Wilson loops} 
\begin{align}
&
\hat{\mathcal{O}}^{(\texttt{j})}_{S,\Gamma^{\,}_{x}}(z)
\:=
\prod_{s^{\,}_{C}\in\Gamma^{\,}_{x}}
\hat{S}^{(\texttt{j})\dag}_{s^{\,}_{C}}(z), 
\label{eq: def Wilson loops a}  
\\
&
\hat{\mathcal{O}}^{(\texttt{j})}_{S,\Gamma^{\,}_{y}}(z)\:=
\prod_{s^{\,}_{C}\in\Gamma^{\,}_{y}} 
\hat{S}^{(\texttt{j})\dag}_{s^{\,}_{C}}(z),
\label{eq: def Wilson loops b}  
\\
&
\hat{\mathcal{O}}^{(\texttt{j})}_{S,\Gamma^{\,}_{z}}\:=
\hat{S}^{(\texttt{j})\dag}_{s^{\,}_{C}}(0,L),
\label{eq: def Wilson loops c} 
\end{align}
and
\begin{align}
&
\hat{\mathcal{O}}^{(\texttt{j})}_{P,\Gamma^{\,}_{\hat{x}}}(z)\:=
\prod_{p^{\,}_{C}\in\Gamma^{\,}_{\hat{x}}} 
\hat{P}^{(\texttt{j})\dag}_{p^{\,}_{C}}(z),
\label{eq: def Wilson loops d} 
\\
&
\hat{\mathcal{O}}^{(\texttt{j})}_{P,\Gamma^{\,}_{\hat{y}}}(z)\:=
\prod_{p^{\,}_{C}\in\Gamma^{\,}_{\hat{y}}} 
\hat{P}^{(\texttt{j})\dag}_{p^{\,}_{C}}(z),
\label{eq: def Wilson loops e} 
\\
&
\hat{\mathcal{O}}^{(\texttt{j})}_{P,\Gamma^{\,}_{\hat{z}}}\:=
\hat{P}^{(\texttt{j})\dag}_{p^{\,}_{C^\prime}}(0,L).
\label{eq: def Wilson loops f} 
\end{align}
\end{subequations}
Here, $\Gamma^{\,}_{x}$ is a non-contractible, directed,
closed path traversing the entire square lattice along the direction $x$, 
while $\Gamma^{\,}_{\hat{x}}$ is a non-contractible, directed, closed path 
traversing the entire dual lattice along the direction $\hat{x}$. 
[A directed path consists of the set of links,
either $s^{\,}_{C}\in\Gamma^{\,}_{x}$
or
$p^{\,}_{C}\in\Gamma^{\,}_{\hat{x}}$,
to be traversed according to the ordering in the product of
vertex operators on the right-hand sides of
Eqs.\ (\ref{eq: def Wilson loops a})
and
(\ref{eq: def Wilson loops b}), respectively.]
Similarly, the non-contractible, directed, closed paths 
$\Gamma^{\,}_{y}$ and $\Gamma^{\,}_{\hat{y}}$
traverse the square lattice
along the $y$- and $\hat{y}$-directions, respectively.
Finally, $\Gamma^{\,}_{z}$ and $\Gamma^{\,}_{\hat{z}}$ 
are non-contractible closed paths traversing a wire
in the $z$ and $\hat{z}$ directions, respectively.
On the right-hand sides of 
Eqs.\
(\ref{eq: def Wilson loops c})
and
(\ref{eq: def Wilson loops f}),
the choice of the link
$s^{\,}_{C}=s^{\,}_{N},s^{\,}_{W},s^{\,}_{S},s^{\,}_{E}$
and
$p^{\,}_{C^\prime}=p^{\,}_{N},p^{\,}_{W},p^{\,}_{S},p^{\,}_{E}$,
respectively,
is of no consequence for the purposes of computing the
topological degeneracy (see below).
For a pictorial representation of these string operators,
see Fig.~\ref{fig: plaquette-type string operators}.  
These operators can be interpreted 
as describing processes in which particle-antiparticle pairs of
different types of star or plaquette defects are created, and where
the particle propagates along a non-contractible loop that
encircles the entire torus before annihilating with its antiparticle.

\begin{figure}[t]
\begin{center}
(a)\includegraphics[width=.45\textwidth]{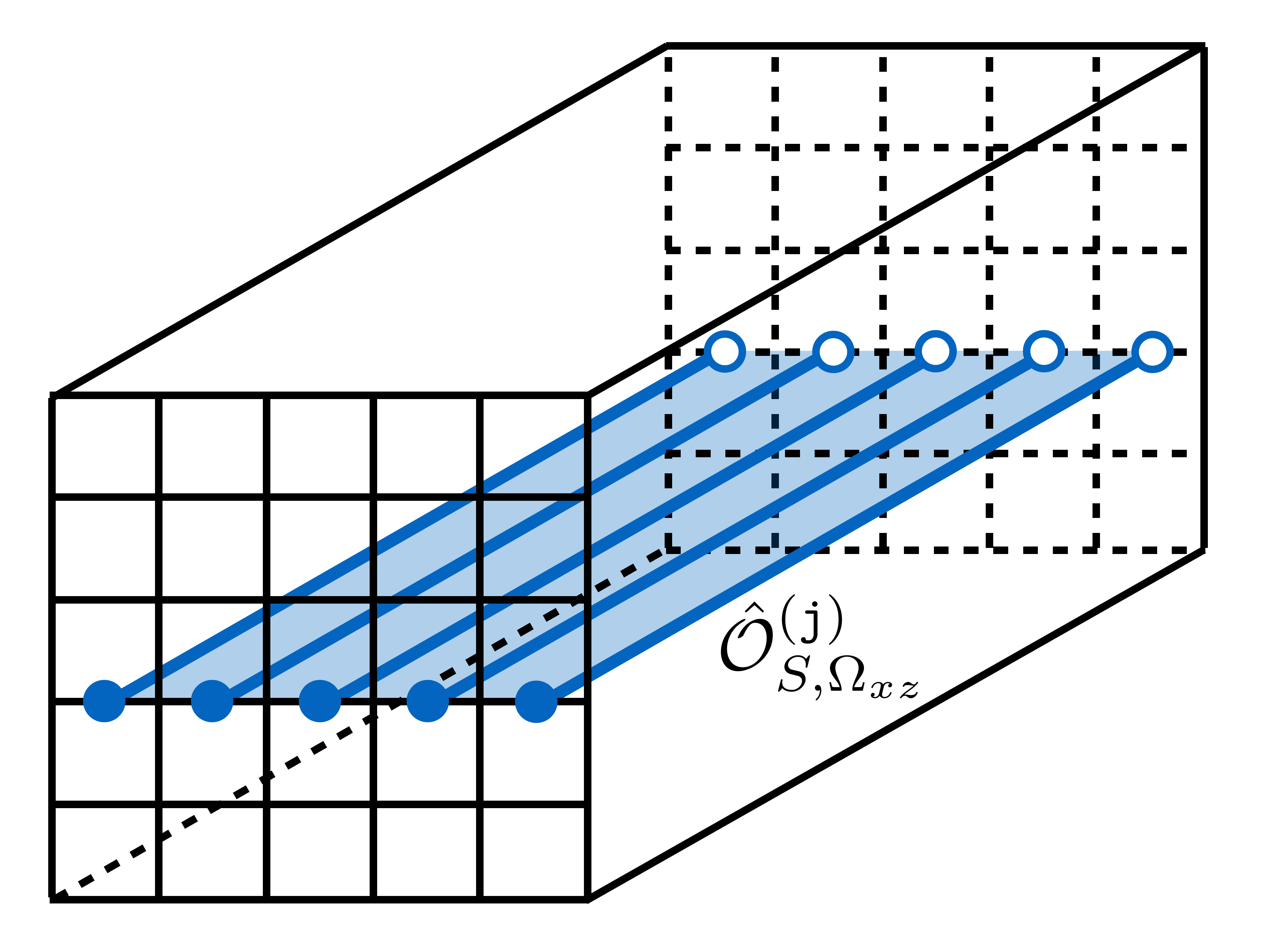}\\
(b)\includegraphics[width=.45\textwidth]{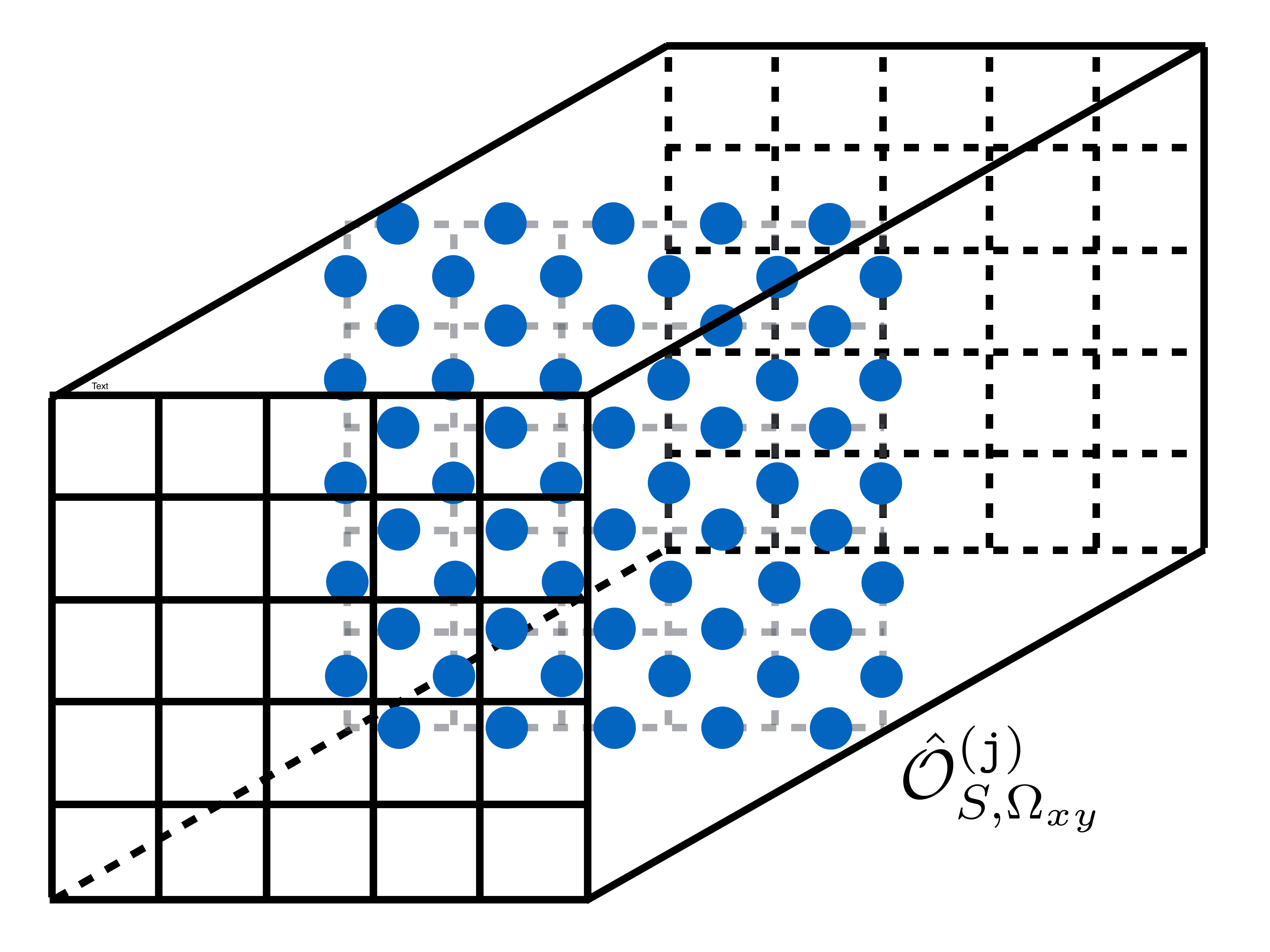}\\
\caption{(Color online)
Pictorial representations of the star-type membrane
operators defined in Eqs.\
\eqref{eq: def Wilson membranes}.  In (a), the membrane
is parallel to the $x$-$z$ plane (there is also a similarly-defined
membrane parallel to the $y$-$z$ plane).  In (b), the membrane is
parallel to the $x$-$y$ plane.  Plaquette-type membrane operators
are represented similarly.
\label{fig: membranes for topological degeneracy}
         }
\end{center}
\end{figure}

Second, for any $\texttt{j}=1,\cdots, M$ and any $0\leq z<L$,
we define the non-local \textit{membrane operators}
\begin{subequations}
\label{eq: def Wilson membranes}
\begin{align}
&
\hat{\mathcal{O}}^{(\texttt{j})}_{S,\Omega^{\,}_{xy}}(z)\:=
\prod_{s^{\,}_{C}\in\Omega^{\,}_{xy}} 
\hat{S}^{(\texttt{j})\dag}_{s^{\,}_{C}}(z),
\label{eq: def Wilson membranes a}
\\
&
\hat{\mathcal{O}}^{(\texttt{j})}_{S,\Omega^{\,}_{xz}}\:=
\prod_{s^{\,}_{C}\in\Gamma^{\,}_{x}}
\hat{S}^{(\texttt{j})\dag}_{s^{\,}_{C}}(0,L),
\label{eq: def Wilson membranes b}
\\
&
\hat{\mathcal{O}}^{(\texttt{j})}_{S,\Omega^{\,}_{yz}}\:=
\prod_{s^{\,}_{C}\in\Gamma^{\,}_{y}}
\hat{S}^{(\texttt{j})\dag}_{s^{\,}_{C}}(0,L),
\label{eq: def Wilson membranes c}
\end{align}
and
\begin{align}
&
\hat{\mathcal{O}}^{(\texttt{j})}_{P,\Omega^{\,}_{\hat{x}\hat{y}}}(z)\:=
\prod_{p^{\,}_{C}\in\Omega^{\,}_{\hat{x}\hat{y}}} 
\hat{P}^{(\texttt{j})\dag}_{p^{\,}_{C}}(z),
\label{eq: def Wilson membranes d}
\\
&
\hat{\mathcal{O}}^{(\texttt{j})}_{P,\Omega^{\,}_{\hat{x}\hat{z}}}\:=
\prod_{p^{\,}_{C}\in\Gamma^{\,}_{\hat{x}}}
\hat{P}^{(\texttt{j})\dag}_{p^{\,}_{C}}(0,L),
\label{eq: def Wilson membranes e}
\\
&
\hat{\mathcal{O}}^{(\texttt{j})}_{P,\Omega^{\,}_{\hat{y}\hat{z}}}\:=
\prod_{p^{\,}_{C}\in\Gamma^{\,}_{\hat{y}}}
\hat{P}^{(\texttt{j})\dag}_{p^{\,}_{C}}(0,L).
\label{eq: def Wilson membranes f}
\end{align}
\end{subequations}
Here, $\Omega^{\,}_{xy}$ is a membrane covering all links 
of the square lattice in the $x$-$y$ plane 
at a constant $z$, while
$\Omega^{\,}_{\hat{x}\hat{y}}$ is a membrane covering all links 
of the dual lattice in the $\hat{x}$-$\hat{y}$ plane  
at a constant $z$. The membranes
$\Omega^{\,}_{xz}$ ($\Omega^{\,}_{\hat{x}\hat{z}}$)
and
$\Omega^{\,}_{yz}$ ($\Omega^{\,}_{\hat{y}\hat{z}}$)
contain the  non-contractible closed paths
$\Gamma^{\,}_{x}$ ($\Gamma^{\,}_{\hat{x}}$)
and
$\Gamma^{\,}_{y}$ ($\Gamma^{\,}_{\hat{y}}$).
Similarly to the string operators, 
these membrane operators can be interpreted as
describing processes in which a linelike defect and its anti-defect 
are created as a pair, before one of the defects propagates along 
a non-contractible loop on the torus and annihilates with its partner.  
For pictorial representations of these membrane operators,
see Fig.~\ref{fig: membranes for topological degeneracy}.

Neither the string operators (\ref{eq: def Wilson loops})
nor the membrane operators (\ref{eq: def Wilson membranes})
create excitations, as understood in Sec.\
\ref{subsubsec: Pointlike and linelike excitations}, 
as the strings and membranes on which these operators act are always closed 
by virtue of the periodic boundary conditions we have imposed.  Consequently,
the string operators (\ref{eq: def Wilson loops})
and the membrane operators (\ref{eq: def Wilson membranes})
commute with the Hamiltonian 
$\tilde{\hat{H}}^{\,}_{\mathrm{int}}\equiv-
\tilde{\hat{L}}^{\,}_{\{\tilde{\mathcal{T}}\}}$
defined in Eq.\ (\ref{def tilde L as sum tilde L0 and tilde LT}).

For any $0\leq z<L$,
the set of $12\,M$ 
string operators (\ref{eq: def Wilson loops})
and membrane operators (\ref{eq: def Wilson membranes})
can be divided into two sets of
$6\,M$, with the equivalent algebras
\begin{subequations}
\label{eq: line-membrane 1}
\begin{align} 
&
\hat{\mathcal{O}}^{(\texttt{j})}_{S,\Gamma^{\,}_{z}}\, 
\hat{\mathcal{O}}^{(\texttt{j}^{\prime})}_{P,\Omega^{\,}_{\hat{x}\hat{y}}}=
\hat{\mathcal{O}}^{(\texttt{j}^{\prime})}_{P,\Omega^{\,}_{\hat{x}\hat{y}}}\, 
\hat{\mathcal{O}}^{(\texttt{j})}_{S,\Gamma^{\,}_{z}}\ 
e^{-\mathrm{i}\, 2\pi\, \tilde{v}^{(\texttt{j})\T}\tilde{K}^{-1}\tilde{w}^{(\texttt{j}^{\prime})}},
\label{xyz membrane1}\\
&
\hat{\mathcal{O}}^{(\texttt{j})}_{S,\Gamma^{\,}_{y}}\, 
\hat{\mathcal{O}}^{(\texttt{j}^{\prime})}_{P,\Omega^{\,}_{\hat{z}\hat{x}}}=
\hat{\mathcal{O}}^{(\texttt{j}^{\prime})}_{P,\Omega^{\,}_{\hat{z}\hat{x}}}\, 
\hat{\mathcal{O}}^{(\texttt{j})}_{S,\Gamma^{\,}_{y}}\,
e^{+\mathrm{i}\, 2\pi\, \tilde{v}^{(\texttt{j})\T}\tilde{K}^{-1}\tilde{w}^{(\texttt{j}^{\prime})}},
\label{xzy membrane1}\\
&
\hat{\mathcal{O}}^{(\texttt{j})}_{S,\Gamma^{\,}_{x}}\, 
\hat{\mathcal{O}}^{(\texttt{j}^{\prime})}_{P,\Omega^{\,}_{\hat{y}\hat{z}}}=
\hat{\mathcal{O}}^{(\texttt{j}^{\prime})}_{P,\Omega^{\,}_{\hat{y}\hat{z}}}\, 
\hat{\mathcal{O}}^{(\texttt{j})}_{S,\Gamma^{\,}_{x}}\ 
e^{+\mathrm{i}\, 2\pi\, \tilde{v}^{(\texttt{j})\T}\tilde{K}^{-1}\tilde{w}^{(\texttt{j}^{\prime})}}
\label{yzx membrane1},
\end{align}
\end{subequations}
and
\begin{subequations}
\label{eq: line-membrane 2}
\begin{align}
&
\hat{\mathcal{O}}^{(\texttt{j})}_{S,\Omega^{\,}_{xy}}\, 
\hat{\mathcal{O}}^{(\texttt{j}^{\prime})}_{P,\Gamma^{\,}_{\hat{z}}}=
\hat{\mathcal{O}}^{(\texttt{j}^{\prime})}_{P,\Gamma^{\,}_{\hat{z}}}\, 
\hat{\mathcal{O}}^{(\texttt{j})}_{S,\Omega^{\,}_{xy}}\, 
e^{+\mathrm{i}\,2\pi\,\tilde{v}^{(\texttt{j})\T}\tilde{K}^{-1}\tilde{w}^{(\texttt{j}^{\prime})}},
\label{xyz membrane2}\\
&
\hat{\mathcal{O}}^{(\texttt{j})}_{S,\Omega^{\,}_{zx}}\, 
\hat{\mathcal{O}}^{(\texttt{j}^{\prime})}_{P,\Gamma^{\,}_{\hat{y}}}=
\hat{\mathcal{O}}^{(\texttt{j}^{\prime})}_{P,\Gamma^{\,}_{\hat{y}}}\, 
\hat{\mathcal{O}}^{(\texttt{j})}_{S,\Omega^{\,}_{zx}}\, 
e^{-\mathrm{i}\,2\pi\,\tilde{v}^{(\texttt{j})\T}\tilde{K}^{-1}\tilde{w}^{(\texttt{j}^{\prime})}},
\label{xzy membrane2}\\
&
\hat{\mathcal{O}}^{(\texttt{j})}_{S,\Omega^{\,}_{yz}}\, 
\hat{\mathcal{O}}^{(\texttt{j}^{\prime})}_{P,\Gamma^{\,}_{\hat{x}}}=
\hat{\mathcal{O}}^{(\texttt{j}^{\prime})}_{P,\Gamma^{\,}_{\hat{x}}}\, 
\hat{\mathcal{O}}^{(\texttt{j})}_{S,\Omega^{\,}_{yz}}\, 
e^{-\mathrm{i}\,2\pi\,\tilde{v}^{(\texttt{j})\T}\tilde{K}^{-1}\tilde{w}^{(\texttt{j}^{\prime})}},
\label{yzx membrane2}
\end{align}
\end{subequations}
respectively.
Note that all string and membrane operators associated with stars commute 
with one another, as do all string and membrane operators associated 
with plaquettes.  Note also that there are, in principle, four equivalent copies of
Eqs.~\eqref{xyz membrane1} and \eqref{xyz membrane2}, one for
each choice of cardinal direction $C$ or $C^\prime$ in
Eqs.~\eqref{eq: def Wilson loops c} and \eqref{eq: def Wilson loops f},
respectively.  However, because we have chosen the vectors
$\tilde{v}^{(\texttt{j})}$ and $\tilde{w}^{(\texttt{j}^{\prime})}$ in an isotropic
way [i.e., by imposing the criterion \eqref{simplifying condition}], these four
copies of Eqs.~\eqref{xyz membrane1} and \eqref{xyz membrane2}
are redundant.  We will henceforth work with fixed cardinalities
$C$ and $C^\prime$ in Eqs.~\eqref{eq: def Wilson loops c} and
\eqref{eq: def Wilson loops f}, respectively.

We are after the minimal topological ground-state degeneracy that
is consistent with Eqs.\ \eqref{eq: line-membrane 1} 
and \eqref{eq: line-membrane 2}.
There are redundancies among the $12\,M$ operators defined in
Eqs.\ \eqref{eq: def Wilson loops} 
and \eqref{eq: def Wilson membranes} 
that reduce the total number of independent 
relations
in Eqs.\ (\ref{eq: line-membrane 1})
and (\ref{eq: line-membrane 2})
to $3\,M$.  For example, observe that
\begin{subequations}
\begin{align}
\hat{\mathcal{O}}^{(\texttt{j})}_{S,\Gamma^{\,}_{x}}\, 
\hat{\mathcal{O}}^{(\texttt{j}^{\prime})}_{P,\Gamma^{\,}_{\hat{z}}}=
\hat{\mathcal{O}}^{(\texttt{j}^{\prime})}_{P,\Gamma^{\,}_{\hat{z}}}\, 
\hat{\mathcal{O}}^{(\texttt{j})}_{S,\Gamma^{\,}_{x}}\, 
e^{+\mathrm{i}\, 2\pi\, \tilde{v}^{(\texttt{j})\T}\tilde{K}^{-1}\tilde{w}^{(\texttt{j}^{\prime})}}.
\end{align}
It is consistent with
Eqs.\ \eqref{yzx membrane1} and \eqref{xyz membrane2}, 
to make either the identification
$\hat{\mathcal{O}}^{(\texttt{j})}_{S,\Gamma^{\,}_{x}}\equiv
\hat{\mathcal{O}}^{(\texttt{j})}_{S,\Omega^{\,}_{xy}}$
or the identification
$\hat{\mathcal{O}}^{(\texttt{j}^{\prime})}_{P,\Gamma^{\,}_{\hat{z}}}\equiv
\hat{\mathcal{O}}^{(\texttt{j}^{\prime})}_{P,\Omega^{\,}_{\hat{y}{\hat{z}}}}$ 
for all $\texttt{j},\texttt{j}'=1,\dots,M$ when acting on the ground-state
subspace. This indicates that one can remove either 
Eq.\ \eqref{yzx membrane1} or Eq.\ \eqref{xyz membrane2} 
from the algebra without changing 
the number of independent degrees of freedom.  
For concreteness, suppose we do away with Eq.\ \eqref{xyz membrane2}.
Then, similarly, using the relations
\begin{align}
&
\hat{\mathcal{O}}^{(\texttt{j})}_{S,\Gamma^{\,}_{z}}\, 
\hat{\mathcal{O}}^{(\texttt{j}^{\prime})}_{P,\Gamma^{\,}_{\hat{x}}}=
\hat{\mathcal{O}}^{(\texttt{j}^{\prime})}_{P,\Gamma^{\,}_{\hat{x}}}\, 
\hat{\mathcal{O}}^{(\texttt{j})}_{S,\Gamma^{\,}_{z}}\, 
e^{-\mathrm{i}\, 2\pi\, \tilde{v}^{(\texttt{j})\T}\tilde{K}^{-1}\tilde{w}^{(\texttt{j}^{\prime})}}, 
\\
&
\hat{\mathcal{O}}^{(\texttt{j})}_{S,\Gamma^{\,}_{z}}\, 
\hat{\mathcal{O}}^{(\texttt{j}^{\prime})}_{P,\Gamma^{\,}_{\hat{y}}}=
\hat{\mathcal{O}}^{(\texttt{j}^{\prime})}_{P,\Gamma^{\,}_{\hat{y}}}\, 
\hat{\mathcal{O}}^{(\texttt{j})}_{S,\Gamma^{\,}_{z}}\, 
e^{-\mathrm{i}\, 2\pi\, \tilde{v}^{(\texttt{j})\T}\tilde{K}^{-1}\tilde{w}^{(\texttt{j}^{\prime})}},
\end{align}
\end{subequations}
we can remove Eqs.\ \eqref{xzy membrane2} and \eqref{yzx membrane2}
from the algebra.  With the redundant operators removed, we are left
with a set of $6\,M$
nonlocal operators obeying the algebra of
Eqs.\ \eqref{eq: line-membrane 1}.

The ground-state degeneracy on the three-torus 
\begin{equation}
\mathbb{T}^{3}\equiv S^{1}\times S^{1}\times S^{1}
\end{equation}
in the strong coupling limit where
the kinetic contribution to the Hamiltonian
$\tilde{\hat{H}}$
defined 
in Eq.\ 
(\ref{def tilde L as sum tilde L0 and tilde LT})
is much smaller than the contribution from
$\tilde{\hat{H}}^{\,}_{\mathrm{int}}\equiv-
\tilde{\hat{L}}^{\,}_{\{\tilde{\mathcal{T}}\}}$
can be deduced from the
algebra \eqref{eq: line-membrane 1} as follows.  
Close to the limit $\tilde{\hat{H}}=\tilde{\hat{H}}^{\,}_{\mathrm{int}}$,
the ground-state manifold must
transform as a representation of the algebra \eqref{eq: line-membrane 1}.
If so, the representation of the algebra
\eqref{eq: line-membrane 1} 
with the smallest dimension
determines the minimal topological ground-state degeneracy.
Equations~\eqref{eq: line-membrane 1} consist of three independent copies of
the generalized ``magnetic algebra," which is ubiquitous in studies of
the ground-state degeneracy of abelian topological states of
matter~\cite{Wen91a,Wen92,Wesolowski94}.  The minimum-dimensional 
representation of any one of the three algebras in 
Eqs.\ \eqref{eq: line-membrane 1} 
has dimension $|\det\varkappa|$, where
\begin{align}\label{kappa def}
\varkappa^{-1}_{\texttt{j}\texttt{j}^{\prime}}\:= 
\tilde{v}^{(\texttt{j})\T}\,
\tilde{K}^{-1}\,
\tilde{w}^{(\texttt{j}^{\prime})}=
\frac{\theta^{\,}_{\texttt{j}\texttt{j}^{\prime}}}{2\pi}
\end{align}
is an $M\times M$-dimensional symmetric matrix~\cite{footnote_kappa_tilde}.
We conclude that the class of coupled wires considered in this work
has a ground-state degeneracy $D^{\,}_{\mathbb{T}^{3}}$ on the three-torus given by
\begin{equation}\label{eq: topo degeneracy on three torus} 
D^{\,}_{\mathbb{T}^{3}}=
|\det\varkappa|^{3}.
\end{equation}

Combining Eq.\ (\ref{eq: topo degeneracy on three torus})
with the definition of the matrix
$\varkappa$ provided in Eq.\ \eqref{kappa def}, one can verify the claim
made earlier in this section, namely that $D^{\,}_{\mathbb{T}^{3}}=1$ if
$|\det\tilde{K}=1|$.  To see this, recall that the inverse of the
matrix $\tilde{K}$ is given by
\begin{align}
\tilde{K}^{-1}= 
\frac{1}{\det\,\tilde{K}}\,C^{\,}_{\tilde{K}},
\end{align}
where $C^{\,}_{\tilde{K}}$ is the cofactor matrix associated with
$\tilde{K}$.  Since $\tilde{K}$ is an integer-valued matrix, it follows
that $C^{\,}_{\tilde{K}}$ is also integer valued, and that $\det\,\tilde{K}$ 
is an integer.  Combining these facts with our assumptions that
$\tilde{v}^{(\texttt{j})}$ and $\tilde{w}^{(\texttt{j})}$ are 
integer-valued and that $\det\,\tilde{K}=\pm1$,
one concludes that
$\varkappa^{-1}_{\texttt{j}\texttt{j}^{\prime}}$ is an integer for all
$\texttt{j}$ and $\texttt{j}^{\prime}=1,\ldots,M$.  Consequently, each
line of Eqs.\ \eqref{eq: line-membrane 1} becomes a trivial commutation
relation for all $\texttt{j}$ and $\texttt{j}^{\prime}$, and we conclude
that $D^{\,}_{\mathbbm{T}^{3}}=1$.

Nontrivial states of matter for which $\det D^{\,}_{\mathbb{T}^{3}}=1$
are examples of short-range entangled (SRE) or symmetry-protected
topological (SPT) states of matter~\cite{Pollmann10,Chen13}.  Although such states of matter do
not yield quasiparticle excitations with fractionalized charges or
statistics, and are therefore not of primary interest to us here, they
are nevertheless readily treated within the formalism developed in
this paper.

\subsubsection{Topological field theory}
\label{subsubsec: Topological field theory}

We close the discussion of the general class of three-dimensional wire
constructions considered in this work by commenting on the topological
field theory characterizing the low-energy behavior of these theories.
In the study of the braiding statistics of quasiparticle excitations
undertaken in Sec.\
\ref{subsubsec: Statistics of point, line, and membrane defects}, 
we found that these wire constructions host both pointlike and 
stringlike excitations of $M$ types, 
labeled by $\texttt{j}=1,\ldots,M$. We also observed that winding
a pointlike defect of type $\texttt{j}$ around a stringlike defect of
type $\texttt{j}^{\prime}$ yields a statistical phase 
$\theta^{\,}_{\texttt{j}\texttt{j}^{\prime}}$, 
and that all other statistical phases were trivial.

We wish to capture this statistical ``interaction" between
quasiparticles with a topological field theory, in a manner similar to
the way in which Chern-Simons (CS) theories in (2+1) dimensions can be
used to encode the statistics of pointlike quasiparticles.  Studies of
topologically-ordered superconductors~\cite{Hansson04} and
(3+1)-dimensional topological
insulators~\cite{Cho11,Chan13} have shown that the
statistics of theories where pointlike excitations acquire a nontrivial
phase when encircling vortex lines can be encoded in so-called BF
theories.  For example, in a (3+1)-dimensional topolgical insulator,
the statistical phase of $\pi$ that a quasiparticle acquires when it
circles a vortex line is encoded in the BF
Lagrangian density~\cite{Hansson04,Cho11,Tiwari14}
\begin{align}
\mathcal{L}^{\,}_{\mathrm{BF}}&\:=
\frac{1}{2\pi}\, 
\epsilon^{\mu\nu\rho\lambda}\, 
a^{\,}_\mu\, 
\partial^{\,}_\nu\, 
b^{\,}_{\rho\lambda},
\end{align}
where $\mu=t,x,y,z$ runs over all spacetime indices,
$\epsilon^{\mu\nu\rho\lambda}$ is the fully antisymmetric Levi-Civita symbol, 
and summation over repeated Greek indices is implied.
Here, the one-form $a^{\,}_{\mu}$ is an emergent gauge field
that couples to the quasiparticle current density, and
$b^{\,}_{\mu\nu}$ is an antisymmetric two-form that couples to the
vortex-line density.  The natural generalization of this BF Lagrangian
to our setting is obtained by introducing $M$ species of one-forms
$a^{(\texttt{j})}_{\mu}$ and $M$ species of two-forms 
$b^{(\texttt{j})}_{\mu\nu}$, 
one for each type of pointlike and stringlike excitation, respectively.  
This results in the multicomponent BF Lagrangian density
\begin{align}\label{eq: def mathcal L BF}
\mathcal{L}^{\,}_{\mathrm{BF}}\:=
\frac{\varkappa^{\,}_{\texttt{j}\texttt{j}^{\prime}}}{4\pi}\, 
\epsilon^{\mu\nu\rho\lambda}\, 
a^{(\texttt{j})}_{\mu}\, 
\partial^{\,}_{\nu}\, 
b^{(\texttt{j}^{\prime})}_{\rho\lambda},
\end{align}
where the $M\times M$ matrix $\varkappa$ is defined in 
Eq.\ \eqref{kappa def}
and summation over repeated Greek and teletype
indices is implied.
This discussion indicates that 
the class of coupled wires considered so far
falls into the same equivalence
class of topological states of matter as the (3+1)-dimensional
fractional topological
insulators~\cite{Swingle11,Maciejko10,Maciejko14,Tiwari14,Sagi15b}.  
This is consistent with the example of $\mathbb{Z}^{\,}_{m}$ topological order in
three spatial dimensions that we discuss in the next section.

Before moving on, we address the question of how this discussion would
have been different if we had instead considered the more general case
$\tilde{v}^{(\texttt{j})}_{1}\neq\tilde{v}^{(\texttt{j})}_{2}$ and
$\tilde{w}^{(\texttt{j})}_{1}\neq\tilde{w}^{(\texttt{j})}_{2}$
[recall 
Eqs. (\ref{eq: 2D criteria}) and (\ref{eq: simplest 2D criteria})].  
As we observed after 
Eq.\ (\ref{eq: chosen case for statistical angle between linelike}), 
this more general case requires us to \textit{impose} the consistency condition
\eqref{eq: general case for statistical angle between linelike} in
order for the statistics of pointlike and linelike excitations to be
well-defined.  However, because, in this case, there is a well-defined
statistical angle $\theta^{\,}_{\texttt{j}\texttt{j}^{\prime}}$, one may
\textit{define} the matrix $\varkappa^{-1}_{\texttt{j}\texttt{j}^{\prime}}$ 
in terms of $\theta^{\,}_{\texttt{j}\texttt{j}^{\prime}}$ 
by making use of the relation \eqref{kappa def}, 
leading again to the multicomponent BF theory defined in
Eq.\ (\ref{eq: def mathcal L BF}).
This observation can be taken as a justification \textit{a posteriori}
for considering from the outset, as we did, the simpler class of
models in which
$\tilde{v}^{(\texttt{j})}_{1}=\tilde{v}^{(\texttt{j})}_{2}=\tilde{v}^{(\texttt{j})}$
and
$\tilde{w}^{(\texttt{j})}_{1}=\tilde{w}^{(\texttt{j})}_{2}=\tilde{w}^{(\texttt{j})}$.

\subsection{Example: $\mathbb{Z}^{\,}_{m}$ topological order in 
three-dimensional space from coupled wires}
\label{subsec: Zm example}

Having developed a toolbox for the construction of a class of
two-dimensional arrays of coupled quantum wires, we now turn to an
illustration of this framework in action.  In this section, we show
how to realize the simplest type of three-dimensional topological
order, namely $\mathbb{Z}^{\,}_{m}$ topological order, within the wire
formalism developed in the previous sections.  This class of examples
includes the three-dimensional toric code, 
which is an example of $\mathbb{Z}^{\,}_{2}$
topological order.

\subsubsection{Definitions and interwire couplings}
\label{subsubsec: Definitions and interwire couplings}

Our starting point is a set of $2N$ decoupled two-component
\textit{bosonic} quantum wires placed on the links of a square
lattice. (We will also discuss momentarily how one can arrive at a
class of $\mathbbm{Z}^{\,}_{2m}$-topologically-ordered states starting from
fermions, although it turns out to be simpler to focus on the bosonic
case.)  We take the decoupled quantum wires to be described by the
Lagrangian \eqref{L0 d=2} with
\begin{subequations}
\label{eq: step 1 to define Zm}
\begin{align}
\mathcal{K}\:= 
\mathbbm 1^{\,}_{2N}\otimes K^{\,}_{\mathrm{b}},
\label{eq: step 1 to define Zm a}
\end{align}
where $K^{\,}_{\mathrm{b}}$ was defined in Eq.~\eqref{eq: def bosonic K b},
and we take $M=1$ so that $K^{\,}_{\mathrm{b}}$ is a $2\times 2$ matrix.
With the $K$-matrix defined in this way, the canonical equal-time
commutation relation for the theory of decoupled wires is given by
\begin{align}
\begin{split}
\[
\partial^{\,}_{z}\hat{\phi}^{\,}_{j,1}(z),
\hat{\phi}^{\,}_{\pri{j},2}(\pri{z})
\]&=
\mathrm{i}\,2\pi\, 
\delta^{\,}_{j \pri{j}}\,  
\delta(z-\pri{z})\\
&=\[
\partial^{\,}_{z}\hat{\phi}^{\,}_{j,2}(z),
\hat{\phi}^{\,}_{\pri{j},1}(\pri{z})
\].
\end{split}
\label{eq: step 1 to define Zm b}
\end{align}
The charge vector that fixes the coupling of the two bosonic fields
$\hat{\phi}^{\,}_{j,1}$ and $\hat{\phi}^{\,}_{j,2}$ to external gauge potentials
is given by
\begin{align}
Q^{\,}_{\mathrm{b}}\:=
2\begin{pmatrix}1 & 0 \end{pmatrix}^{\T},
\label{eq: step 1 to define Zm c}
\end{align}
\end{subequations}
so that $\hat{\phi}^{\,}_{j,1}$ can be interpreted as the ``charge" mode and 
$\hat{\phi}^{\,}_{j,2}$
can be interpreted as the ``spin" mode.

It is convenient to write down the interwire couplings for this model
in the new basis defined by the transformation 
\eqref{eq: def W transformation} with
\begin{subequations}
\label{eq: Zm K and Q tilde}
\begin{align}
W\:=\mathrm{diag}(1,m),
\label{eq: Zm W to tilde}
\end{align}
so that the transformed $K$-matrix and charge-vector are given by
\begin{align}
&
\tilde{K}^{\,}_{m}\:=
\begin{pmatrix}
0&m
\\
m&0
\end{pmatrix},
\label{eq: Zm K tilde}
\\
&
\tilde{Q}\:= 
Q^{\,}_{\mathrm{b}}.
\label{eq: Zm Q tilde}
\end{align}
\end{subequations}
In this example, we will impose time-reversal symmetry (TRS), which
constrains the allowed interwire couplings.  TRS acts on the bosonic
fields as
\begin{align}\label{2D Zm TRS action}
\tilde{\hat{\phi}}_{j,\alpha}(t,z)\mapsto 
(-1)^{\alpha-1}\,\tilde{\hat{\phi}}_{j,\alpha}(-t,z)
\end{align}
for all $j = 1,\dots 2N$ and $\alpha = 1,2$.  Note that this is not the
only possible choice for the action of TRS (see, e.g., Ref.~\cite{Neupert11b}),
but that this representation of TRS squares to unity, as expected for
bosons.

Before proceeding to write down the interwire couplings, we first
point out that a theory similar to the one defined by the universal
data \eqref{eq: Zm K and Q tilde} can also be reached starting from 
wires supporting spinless fermions
defined by the data
\begin{subequations}
\begin{align}
&
K^{\,}_{\mathrm{f}}\:=
\begin{pmatrix}
+1&0
\\
0&-1
\end{pmatrix},
\\
&
Q^{\,}_{\mathrm{f}}\:= 
\begin{pmatrix}1 & 1\end{pmatrix}^{\T},
\end{align}
\end{subequations}
using the transformation
\begin{align}\label{W prime}
W^{\prime} 
\:= 
\begin{pmatrix}
-1 & -m
\\
-1 & +m
\end{pmatrix}.
\end{align}
In this alternative interpretation of Eqs.~\eqref{eq: Zm K and Q tilde}, 
we view the original bosons as being composite objects consisting of
paired fermions, since the transformed $K$-matrix and charge vector read 
\begin{subequations}
\label{eq: Zm K and Q tilde prime}
\begin{align}
&
\tilde{K}^{\prime}_{m}\:=
W^{\prime\T}\, K^{\,}_{\mathrm{f}}\,W^{\prime}=
2\,\tilde{K}^{\,}_{m},
\label{eq: Zm K and Q tilde prime a}
\\
&
\tilde{Q}^{\prime}_{m}\:=
W^{\prime\T}\,Q^{\,}_{\mathrm{f}} = 
-Q^{\,}_{\mathrm{b}}.
\label{eq: Zm K and Q tilde prime b}
\end{align}
\end{subequations}
The additional multiplicative factors of $2$ 
on the right-hand sides of the above equations
can be seen as evidence of this pairing.  Furthermore,
the action of TRS on the bosonic fields after performing the
transformation \eqref{W prime} is still given by
\eqref{2D Zm TRS action}, indicating that the theory
defined by the data \eqref{eq: Zm K and Q tilde prime} and the
theory defined by the data \eqref{eq: Zm K and Q tilde} transform
in the same way under TRS.

Hence, although we choose to focus here on the bosonic case, with universal
data given by Eqs.~\eqref{eq: Zm K and Q tilde}, all results that follow
could be interpreted as arising from paired fermions, so long as $m$ is taken
to be even.

\begin{figure}[t]
\begin{center}
\begin{flushleft}
\hspace{.5cm} (a)
\end{flushleft}
\includegraphics[width=.4\textwidth]{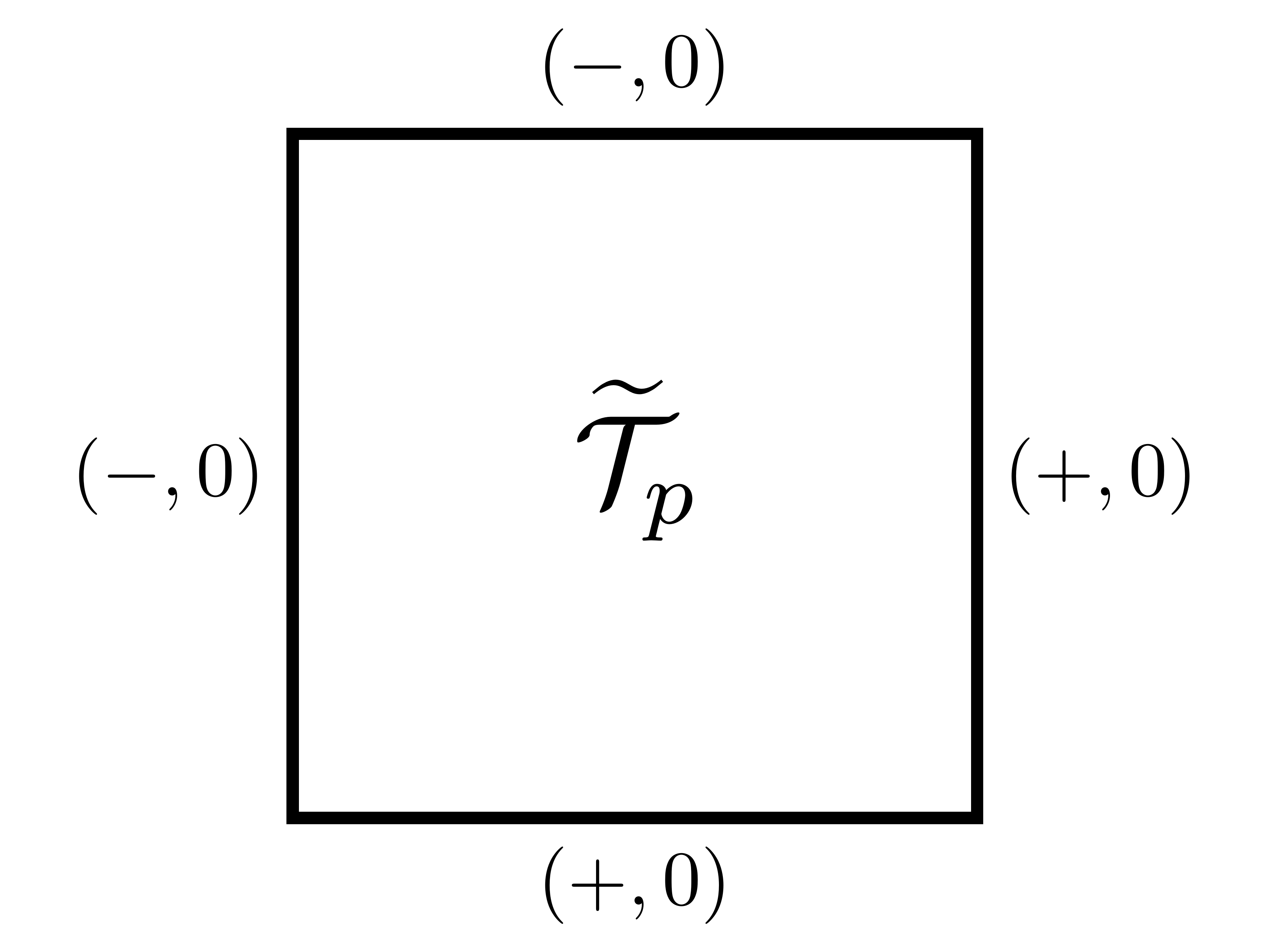}\\
\begin{flushleft}
\hspace{.5cm} (b)
\end{flushleft}
\includegraphics[width=.4\textwidth]{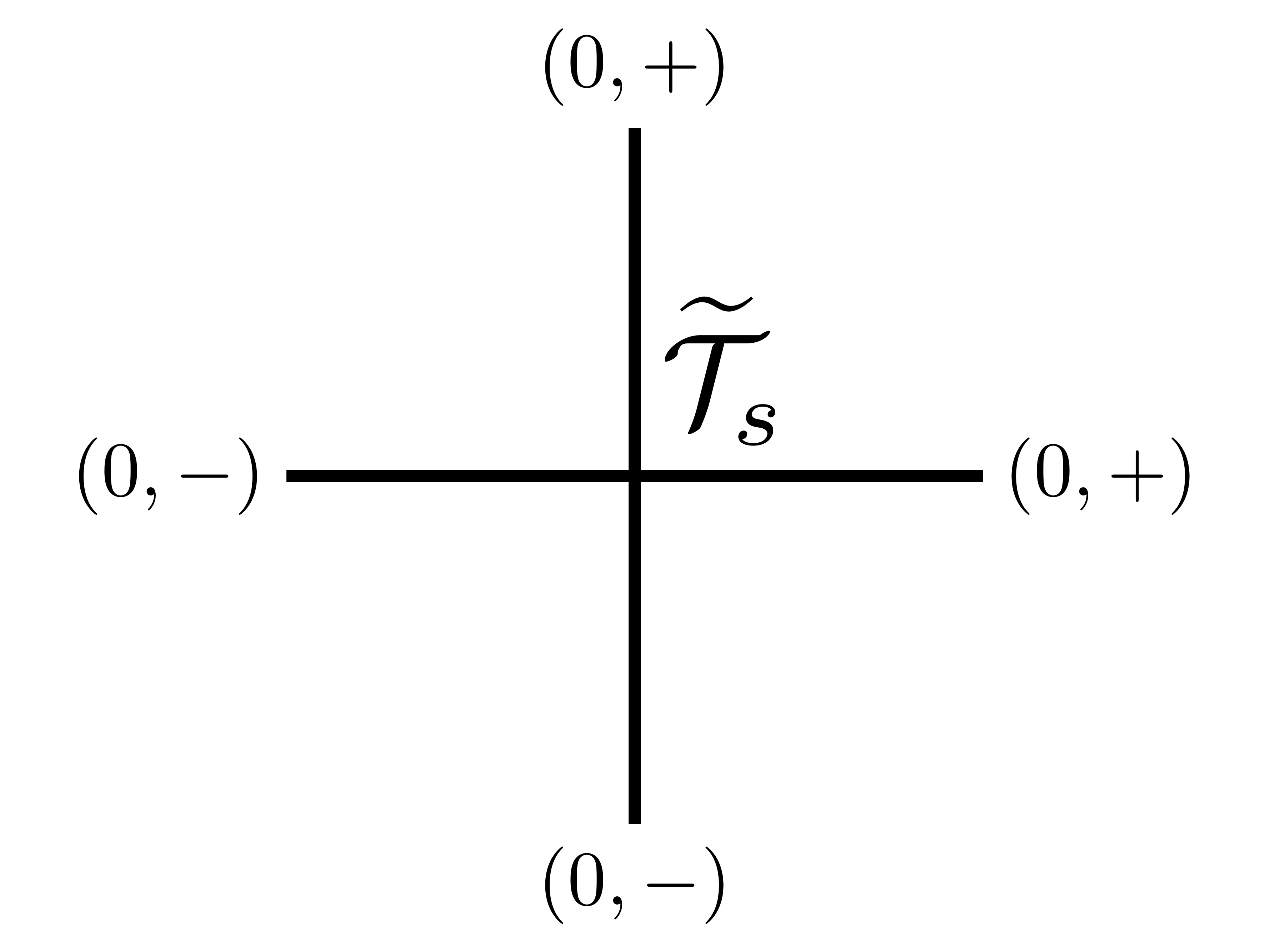}
\caption{
Pictorial representations of the tunneling vectors 
(a) $\tilde{\mathcal{T}}^{\,}_{s}$ 
and
(b) $\tilde{\mathcal{T}}^{\,}_{p}$ 
built using the vectors $\tilde{v}$ and $\tilde{w}$ 
defined in Eqs.\ (\ref{eq: Zm 2D tunneling vecs}).
The signs $\pm$ indicate whether a $\pm 1$
appears in the tunneling vector associated with that link.
\label{fig: Zm 2D tunneling vecs}
         }
\end{center}
\end{figure}

We couple the $2N$ quantum wires  with a Lagrangian
$\tilde{\hat{L}}^{\,}_{\{\tilde{\mathcal{T}}\}}$, defined as in
Eq.\ \eqref{Lint}, for tunneling vectors $\tilde{\mathcal{T}}$ defined
as in Eqs.\ \eqref{2d T-vecs charge conserving} with 
(see Fig.\ \ref{fig: Zm 2D tunneling vecs})
\begin{subequations}
\label{eq: Zm 2D tunneling vecs}
\begin{align}
&
\tilde{v}^{\,}_{1}\equiv
\tilde{v}^{\,}_{2}\equiv
\tilde{v}\:=
\begin{pmatrix}0 & +1\end{pmatrix}^{\T},
\\
&
\tilde{w}^{\,}_{1}\equiv
\tilde{w}^{\,}_{2}\equiv
\tilde{w}\:=
\begin{pmatrix}-1 & 0\end{pmatrix}^{\T}.
\end{align}
\end{subequations}
It is readily verified that these tunneling vectors satisfy the
criteria \eqref{eq: simplest 2D criteria}, which ensure that the
interaction terms in $\tilde{\hat{L}}^{\,}_{\{\tilde{\mathcal{T}}\}}$
are sufficient to gap out the array of quantum wires
when periodic boundary conditions are imposed.  Furthermore,
the cosine terms associated with the tunneling vectors 
\eqref{eq: Zm 2D tunneling vecs}
are even under TRS, as desired.

\subsubsection{Excitations}
\label{subsubsec: Excitations}

Excitations of the array of coupled wires can be constructed using
the procedure outlined in Sec.\
\ref{subsubsec: Pointlike and linelike excitations}. 
 
First, we define the local vertex operators
\begin{subequations}
\label{eq: 2D Zm defect hopping}
\begin{align}\label{eq: 2D Zm defect hopping a}
\hat{S}^{\dag}_{s^{\,}_{C}}(z)\:=
\exp
\(
-\mathrm{i}\, 
\tilde{\hat{\phi}}^{\,}_{s^{\,}_{C},2}(z)
\)
\end{align}
and
\begin{align}\label{eq: 2D Zm defect hopping b}
\hat{P}^{\dag}_{p^{\,}_{C}}(z)\:=
\exp
\(
+\mathrm{i}\, 
\tilde{\hat{\phi}}^{\,}_{p^{\,}_{C},1}(z)
\).
\end{align}
\end{subequations}
These vertex operators are eigenstates
of the charge operator
$\tilde{\hat{Q}}^{\,}_{j,\alpha}$ defined in Eq.\ \eqref{Q tilde}
with the $\tilde{K}$ matrix 
(\ref{eq: Zm K and Q tilde prime a})
and the charge vector 
(\ref{eq: Zm K and Q tilde prime b}),
respectively. Indeed,
following the derivation of
Eq.\ (\ref{eq: q is eigenstate of charge Q}),
we find the equal-time commutators
\begin{subequations}
\label{eq: Zm charges for star and plaquette operators}
\begin{align}
&
\[\tilde{\hat{Q}}^{\,}_{j,\alpha},\hat{S}^{\dag}_{s^{\,}_{C}}(z)\]=
+
\frac{2}{m}\, 
\delta^{\,}_{j,s^{\,}_{C}}\, 
\delta^{\,}_{\alpha,1}\,
\hat{S}^{\dag}_{s^{\,}_{C}}(z),
\label{eq: Zm charges for star and plaquette operators a}
\\
&
\[\tilde{\hat{Q}}^{\,}_{j,\alpha},\hat{P}^{\dag}_{p^{\,}_{C}}(z)\]=
0.
\label{eq: Zm charges for star and plaquette operators b}
\end{align}
\end{subequations}
The meaning of Eq.\
(\ref{eq: Zm charges for star and plaquette operators a})
is that the vertex operator $\hat{S}^{\dag}_{s^{\,}_{C}}(z)$ creates along the wire 
$j$ piercing the midpoint of the bond $s^{\,}_{C}$ 
($C=N,W,S,E$)
belonging to the star $s$ an excitation with charge $2/m$ for the flavor
$\alpha=1$.
The meaning of Eq.\
(\ref{eq: Zm charges for star and plaquette operators b})
is that $\hat{P}^{\dag}_{s^{\,}_{C}}(z)$ creates a charge-neutral 
excitation.

A second attribute of these quasi-particle operators is that they 
create fractional kinks in the charge-neutral operators
\begin{subequations}
\label{eq: pinned fields if Zm topoorder}
\begin{equation}
\begin{split}
\hat{T}^{\,}_{s}(z)\:=&\,
\frac{1}{m}\,
\tilde{\mathcal{T}}^{\T}_{s}\,
\tilde{\mathcal{K}}^{\,}_{m}\tilde{\hat\Phi}(z)
\\
=&\,
\tilde{\hat{\phi}}^{\,}_{s^{\,}_{E},1}(z)
-
\tilde{\hat{\phi}}^{\,}_{s^{\,}_{W},1}(z)
+
\tilde{\hat{\phi}}^{\,}_{s^{\,}_{N},1}(z)
-
\tilde{\hat{\phi}}^{\,}_{s^{\,}_{S},1}(z)
\end{split}
\label{eq: pinned fields if Zm topoorder a}
\end{equation}
and
\begin{equation}
\begin{split}
\hat{T}^{\,}_{p}(z)\:=&\,
\frac{1}{m}
\tilde{\mathcal{T}}^{\T}_{p}\,
\tilde{\mathcal{K}}^{\,}_{m}
\tilde{\hat\Phi}(z)
\\
=&\,
\tilde{\hat{\phi}}^{\,}_{p^{\,}_{E},2}(z)
-
\tilde{\hat{\phi}}^{\,}_{p^{\,}_{W},2}(z)
+
\tilde{\hat{\phi}}^{\,}_{p^{\,}_{S},2}(z)
-
\tilde{\hat{\phi}}^{\,}_{p^{\,}_{N},2}(z),
\end{split}
\label{eq: pinned fields if Zm topoorder b}
\end{equation}
\end{subequations} 
respectively. Indeed, application of
Eqs.\ (\ref{eq: 2D pinning soliton star})
and
(\ref{eq: 2D pinning soliton star prime})
in combination with 
Eq.\ (\ref{eq: Zm 2D tunneling vecs})
delivers
\begin{subequations}
\label{eq: S/P of sC/pC z' T of s'/p' z Sdag/Pdag of sC/pC z'} 
\begin{equation}
\hat{S}^{\,}_{s^{\,}_{C}}(z')\,
\hat{T}^{\,}_{s'}(z)\,
\hat{S}^{\dag}_{s^{\,}_{C}}(z')=
\hat{T}^{\,}_{s'}(z)
+
\sigma^{\,}_{s,s';C}\,
\frac{2\pi}{m}\,
\Theta(z-z'),
\label{eq: S/P of sC/pC z' T of s'/p' z Sdag/Pdag of sC/pC z' a} 
\end{equation}
where we have introduced the function
$
\sigma^{\,}_{s,s';C}
$
that returns the signs multiplying the Heaviside step functions
on the right-hand side of 
Eq.\ (\ref{eq: 2D pinning soliton star}) 
if $s=s'$,
the signs multiplying the Heaviside step functions
on the right-hand side of 
Eq.\ (\ref{eq: 2D pinning soliton star prime})
if $s$ and $s'$ share $s^{\,}_{C}$,
and zero otherwise.
Similarly, application of
Eqs.\ (\ref{eq: 2D pinning soliton plaquette})
and
(\ref{eq: 2D pinning soliton plaquette prime}),
in combination with Eq.\ (\ref{eq: Zm 2D tunneling vecs})
delivers
\begin{equation}
\label{eq: S/P of sC/pC z' T of s'/p' z Sdag/Pdag of sC/pC z' b} 
\hat{P}^{\,}_{p^{\,}_{C}}(z')\,
\hat{T}^{\,}_{p'}(z)\,
\hat{P}^{\dag}_{p^{\,}_{C}}(z')=
\hat{T}^{\,}_{p'}(z)
+
\sigma^{\,}_{p,p';C}\,
\frac{2\pi}{m}\,
\Theta(z-z'),
\end{equation}
\end{subequations}
where we have introduced the function
$
\sigma^{\,}_{p,p';C}
$
that returns the signs multiplying the Heaviside step functions
on the right-hand side of 
Eq.\ (\ref{eq: 2D pinning soliton plaquette}) 
if $p=p'$,
the signs multiplying the Heaviside step functions
on the right-hand side of 
Eq.\ (\ref{eq: 2D pinning soliton plaquette prime}) 
if $p$ and $p'$ share $p^{\,}_{C}$,
and zero otherwise.
    
If we define the soliton density operator 
for any star $s$ in the square lattice by
\begin{subequations}
\begin{equation}
\hat{\rho}^{\mathrm{sol}}_{s}(z)\:=
\frac{1}{2\pi}
\left(
\partial^{\,}_{z}
\hat{T}^{\,}_{s}
\right)(z)
\end{equation}
and do the same with
\begin{equation}
\hat{\rho}^{\mathrm{sol}}_{p}(z)\:=
\frac{1}{2\pi}
\left(
\partial^{\,}_{z}
\hat{T}^{\,}_{p}
\right)(z)
\end{equation}
\end{subequations}
for any plaquette $p$ in the square lattice, we can then make the
substitutions $\hat{T}^{\,}_{s'}\to\hat{\rho}^{\mathrm{sol}}_{s'}$,
$\hat{T}^{\,}_{p'}\to\hat{\rho}^{\mathrm{sol}}_{p'}$, and
$2\pi\,\Theta(z-z')\to\delta(z-z')$ in Eqs.\ 
(\ref{eq: S/P of sC/pC z' T of s'/p' z Sdag/Pdag of sC/pC z' a}) 
and 
(\ref{eq: S/P of sC/pC z' T of s'/p' z Sdag/Pdag of sC/pC z' b}), 
respectively. The resulting
pair of equations is interpreted as the fact that any one of the pair
of operators $\hat{S}^{\dag}_{s^{\,}_{C}}(z)$ and
$\hat{P}^{\dag}_{p^{\,}_{C}}(z)$ creates a dipole with a soliton
charge of magnitude $1/m$ straddling the link $s^{\,}_{C}$ or
$p^{\,}_{C}$ with the cardinality $C=N,W,S,E$ belonging to the star
$s$ and plaquette $p$, respectively.  Upon multiplying
$\hat{\rho}^{\mathrm{sol}}_{s}$ and $\hat{\rho}^{\mathrm{sol}}_{p}$ by
the electric charges $\tilde{Q}^{\,}_{1}=2$ and
$\tilde{Q}^{\,}_{2}=0$, respectively, we conclude that
$\hat{S}^{\dag}_{s^{\,}_{C}}(z)$ creates an electric dipole with a
charge of magnitude $2/m$ straddling the link$s^{\,}_{C}$ with the
cardinality $C=N,W,S,E$ belonging to the star $s$.  
On the other hand, the operator 
$\hat{P}^{\dag}_{p^{\,}_{C}}(z)$ creates an
electrically neutral dipole.  Hence, anticipating a connection to 3D
toric code models that we will demonstrate shortly, we refer to the
charged constituents of the electric dipole created by the operator
$\hat{S}^{\dag}_{s^{\,}_{C}}(z)$ as ``electric" excitations, and to
the constituents of the neutral dipole greated by the operator
$\hat{P}^{\dag}_{p^{\,}_{C}}(z)$ as ``magnetic" excitations.

Second, the bilocal operators
\begin{subequations}
\label{eq: def 2D Zm z-string defs}
\begin{align}
\begin{split}
\hat{S}^{\dag}_{s^{\,}_{C}}(z^{\,}_{1},z^{\,}_{2})&\:=
\hat{S}^{\dag}_{s^{\,}_{C}}(z^{\,}_{2})\,
\hat{S}^{\phantom\dag}_{s^{\,}_{C}}(z^{\,}_{1})
\\
&=
\exp
\bigg(
-\mathrm{i}\, 
\int\limits^{z^{\,}_{2}}_{z^{\,}_{1}}\mathrm{d}z\ 
\partial^{\,}_{z}\tilde{\hat{\phi}}^{\,}_{s^{\,}_{C},2}(z)
\bigg),
\end{split}
\label{eq: def 2D Zm z-string defs a}
\end{align}
and
\begin{align}
\begin{split}
\hat{P}^{\dag}_{p^{\,}_{C}}(z^{\,}_{1},z^{\,}_{2})&\:=
\hat{P}^{\dag}_{p^{\,}_{C}}(z^{\,}_{2})\,
\hat{P}^{\phantom\dag}_{p^{\,}_{C}}(z^{\,}_{1})
\\
&=
\exp
\bigg(
+\mathrm{i}\, 
\int\limits^{z^{\,}_{2}}_{z^{\,}_{1}}\mathrm{d}z\ 
\partial^{\,}_{z}\tilde{\hat{\phi}}^{\,}_{p^{\,}_{C},1}(z)
\bigg),
\end{split}
\label{eq: def 2D Zm z-string defs b}
\end{align}
\end{subequations}
can be used to create and propagate linelike defects that extend in
the $z$-direction, as in Fig.~\ref{fig: xy line and membrane defects}(a) 
and (b).  Linelike defects lying in the $x$-$y$ plane
can be created and propagated by repeated application of the vertex
operators in Eqs.~\eqref{eq: 2D Zm defect hopping}, as in the example
of Fig.~\ref{fig: xy line and membrane defects}(c) and (d).

The statistical angle $\theta$ obtained upon winding of the pointlike
and linelike excitations created by these operators can be computed
from Eq.~\eqref{eq: chosen case for statistical angle between linelike}, 
which gives
\begin{align}\label{Zm statistics}
\theta = -2\pi/m.
\end{align}
The case $m=2$ produces the expected statistical phase of $\pi$
between ``electric" quasiparticles and ``magnetic" strings in the 3D
toric code.  We will see this resemblance borne out in the next
section, where we compute the ground state degeneracy.

\subsubsection{Ground state degeneracy on the three-torus}
\label{subsubsec: Ground state degeneracy}

The nonlocal string and membrane operators used to obtain the ground
state degeneracy on the three-torus $\mathbb{T}^{3}$ for this example can
be assembled from the vertex operators defined in Eqs.\
\eqref{eq: 2D Zm defect hopping} and the bilocal operators defined in
Eqs.\ \eqref{eq: def 2D Zm z-string defs}, as outlined in
Sec.~\ref{subsubsec: Topological ground state degeneracy}.  
As discussed in Sec.\
\ref{subsubsec: Topological ground state degeneracy}, 
it is sufficient to consider the algebra of star-type
string operators and plaquette-type membrane operators to deduce the
degeneracy. This is given by
\begin{subequations}\label{Zm line-membrane 1}
\begin{align} 
&
\hat{\mathcal{O}}^{\,}_{S,\Gamma^{\,}_{z}}\, 
\hat{\mathcal{O}}^{\,}_{P,\Omega^{\,}_{\hat{x}\hat{y}}}=
\hat{\mathcal{O}}^{\,}_{P,\Omega^{\,}_{\hat{x}\hat{y}}}\, 
\hat{\mathcal{O}}^{\,}_{S,\Gamma^{\,}_{z}}\ 
e^{+\mathrm{i}\, 2\pi/m},
\label{Zm xyz membrane1}\\
&
\hat{\mathcal{O}}^{\,}_{S,\Gamma^{\,}_{y}}\, 
\hat{\mathcal{O}}^{\,}_{P,\Omega^{\,}_{\hat{z}\hat{x}}}=
\hat{\mathcal{O}}^{\,}_{P,\Omega^{\,}_{\hat{z}\hat{x}}}\, 
\hat{\mathcal{O}}^{\,}_{S,\Gamma^{\,}_{y}}\,
e^{-\mathrm{i}\, 2\pi/m},
\label{Zm xzy membrane1}\\
&
\hat{\mathcal{O}}^{\,}_{S,\Gamma^{\,}_{x}}\, 
\hat{\mathcal{O}}^{\,}_{P,\Omega^{\,}_{\hat{y}\hat{z}}}=
\hat{\mathcal{O}}^{\,}_{P,\Omega^{\,}_{\hat{y}\hat{z}}}\, 
\hat{\mathcal{O}}^{\,}_{S,\Gamma^{\,}_{x}}\ 
e^{-\mathrm{i}\, 2\pi/m}.
\label{Zm yzx membrane1}
\end{align}
\end{subequations}
[See Eqs.~\eqref{eq: def Wilson loops} and \eqref{eq: def Wilson membranes} 
for definitions of these
operators.]  Each line of Eqs.\ \eqref{Zm line-membrane 1}
contributes an $m$-fold topological degeneracy, for a total degeneracy
on the three-torus
\begin{align}\label{Zm torus degeneracy}
D^{\,}_{\mathbb{T}^{3}}=m^{3}.
\end{align}
Note that for $m=2$, which corresponds to the case of $\mathbb{Z}^{\,}_{2}$
topological order, the ground-state degeneracy is 8-fold.  This is the
expected topological degeneracy of the three-dimensional toric
code~\cite{Castelnovo08,Mazac12}, which is an important sanity
check.

\subsubsection{Surface states}
\label{subsubsec: Surface theory}

All properties that we have discussed so
far pertain to the bulk of the array of coupled wires, 
as we have always imposed
periodic boundary conditions in all spatial directions.  
However, the wire formalism provides 
means to address the surface states as well.  
We first illustrate this fact with the example of the 
$\mathbb{Z}^{\,}_{m}$
theories discussed in this section, 
before commenting on surface
states in more generality.

\begin{figure}[t]
\begin{center}
\includegraphics[width=.45\textwidth]{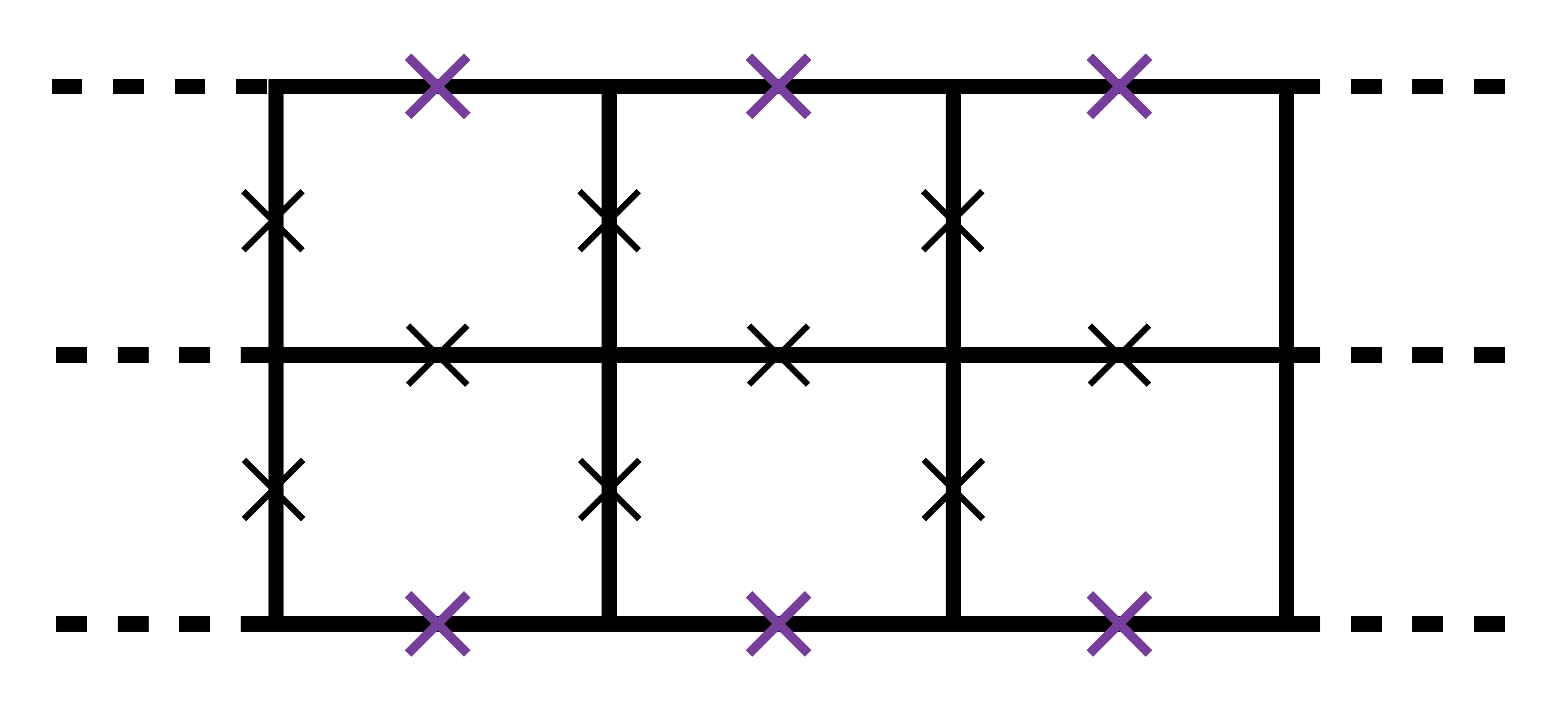}\\
\caption{ (Color online)
Example of an array of quantum wires with open boundary
conditions in the $y$-direction and periodic boundary conditions
along all other directions.  The dashed links indicate the presence
of periodic boundary conditions in the $x$-direction.  The crosses
represent quantum wires on the links of the square lattice that are
inequivalent modulo the periodic boundary conditions.  In this
example, $N^{\,}_{x}=3$ and $N^{\,}_{y}=2$.  Consequently, there are
$2\, N^{\,}_{x}\,N^{\,}_{y}+ N^{\,}_{x}=15$ wires in the array, and 
$2\,N^{\,}_{x}\,N^{\,}_{y}-N^{\,}_{x}=9$ wires are gapped by the allowed
tunneling vectors.  Consequently, there are $6$ wires in the array
that remain gapless when these tunneling vectors are included (3 on
the top face and 3 on the bottom face, represented by the purple
crosses).
\label{fig: surface counting}
         }
\end{center}
\end{figure}

Let us begin by relaxing the constraint of periodic boundary
conditions that we have imposed until now. We choose 
open boundary conditions in the $y$-direction, while leaving periodic
boundary conditions in the $x$ and $z$-directions.  In this case, the
surface of the system has the same topology as the two-torus
\begin{equation}\label{eq: def two torus}
\mathbb{T}^{2}\:=S^{1}\times S^{1}.
\end{equation} 
The latter can be viewed as a plane parallel to the $x$-$z$ plane
whose adjacent sides have been identified. There are two types of
surface terminations of the square lattice whose links host the
constituent quantum wires in the array.
These are ``rough" boundaries, which consist of stars,
and ``smooth" boundaries, which consist of plaquettes.  
For the sake of specificity, we will focus on ``smooth"
boundaries, as in Fig.~\ref{fig: surface counting}, for the time
being.  All statements that we make about ``smooth" boundaries below
have analogs for the case of rough boundaries.  However, the
differences between the two types of boundary are not always
physically insignificant,
as we will provide shortly an
example of a difference between rough and smooth boundaries.

The effects of imposing these semi-open boundary conditions are twofold.
First, they \textit{increase} the number of gapless degrees of freedom
in the array of coupled wires, as the wires along the terminating
surfaces of the wire array are no longer identified with each other.
Second, they \textit{decrease} the number of tunneling vectors in the
Haldane set $\mathbb{H}$, as any stars or plaquettes that were formerly
completed by virtue of the periodicity of the array of wires are now
nonlocal, and therefore cannot be included.  This results in a number,
which we will determine momentarily, of ``extra" gapless modes on the
terminating surfaces of the array of coupled wires.

We can determine the existence of gapless surface states for the
coupled-wire theory defined in Sec.\
\ref{subsubsec: Definitions and interwire couplings} 
by the following counting argument.  
First, recall that, when periodic boundary conditions are imposed, 
the square lattice contains $2N$ quantum wires, placed on its links.  
Let us write $N\equiv N^{\,}_{x}\times N^{\,}_{y}$,
where $N^{\,}_{x}$ counts either the number of stars or 
the number of plaquettes along the $x$-direction.
The number $N^{\,}_{y}$ does the same along the $y$-direction.  
When periodic boundary conditions are relaxed along the $y$-direction, 
the wires along the bottom and top faces of the array of wires
(see Fig.~\ref{fig: surface counting})
are no longer identified with one another, which adds $N^{\,}_{x}$ wires
to the array. The total number of wires in the array
with the topology (\ref{eq: def two torus}) is therefore
\begin{subequations}\label{number of wires}
\begin{equation}
2\,N^{\,}_{x}\,N^{\,}_{y}+\,N^{\,}_{x},
\end{equation}
and the associated number of gapless degrees of freedom is
\begin{equation}
4\,N^{\,}_{x}\,N^{\,}_{y}+2\,N^{\,}_{x}.
\end{equation}
\end{subequations}

Next, we count the number of available tunneling vectors
in the array of wires when the topology (\ref{eq: def two torus}) 
is imposed. Before relaxing periodic boundary conditions, 
there are $2\,N^{\,}_{x}\,N^{\,}_{y}$ tunneling
vectors in the Haldane set $\mathbb{H}$, which is sufficient to gap out
all $4\,N^{\,}_{x}\,N^{\,}_{y}$ degrees of freedom when periodic boundary
conditions are imposed. However, when periodic boundary conditions are
relaxed in the $y$-direction, $N^{\,}_{x}$ tunneling vectors must be removed
from the set $\mathbb{H}$. Consequently, 
the total number of degrees of freedom
left once all allowed tunneling vectors are included is given by
\begin{align}\label{number of gapless modes left over}
4\,N^{\,}_{x}\,N^{\,}_{y}+2\,N^{\,}_{x} 
- 
(4\,N^{\,}_{x}\,N^{\,}_{y}-2\,N^{\,}_{x})= 
4\,N^{\,}_{x}.
\end{align}
Since the remaining degrees of freedom must live on the boundary,
where we have deleted tunneling vectors from the set $\mathbb{H}$, 
we can split the remaining $4\,N^{\,}_{x}$ degrees of freedom evenly
among the top and bottom edges of the array of wires.  
This simply leaves $N^{\,}_{x}$ gapless quantum wires on each 
exposed surface, i.e., $2\,N^{\,}_{x}$
gapless degrees of freedom on each of the top and bottom surfaces, 
respectively.
(An example of this counting procedure is shown in Fig.\
\ref{fig: surface counting}.)

It is a nontrivial task to determine the exact surface Lagrangian
governing the remaining $2\,N^{\,}_{x}$ gapless degrees of freedom on
each terminating surface of the array of wires.  For example, in the
case of Fig.~\ref{fig: surface counting}, it is tempting to 
deduce
that the surface Lagrangian describes a theory of decoupled quantum
wires built out of the fields $\tilde{\hat\phi}^{\,}_{i,1}$
that no
longer enter any cosine terms due to the removal of the
``three-legged" stars that lie on the terminating surfaces, and their
conjugate fields $\tilde{\hat\phi}^{\,}_{i,2}$.  However, the latter
fields couple to the bulk of the array of quantum wires via cosine
terms associated with the plaquettes that lie along the terminating
surfaces.  Consequently, the fields  $\tilde{\hat\phi}^{\,}_{i,1}$ and
$\tilde{\hat\phi}^{\,}_{i,2}$ do not provide the right basis for the gapless
surface states.

However, despite the difficulty of determining a Lagrangian description of
these gapless surface states, the determination of the stability of
these surface states and the characterization of any proximal gapped
phases are readily feasible with the tools already developed in this
work.

The stability of the gapless surfaces
can be addressed by seeking out a set of $2\, N^{\,}_{x}$ tunneling vectors,
i.e., $N^{\,}_{x}$ tunneling vectors for each terminating surface, 
to complete the Haldane set $\mathbb{H}$.  
These surface tunneling vectors must be chosen
to comply with all symmetries of the problem, in this case TRS and charge
conservation, and must be compatible with the bulk tunneling vectors in the
sense of the Haldane criterion \eqref{Haldane criterion}.  If any number less
than $N^{\,}_{x}$ tunneling vectors for each
terminating surface is found, then the gapless surface states are stable,
since it is impossible to localize all gapless degrees of freedom in the array
of quantum wires with the topology \eqref{eq: def two torus}, 
while simultaneously preserving all symmetries.  
If, instead, the necessary number of compatible
tunneling vectors is found, then the gapless surface states are unstable.

Each distinct set of tunneling vectors that completes the Haldane set 
$\mathbb{H}$
realizes a two-dimensional gapped state of matter on each exposed surface of
the array of quantum wires.  The resulting gapped surface states can be
characterized, as in Sec.\
\ref{subsec: Fractionalization in the coupled wire array},
by the set of deconfined quasiparticle excitations defined on the surface.

In the remainder of this discussion, we will show that the class of
$\mathbb{Z}^{\,}_{m}$-topologically-ordered states realized by the wire
construction defined in Sec.\ 
\ref{subsubsec: Definitions and interwire couplings} 
has unstable surface states that can be gapped while
maintaining TRS and charge conservation.  We will further show that,
if the surface termination is ``rough" (i.e., if it consists of
stars), one can obtain a charge-conserving gapped surface state with
Laughlin topological order, at the expense of explicit TRS-breaking at
the surface.

We first show that the gapless surface states are unstable in the
present example of a $\mathbbm{Z}^{\,}_{m}$-topologically-ordered bulk.  
To do this, consider the following two sets of tunneling vectors,
\begin{subequations}
\begin{align}
\mathcal{T}^{\,}_{1,j}\:=
\begin{pmatrix}
\cdots | \, 0 & 0 \, | \, +1 & 0 \, | \, -1 & 0 \, | \, 0 & 0 \, | \cdots
\end{pmatrix}^{\T},
\end{align}
and
\begin{align}
\mathcal{T}^{\,}_{2,j}\:=
\begin{pmatrix}
\cdots | \, 0 & 0 \, | \, 0 & +1 \, | \, 0 & -1 \, | \, 0 & 0 \, | \cdots
\end{pmatrix}^{\T},
\end{align}
\end{subequations}
where $j=1,\dots,N^{\,}_{x}$ indexes the gapless wires on the top
surface of the wire array (there is a similar set of tunneling vectors
that can be defined for the other surface to complete each set). 
Each set of tunneling vectors generates terms that allow bosons to hop
between wires on the surface.  These two sets of tunneling vectors
each satisfy the Haldane criterion \eqref{Haldane criterion}
with the $K$-matrix (\ref{eq: step 1 to define Zm a}), both
among themselves and with the plaquettes lining each smooth surface.
(One can verify that this is equally true for rough boundaries, where
the lattice terminates with stars rather than plaquettes.)
Furthermore, the cosine terms that they generate preserve TRS, defined
as in Eq.\ \eqref{2D Zm TRS action}, and charge conservation, defined
as in Eq.\ \eqref{charge conservation} with the charge vector
(\ref{eq: step 1 to define Zm c}).  
They therefore generate two
distinct two-dimensional gapped states of matter that preserve all
symmetries of the bulk: $\{\mathcal{T}^{\,}_{1,j}\}$ generates one
with deconfined ``magnetic" excitations, while $\{\mathcal{T}^{\,}_{2,j}\}$
and one with deconfined ``electric" excitations.

We now demonstrate that, in the presence of a set of surface tunneling
vectors that break TRS, a rough terminating surface can be made into a
fractional-quantum-Hall-like state of matter with Laughlin topological
order, while preserving charge conservation.  In this case, we can use
another set of $N^{\,}_{x}$ tunneling vectors, given by
(for any $j=1,\dots,N^{\,}_{x}$)
\begin{align}\label{eq: TRS breaking surface T-vecs}
\mathcal{T}^{\,}_{3,j}\:=
\begin{pmatrix}
\cdots | \, 0 & 0 \, | \, +1 & +1 \, | \, -1 & +1 \, | \, 0 & 0 \, | 
\cdots
\end{pmatrix}^{\T},
\end{align}
which both conserve charge and satisfy the Haldane criterion among
themselves and with the stars lying along the terminating
surface, to gap the surface. 
Observe that these tunneling vectors pin the fields
(for any $j=1,\dots,N^{\,}_{x}$)
\begin{align}
\mathcal{T}^{\T}_{3,j}\,
\tilde{\mathcal{K}}^{\,}_{m}\,
\tilde{\hat{\Phi}}=
m
\(
\tilde{\hat{\phi}}^{\,}_{j,1}
+
\tilde{\hat{\phi}}^{\,}_{j,2}
\)
+
m
\(
\tilde{\hat{\phi}}^{\,}_{j+1,1}
-
\tilde{\hat{\phi}}^{\,}_{j+1,2}\),
\end{align}
which are neither even nor odd under the definition of TRS given in
Eq.\ \eqref{2D Zm TRS action}. Therefore, the associated cosine
potentials break TRS explicitly.  We will now show that the gapless
surface in the presence of the cosine terms generated 
by the tunneling vectors of the form 
\eqref{eq: TRS breaking surface T-vecs}, 
in addition to being gapped, supports pointlike
excitations with fractional statistics, consistent with a (fractional)
quantum Hall effect on each two-dimensional surface.

The excitations of the surface theory are defined, 
as they are in the bulk, to be solitons in the pinned field
$\mathcal{T}^{\T}_{3,j}\,
\tilde{\mathcal{K}}^{\,}_{m}\,
\tilde{\hat{\Phi}}$
for any $j=1,\dots,N^{\,}_{x}$.  
Define
\begin{align}
\tilde{\hat\phi}^{\,}_{j,\pm}\:=
\tilde{\hat\phi}^{\,}_{j,1}\pm\tilde{\hat\phi}^{\,}_{j,2}.
\end{align}
We begin by observing that the equal-time commutators
\begin{align}
\begin{split}
&
\[
\partial^{\,}_{z}\tilde{\hat{\phi}}^{\,}_{j,\pm}(z),
\tilde{\hat{\phi}}^{\,}_{\pri{j},\pm}(\pri{z})
\]
=
\pm
\mathrm{i}\, 
\frac{4\pi}{m}\, 
\delta^{\,}_{j\pri j}\, 
\delta(z-\pri{z}),
\\
&
\[
\partial^{\,}_{z}\tilde{\hat{\phi}}^{\,}_{j,\pm}(z),
\tilde{\hat{\phi}}^{\,}_{\pri{j},\mp}(\pri{z})
\]=0,
\end{split}
\end{align}
hold for any $j,j'=1,\dots,N^{\,}_{x}$. 
One deduces from this algebra
[recall Eqs.\
\eqref{eq: 2D pinning soliton star} 
and
\eqref{eq: 2D pinning soliton plaquette}]
that the local operator
\begin{align}\label{eq: surface fqhe x-hopping}
\hat{q}^{\dag}_{j-1,j}(z)&\:=
e^{
-\mathrm{i}
\[
\tilde{\hat{\phi}}^{\,}_{j,+}(z)
+
\tilde{\hat{\phi}}^{\,}_{j,-}(z)
\]\big/2
  }=
e^{
-\mathrm{i}\,\tilde{\hat{\phi}}^{\,}_{j,1}(z)
  }
\end{align}
creates a $-2\pi$-soliton in 
$\mathcal{T}^{\T}_{3,j-1}\,\tilde{\mathcal{K}}^{\,}_{m}\,\tilde{\hat{\Phi}}$ 
and a $+2\pi$-soliton in
$\mathcal{T}^{\T}_{3,j}\,\tilde{\mathcal{K}}\,\tilde{\hat{\Phi}}$
for any $j=1,\dots,N^{\,}_{x}$.
Consequently, the operator $q^{\dag}_{j-1,j}$ 
can be interpreted as hopping 
a quasiparticle 
from the link connecting
wires $j-1$ and $j$ to the link connecting wires $j$ and $j+1$.
We will see below that this quasiparticle has fractional statistics.
Repeated application of this operator on successive
wires hops the
fractionalized quasiparticle
along the $x$-direction, perpendicular to the wires.
(Note that the vertex operator associated with the other independent linear
combination of the fields $\tilde{\hat\phi}^{\,}_{j,\pm}$, namely 
$\tilde{\hat{\phi}}^{\,}_{j,+}-\tilde{\hat{\phi}}^{\,}_{j,-}$,
does not create a deconfined quasiparticle because repeated application
of this vertex operator generates additional defects with each application.
We therefore choose to ignore this quasiparticle, as it is confined.)  

A fractionalized quasiparticle can be moved along the
$z$-direction, parallel to the wires, by applying the bilocal operator
\begin{align}\label{eq: surface fqhe z-hopping}
\begin{split}
\hat{q}^{\dag}_{j}(z^{\,}_{1},z^{\,}_{2})&\:=
e^{-\mathrm{i}\,\tilde{\hat\phi}^{\,}_{j,+}(z^{\,}_{2})/2}\, 
e^{+\mathrm{i}\,\tilde{\hat\phi}^{\,}_{j,+}(z^{\,}_{1})/2}
\\
&= 
e^{
-
\frac{\mathrm{i}}{2}
\int\limits_{z^{\,}_{1}}^{z^{\,}_{2}}\mathrm{d}z\, 
\partial^{\,}_{z}\tilde{\hat\phi}^{\,}_{j,+}(z)
  }.
\end{split}
\end{align}
for any $j=1,\dots,N^{\,}_{x}$.
Acting with $q^{\dag}_{j}(z^{\,}_{1},z^{\,}_{2})$ on a ground state
transfers a quasiparticle from point $z^{\,}_{1}$ to
point $z^{\,}_{2}$ along wire $j=1,\dots,N^{\,}_{x}$.

This (pointlike) surface quasiparticle is an anyon whose
self-statistics is defined by the statistical angle 
$\theta=\pi/m=2\pi/2m$, 
which is half the statistical angle acquired when a pointlike excitation
winds around a linelike excitation in the bulk. 
This quasiparticle is therefore only supported on the surface.
The statistical angle can be determined, as it was in 
Sec.~\ref{subsubsec: Statistics of point, line, and membrane defects}, 
by the algebra between the vertex operators
(\ref{eq: surface fqhe x-hopping})
and
(\ref{eq: surface fqhe z-hopping})
that 
allow for the propagation of
this quasiparticle along any non-contractible
loop of the toroidal terminating surface.  
This algebra is given by
\begin{align}\label{eq: algebra between local and bilocal operators}
\hat{q}^{\dag}_{j-1,j}(z)\,\hat{q}^{\dag}_{j}(z^{\,}_{1},z^{\,}_{2})=
\hat{q}^{\dag}_{j}(z^{\,}_{1},z^{\,}_{2})\,\hat{q}^{\dag}_{j-1,j}(z)\, 
e^{-\mathrm{i}\, \pi/m},
\end{align}
where it is assumed that $z^{\,}_{1}<z<z^{\,}_{2}$
and $j=1,\dots,N^{\,}_{x}$.  
Accordingly, the
excitation spectrum of the surface in the presence of the
correlated tunneling processes generated by the tunneling vectors
\eqref{eq: TRS breaking surface T-vecs} 
consists of a single quasiparticle
type with statistics $\pi/m$.  Combining 
Eq.\ (\ref{eq: algebra between local and bilocal operators})
with the fact that TRS
is broken on the surface while charge is conserved, we conclude that
the gapped surface state selected by the many-body
interaction encoded by the tunneling vectors 
\eqref{eq: TRS breaking surface T-vecs}
is a fractional quantum Hall liquid with
Laughlin topological order.  The Hall conductivity of this surface
fractional quantum Hall liquid is given by $[(2e)^2/h]\times(1/2m)$,
consistent with the $2\pi/2m$ self-statistics of the surface quasiparticle
and the fundamental charge $2e$ of the underlying bosonic quantum wires.

Finally, let us point out that the above discussion of TRS breaking on
the surface applies also to smooth boundaries, although one must use
the surface tunneling vectors
\begin{align}\label{TRS breaking surface T-vecs smooth}
\tilde{\mathcal{T}}^{\,}_{4,j}\:=
\begin{pmatrix}
\cdots | \, 0 & 0 \, | \, +1 & +1 \, | \, +1 & -1 \, | \, 0 & 0 \, | 
\cdots
\end{pmatrix}^{\T},
\end{align}
instead of the ones defined in Eq.\
\eqref{eq: TRS breaking surface T-vecs}.  
This is necessary in order to ensure Haldane-compatibility
with the plaquettes lining the smooth surface.  However, observe that
this choice of surface tunneling vectors breaks charge conservation as
well as TRS on the smooth surface.  The only remaining symmetry of the
smooth surface is then number-parity conservation, as defined in
Eq.~\eqref{parity conservation}.  However, the analysis of the
excitations of the surface theory in this case proceeds similarly to
the case of the rough surface, and the conclusion that the surface
supports a single deconfined quasiparticle with self-statistics
$\pi/m$ remains.

The methods used in this section to address the surface physics
of the array of coupled quantum wires generalizes readily from the
example discussed here to any array of coupled quantum wires constructed
in Sec.~\ref{sec: Three-dimensional wire constructions}.  One can determine
the existence of gapless surface states using the counting
argument presented at the beginning of this section, with slight modifications
to account for the $M$ ``flavors" of stars and plaquettes that are allowed in the
general case.  One can then determine the stability of these gapless surfaces
by searching for a set of $M\,N^{\,}_{x}$ tunneling vectors for each 
terminating surface that are compatible with the bulk couplings.  
The process of characterizing any symmetry-preserving or
symmetry-breaking gapped surface states that descend from these
gapless states is also the same. For every admissible set of tunneling
vectors satisfying the necessary compatibility requirements, there is an
associated gapped surface. The excitation spectrum of each gapped
surface can be studied using the methods of Sec.\
\ref{subsec: Fractionalization in the coupled wire array}.

\section{Higher-dimensional wire constructions}
\label{sec: Higher-dimensional wire constructions}

The strategy developed in Sec.\ 
\ref{sec: Three-dimensional wire constructions} 
for constructing fully gapped three-dimensional
Abelian topological states of matter from coupled quantum wires owes
its success to several factors.  First, placing quantum wires on the
links of a square lattice in two spatial dimensions allows for a
simple enumeration of the number of gapless degrees of freedom in the
system.  Second, the ability to encode many-body interactions in
tunneling vectors associated with stars and plaquettes makes
straightforward the determination, via the Haldane criterion
\eqref{Haldane criterion}, 
of the number of gapless degrees of freedom that can be gapped out by 
these interactions.  Third, the fact that stars and plaquettes can share 
at most two wires allows one to derive simple conditions, 
like those of Eqs.\ \eqref{eq: 2D criteria}, 
to determine whether the Haldane criterion is satisfied.
Finally, the existence of a subextensive number of nonlocal
constraints, given in Eqs.\ \eqref{2D constraints}, allows for the
existence of nonlocal operators that can encode topological
ground-state degeneracy, if such a degeneracy is allowed by the chosen
many-body interactions.

These four advantageous properties all arose because we chose to
arrange the wires and their couplings in a manner reminiscent of the
qubits and commuting projectors of the toric code. While the toric
code is an archetypal example of topological order in two spatial
dimensions, it can also be defined on hypercubic lattices of dimension
greater than two. In fact, in spatial dimensions four and higher,
there are multiple toric codes that are distinguished from one another
by the number of nonlocal constraints that give rise to the
topological degeneracy. It is therefore natural to ask the question
of whether or not it is possible to build Abelian topological phases
in spatial dimension $D\geq3$ by arranging quantum wires on a
hypercubic lattice of dimension
\begin{align}
d\:=D-1
\end{align}
and coupling them in a manner reminiscent of a $d$-dimensional toric code.

We will
answer this question affirmatively.  
In Sec.\ \ref{subsec: Review of toric codes in arbitrary dimensions},
we describe a family of hypercubic arrays of quantum wires, 
and review some basic geometric facts about such arrays.  
In light of these facts, we generalize in 
Sec.\ \ref{subsec: Generalizing the results of Sec. ...} 
the prescriptions of Sec.\ \ref{subsec: 2D coupling discussion}
for defining compatible interwire couplings for 
a $d$-dimensional hypercubic lattice of quantum wires that yield
gapped $D$-dimensional phases of matter in the strong-coupling limit.  
Finally, in Sec.\ \ref{subsec: Example: 4D toric codes},
we provide explicit examples of four-dimensional phases of matter 
constructed according to these prescriptions.

\subsection{Hypercubic arrays of quantum wires}
\label{subsec: Review of toric codes in arbitrary dimensions}

Consider a $d$-dimensional hypercubic lattice.  We will view this
lattice as being composed of elementary objects called $k$-cells,
where $k=0,\dots,d$ is an integer.  For example, a 3-dimensional cubic
lattice can be decomposed as a set of 0-cells (sites), 1-cells (bonds
with sites at either end), 2-cells (square plaquettes with four sites
at their corners), or 3-cells (cubic plaquettes with eight sites at
their corners).  Any of these decompositions of the lattice covers all
sites of the lattice at least once.

We now consider hypercubic arrays of quantum wires labeled by a pair
of integers $(d,k^{\,}_{0})$.  Such an array consists of a
$d$-dimensional hypercubic lattice, embedded in $d+1=D$-dimensional
space, with quantum wires placed on the centers of the elementary
$k^{\,}_{0}$-cells of the lattice, for $1\leq k^{\,}_{0}\leq d-1$.  For
example, the arrays of quantum wires considered in Sec.\
\ref{sec: Three-dimensional wire constructions} 
are all of type $(2,1)$, 
since the array consists of quantum wires placed on the links of a square
lattice.  (Notice that this pair is the only one allowed for $d=2$.)
We take the wires to extend along a direction orthogonal to the
$d$-dimensional subspace occupied by the hypercubic lattice.  In the
array of quantum wires labeled by the pair $(2,1)$, for example, the
square lattice can be chosen to lie in a plane parallel to the $x$-$y$
plane, and the wires can be chosen to extend along the $z$-direction.
A hypercubic array of type $(d,k^{\,}_{0})$ contains
\begin{align}\label{N wire}
N^{\,}_{\mathrm{w}} = 
\binom{d}{k^{\,}_{0}}\,N
\end{align}
quantum wires, where $N$ is the number of vertices (i.e. 0-cells) in the
hypercubic lattice that hosts the array of wires.

\begin{figure}[t]
\begin{center}
(a)\includegraphics[width=.45\textwidth]{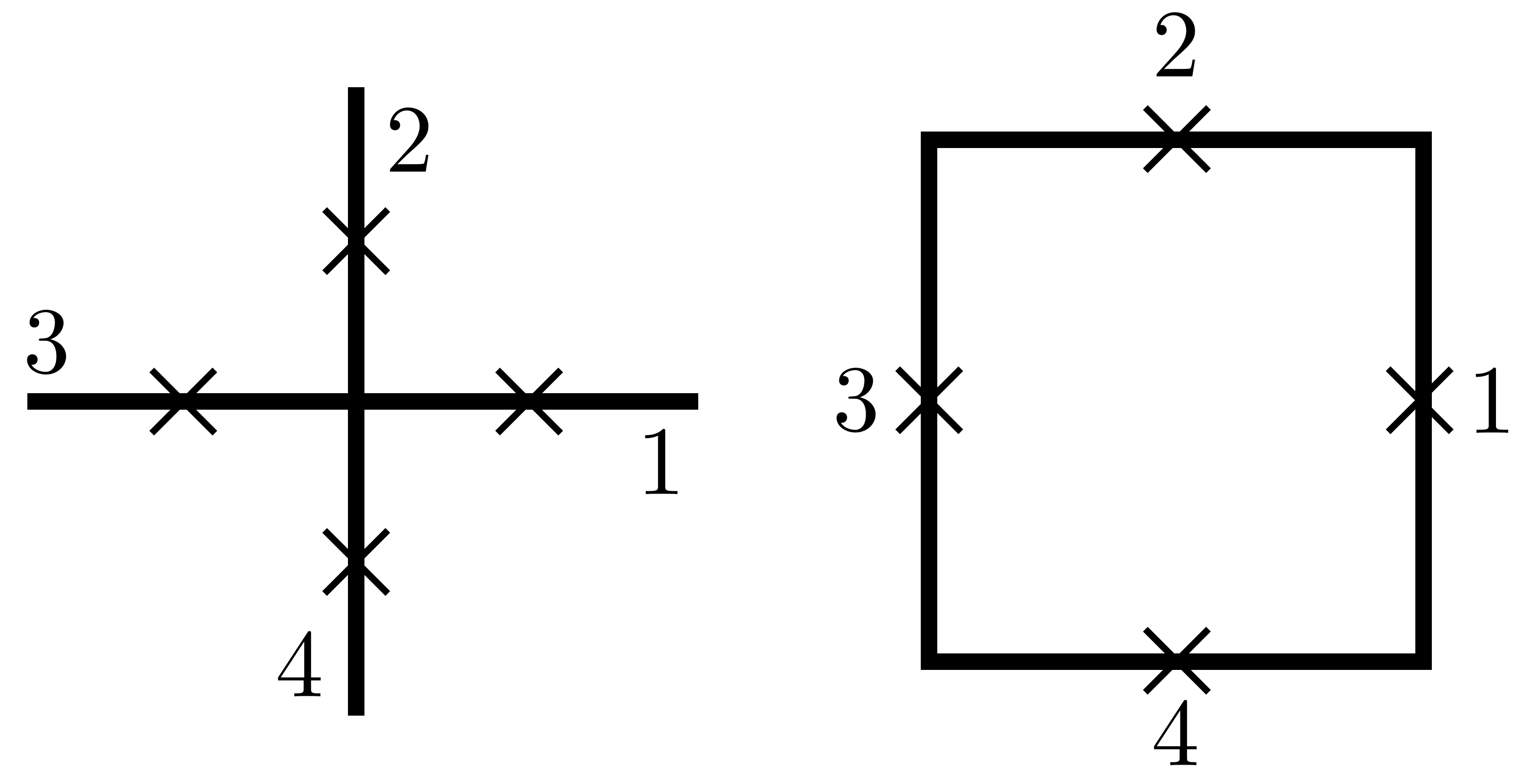}\\
(b)\includegraphics[width=.45\textwidth]{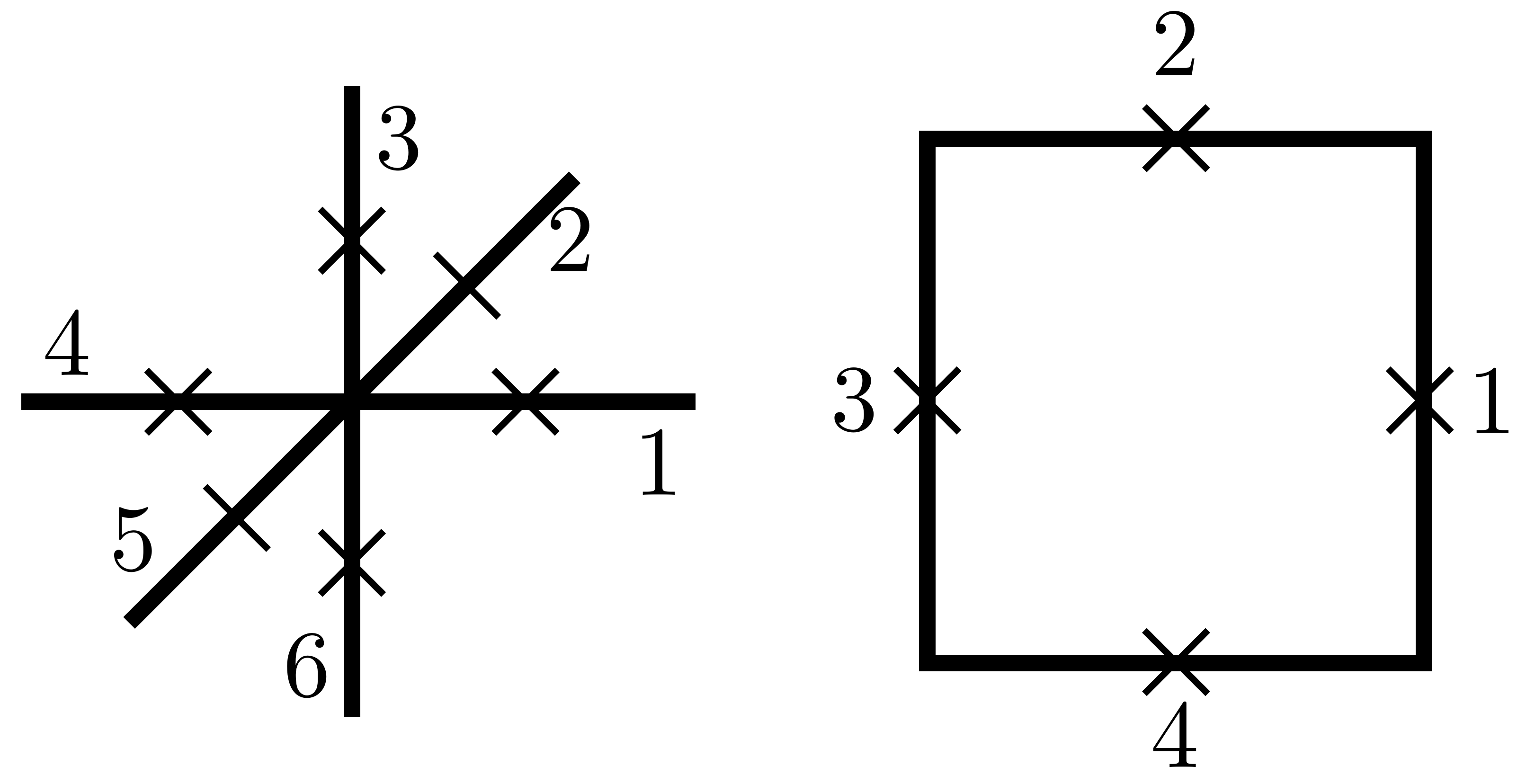}\\
(c)\includegraphics[width=.45\textwidth]{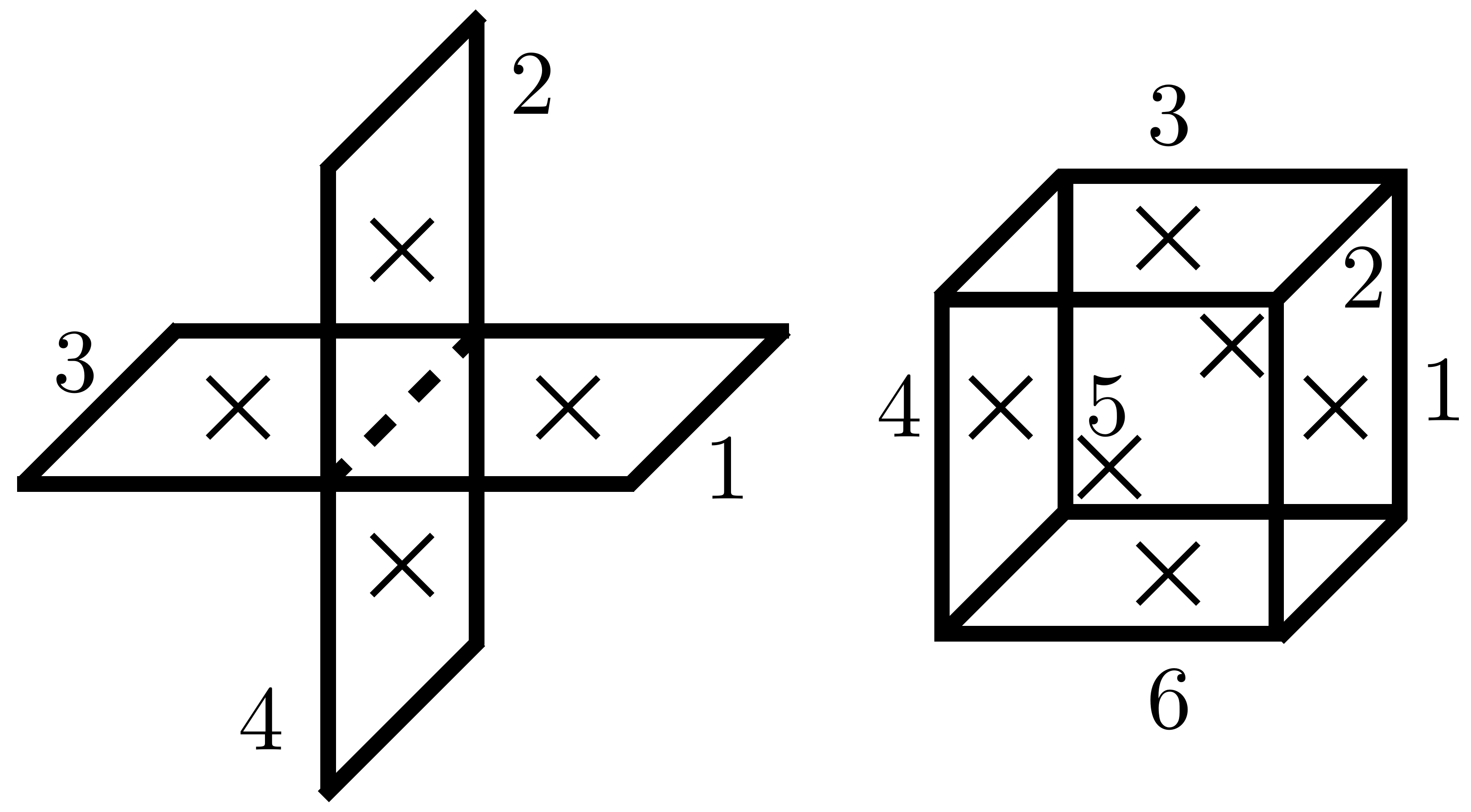}\\
\caption{Examples of hypercubic stars and plaquettes for arrays of
quantum wires of types (a) (2,1), (b) (3,1), and (c) (3,2).  Black
crosses represent wires extending perpendicular to all principal
directions of the respective hypercubic lattices. The numbers label
generalized cardinal directions $C^{\,}_{\mathrm{s}}$ and
$C^{\,}_{\mathrm{p}}$ defined in Eqs.~\eqref{Cs definition} and
\eqref{Cp definition}.
\label{fig: hypercubic star and plaquette}
         }
\end{center}
\end{figure}

With the hypercubic array of quantum wires defined in this way, we now
define hypercubic analogs of stars and plaquettes.  Examples of these
hypercubic stars and plaquettes are shown in Fig.~\ref{fig: hypercubic star and plaquette}.

Hypercubic ``stars" $s$ are centered on each 
$k^{\,}_{\mathrm{s}}$-cell of the $d$-dimensional
lattice with 
\begin{subequations}
\begin{equation}
k^{\,}_{\mathrm{s}}\:=
k^{\,}_{0}-1.
\end{equation} 
They consist of the
$2[d-(k^{\,}_{0}-1)]$ nearest-neighbor $k^{\,}_{0}$-cells 
(and the wires centered on these cells) that border the $(k^{\,}_{0}-1)$-cell
$s$. (See Fig.~\ref{fig: hypercubic star and plaquette} for examples.) 
We label the $2[d-(k^{\,}_{0}-1)]$ quantum wires belonging to a
hypercubic star $s$ by the generalized cardinal direction
\begin{align}\label{Cs definition}
C^{\,}_{\mathrm{s}}\:=1,\dots,2[d-(k^{\,}_{0}-1)].
\end{align}
There are
\begin{align}\label{N star}
N^{\,}_{\mathrm{s}}&=\binom{d}{k^{\,}_{0}-1}\, N
\end{align}
such hypercubic stars in the array of wires labeled by $(d,k^{\,}_{0})$.
\end{subequations}

Hypercubic ``plaquettes" $p$ are centered on each 
$k^{\,}_{\mathrm{p}}$-cell of the $d$-dimensional lattice with 
\begin{subequations}
\begin{equation}
k^{\,}_{\mathrm{p}}\:=k^{\,}_{0}+1.
\end{equation}
They consist of
the $2(k^{\,}_{0}+1)$ nearest-neighbor $k^{\,}_{0}$-cells that border
the $(k^{\,}_{0}+1)$-cell $p$.  (See Fig.~\ref{fig: hypercubic star and plaquette} for examples.)
We label the $2(k^{\,}_{0}+1)$ quantum
wires belonging to a hypercubic plaquette $p$ by the generalized
cardinal direction
\begin{align}\label{Cp definition}
C^{\,}_{\mathrm{p}}\:=1,\dots,2(k^{\,}_{0}+1).
\end{align}
There are
\begin{align}\label{N plaquette}
N^{\,}_{\mathrm{p}}&=\binom{d}{k^{\,}_{0}+1}\, N
\end{align}
such hypercubic plaquettes in the array of wires labeled by $(d,k^{\,}_{0})$.
\end{subequations}

For the square array of quantum wires studied in Sec.\
\ref{sec: Three-dimensional wire constructions}, 
which has $d=2$ and $k^{\,}_{0}=1$, 
the generalized cardinalities $C^{\,}_{\mathrm{s}}$ and
$C^{\,}_{\mathrm{p}}$ each take values $1,\dots, 4$. 
We identify these with
the traditional cardinal directions $C=N,W,S,E$ used in Sec.\ 
\ref{sec: Three-dimensional wire constructions}.

Note that the substitution $k^{\,}_{0}\to d-k^{\,}_{0}$ exchanges
$d-(k^{\,}_{0}-1) \leftrightarrow k^{\,}_{0}+1$.  Consequently, the
hypercubic array of quantum wires labeled by the pair $(d,k^{\,}_{0})$
is dual to the array labeled by the pair $(d,d-k^{\,}_{0})$, in the
sense that the stars of the former are the plaquettes of the latter,
and the plaquettes of the former are the stars of the latter.  In even
dimensions $d$, the hypercubic array of wires labeled by $(d,d/2)$ is
therefore self-dual.  Consequently, modulo dualities, there is only
one such array for $d=3$, as the pairs labeled by $(3,1)$ and $(3,2)$
are dual to one another. The first case where there are multiple
hypercubic arrays of quantum wires is therefore $d=4$, which has the
dual arrays $(4,1)$ and $(4,3)$, and one self-dual array $(4,2)$.

\subsection{Generalizing the results of Sec.~\ref{subsec: 2D coupling discussion}}
\label{subsec: Generalizing the results of Sec. ...}

We now turn to the problem of choosing a compatible set of tunneling
vectors to gap the bulk of an array of $N^{\,}_{\mathrm{w}}$ quantum
wires like those defined in Sec.\
\ref{subsec: Review of toric codes in arbitrary dimensions}.  
As in Sec.\ 
\ref{subsec: 2D coupling discussion}, 
the starting point is an array of decoupled quantum
wires described by the quadratic Lagrangian $\hat{L}^{\,}_{0}$ 
defined in Eq.\ \eqref{L0 d=2}, 
except that the matrices $\mathcal{K}$ and $\mathcal{V}$ are now
of dimension $2MN^{\,}_{\mathrm{w}}$, and the vector of scalar fields
\begin{align}
\begin{split}
\hat{\Phi}(t,z)&\:=
\Big(
\hat{\phi}^{\,}_{1,1}(t,z)\ 
\dots\ 
\hat{\phi}^{\,}_{1,2M}(t,z) 
\mid 
\\
&\qquad
\dots
\mid  
\hat{\phi}^{\,}_{N^{\,}_{\mathrm{w}},1}(t,z)\ 
\dots\ 
\hat{\phi}^{\,}_{N^{\,}_{\mathrm{w}},2M}(t,z) 
\Big)^{\T}.
\end{split}
\end{align}
This reflects that the quantum wires are now placed on the elementary
$k^{\,}_{0}$ cells of a $d$-dimensional hypercubic lattice embedded in
$D=d+1$-dimensional Euclidean space. We then add to the free theory
the interaction terms $\hat{L}^{\,}_{\{\mathcal{T}\}}$ given in
Eq.\ \eqref{Lint}, and set ourselves the challenge of finding a set of
$MN^{\,}_{\mathrm{w}}$ tunneling vectors $\mathcal{T}$ satisfying the
Haldane criterion \eqref{Haldane criterion}.  We also demand that
these tunneling vectors respect some set of symmetries---here, we will
enforce only charge conservation [Eq.\ \eqref{charge conservation}],
but others, such as TRS or particle-hole symmetry 
(see Ref.\ \onlinecite{Neupert14}), 
may also be relevant.  If we can find such a
set of tunneling vectors, then, in the strong-coupling limit,
$U^{\,}_{\mathcal{T}}\to\infty$ for all $\mathcal{T}$, the array of
wires acquires a gap.

As in Sec.~\ref{sec: Three-dimensional wire constructions}, 
we reserve the Greek index $\alpha=1,\ldots,2M$
for labeling the bosonic fields within each wire.
We reserve the Latin index $j=1,\ldots,N^{\,}_{\mathrm{w}}$ for
labeling the wires. A component of the vector of scalar fields
$\hat{\Phi}(t,z)$ is then $\hat{\phi}^{\,}_{j,\alpha}(t,z)$.

We claim that the following set of $M\, N^{\,}_{\mathrm{s}}$ 
integer-valued vectors of dimension $2MN^{\,}_{\mathrm{w}}$,
\begin{subequations}\label{arbitrary dimension tunneling vectors}
\begin{align}\label{arbitrary dimension star}
(\mathcal{T}^{(\texttt{j})}_{s})^{\,}_{j,\alpha}\:= 
v^{(\texttt{j})}_{\alpha}
\sum^{d-(k^{\,}_{0}-1)}_{C^{\,}_{\mathrm{s}}=1}
\Big(
\delta^{\,}_{j,s^{\,}_{C^{\,}_{\mathrm{s}}}}\!
-\,
\delta^{\,}_{j,s^{\,}_{C^{\,}_{\mathrm{s}}+d-(k^{\,}_{0}-1)}}
\Big),
\end{align}
and the following set of 
$M\,N^{\,}_{\mathrm{p}}$ integer-valued vectors 
of dimension $2MN^{\,}_{\mathrm{w}}$,
\begin{align}\label{arbitrary dimension plaquette}
\begin{split}
(\mathcal{T}^{(\texttt{j})}_{p})^{\,}_{j,\alpha}&\:= 
-w^{(\texttt{j})}_{\alpha}
\sum^{k^{\,}_{0}}_{C^{\,}_{\mathrm{p}}=1}
\Big(
\delta^{\,}_{j,p^{\,}_{C^{\,}_{\mathrm{p}}}}
\!-\,
\delta^{\,}_{j,p^{\,}_{C^{\,}_{\mathrm{p}}+k^{\,}_{0}+1}}
\Big)
\\
&\qquad\qquad 
+w^{(\texttt{j})}_{\alpha}
\Big(
\delta^{\,}_{j,p^{\,}_{k^{\,}_{0}+1}}
\!-\,
\delta^{\,}_{j,p^{\,}_{2(k^{\,}_{0}+1)}}
\Big),
\end{split}
\end{align}
\end{subequations}
does the job, so long as the criteria \eqref{eq: simplest 2D criteria}
are satisfied.  Here, $v^{(\texttt{j})}$ and $w^{(\texttt{j})}$ are
$2M$-dimensional vectors that specify the linear combinations of the
fields $\hat{\phi}^{\,}_{j,\alpha}$ in each wire $j$ that enter the
cosine terms associated with the tunneling vectors 
$\mathcal{T}^{(\texttt{j})}_{s}$ and
$\mathcal{T}^{(\texttt{j})}_{p}$, respectively.
(Their meaning is thus identical to the vectors of the same names
presented in Sec.\ \ref{subsec: 2D coupling discussion}.)  Stars and plaquettes 
are themselves labeled by the indices
$s=1,\ldots,N^{\,}_{\mathrm{s}}$ and $p=1,\ldots,N^{\,}_{\mathrm{p}}$,
respectively.   The teletype
index $\texttt{j}=1,\dots,M$ labels $M$ ``flavors" of stars and
plaquettes. These flavors are necessary, 
as they were in Sec.\ 
\ref{subsec: 2D coupling discussion}, 
to produce a number of tunneling vectors that
is sufficient to gap out all $2M$ gapless degrees of freedom in each
wire.  (More on counting gapless degrees of freedom in a moment.)

The tunneling vectors defined in Eqs.\
\eqref{arbitrary dimension tunneling vectors} 
conserve charge in the sense of Eq.\ \eqref{charge conservation} 
for \textit{any} $2M\,N^{\,}_{\mathrm{w}}$-dimensional
charge vector 
$\mathcal{Q}=(Q\mid Q\mid\dots\mid Q)^{\T}$  
[recall Eq.~\eqref{charge vector definition}].
To see that this is the case, 
it suffices to note that the vectors $v^{(\texttt{j})}$ and $w^{(\texttt{j})}$
each enter their respective tunneling vectors 
with an equal number of $+$ and $-$ signs. Consequently,
no matter the values of $Q^{\T} v^{(\texttt{j})}$ and $Q^{\T} w^{(\texttt{j})}$,
this value is added and subtracted an equal number of times.  
This fact provides a direct parallel with the construction
of Sec.\ \ref{subsec: 2D coupling discussion}, 
where the tunneling vectors defined in Eqs.\
\eqref{2d T-vecs charge conserving}
conserve charge independently of the form of the 
$2M$-dimensional charge vector $Q$ of a single wire.

One can verify that the tunneling vectors 
\eqref{arbitrary dimension tunneling vectors} 
satisfy the Haldane criterion \eqref{Haldane criterion}, 
as expressed in Eqs.\ \eqref{Haldane 2D}, 
if Eqs.\ \eqref{eq: simplest 2D criteria} hold, 
with the help of the following observations.  
First, note that these tunneling vectors coincide with the tunneling vectors 
\eqref{2d T-vecs charge conserving} defined in 
Sec.\ \ref{subsec: 2D coupling discussion} 
in the case $(d,k^{\,}_{0})=(2,1)$, which was studied there.  
Second, note that Eqs.~\eqref{star-star Haldane} and 
\eqref{plaquette-plaquette Haldane} hold if 
Eqs.\ \eqref{eq: simplest 2D criteria} hold.  
Third, recall that if a hypercubic star $s$ and a hypercubic plaquette $p$ 
overlap with one another, then they share two wires (see Ref.~\cite{Mazac12}).
With this in mind, we can see that Eq.\ \eqref{star-plaquette Haldane} 
holds for the tunneling vectors \eqref{arbitrary dimension tunneling vectors} 
by focusing on the case where the star $s$ and plaquette $p$ overlap 
[since Eq.\ \eqref{star-plaquette Haldane} holds trivially otherwise].  
This can be seen by looking only at the parts of the tunneling vectors 
\eqref{arbitrary dimension tunneling vectors} 
that lie in the crystal plane that contains the two wires in the union of 
$s$ and $p$.  The projection of the tunneling vectors 
\eqref{arbitrary dimension tunneling vectors} into this plane is, 
by construction, precisely the set of tunneling vectors 
\eqref{2d T-vecs charge conserving} defined in 
Sec.\ \ref{subsec: 2D coupling discussion} 
(but specialized from the outset to the case 
$v^{(\texttt{j})}_{1}=v^{(\texttt{j})}_{2}=v^{(\texttt{j})}$ 
and 
$w^{(\texttt{j})}_{1}=w^{(\texttt{j})}_{2}=w^{(\texttt{j})}$).  
Equation \eqref{star-plaquette Haldane} then follows.

Having seen that the tunneling vectors 
\eqref{arbitrary dimension tunneling vectors} 
encode charge-conserving many-body interactions
and are Haldane-compatible, one must next determine that these
tunneling vectors are sufficient in number to gap out all
$2M\,N^{\,}_{\mathrm{w}}$ gapless degrees of freedom in the array 
of quantum wires. 

Recall from the discussion in Sec.\
\ref{subsec: 2D coupling discussion} 
that, in order to produce a gapped array of quantum
wires, one requires $M\,N^{\,}_{\mathrm{w}}$ admissible tunneling vectors
(since each admissible tunneling vector gaps out two gapless modes).
It is therefore necessary to compare the number of tunneling vectors
in the set defined in Eqs.\
\eqref{arbitrary dimension tunneling vectors} 
with the number of wires in the array.  From Eqs.\ 
\eqref{N wire}, \eqref{N star}, and \eqref{N plaquette}, 
we see that
\begin{align}\label{arbitrary d counting}
N^{\,}_{\mathrm{s}}+N^{\,}_{\mathrm{p}}\geq 
N^{\,}_{\mathrm{w}},
\end{align}
with strict equality occurring in arrays of type $(2,1)$, 
which were studied in 
Sec.\ \ref{sec: Three-dimensional wire constructions}.

In cases where Eq.\ \eqref{arbitrary d counting} is an inequality, 
the question arises of how one can account for the extra 
gapless degrees of freedom.  
In this case, we can appeal to intuition developed from the study of 
toric codes in arbitrary dimensions (see, e.g., Ref.\ \cite{Mazac12}).  
In a toric code on a hypercubic lattice of type $(d,k^{\,}_{0})$ 
(with spin-1/2 degrees of freedom, rather than quantum wires, 
placed on the centers of elementary $k^{\,}_{0}$-cells of the 
$d$-dimensional hypercubic lattice), 
the same inequality shown in Eq.\ \eqref{arbitrary d counting} 
holds (with the number of spins now given by $N^{\,}_{\mathrm{w}}$).  
However, in the toric code of type $(d,k^{\,}_{0})$, 
there are precisely $N^{\,}_{\mathrm{s}}+N^{\,}_{\mathrm{p}}-N^{\,}_{\mathrm{w}}$ 
\textit{local constraints} that account for the discrepancy between the 
number of spin-1/2 degrees of freedom and the total number of stars and 
plaquettes.  
For example, in the three-dimensional toric code labeled by $(3,1)$, 
the product of all two-dimensional plaquettes on the surface of 
a cubic unit cell of the lattice is equal to 1.  
This introduces $N$ local constraints, 
since there are $N$ such cubes in the lattice. From Eqs.\ \eqref{N wire}, 
\eqref{N star}, and $\eqref{N plaquette}$, we have 
$N^{\,}_{\mathrm{w}}= 3N$, 
$N^{\,}_{\mathrm{s}}=N$, 
and 
$N^{\,}_{\mathrm{p}}=3N$.  
The $N$ local constraints thus account for the 
$N^{\,}_{\mathrm{s}}+N^{\,}_{\mathrm{p}}-N^{\,}_{\mathrm{w}}=N$ 
missing degrees of freedom.  (Note that, similarly to the toric code,
there are also further nonlocal constraints among the tunneling vectors.
These are important for determining the topological ground-state degeneracy.)

In the corresponding array of coupled quantum wires, 
the local constraints described above translate into linear dependencies 
within the sets 
$\{\mathcal{T}^{(\texttt{j})}_{s}\}$ 
and 
$\{\mathcal{T}^{(\texttt{j})}_{p}\}$ for each flavor 
$\texttt{j} = 1,\dots,M$.  
In other words, if there are 
$N^{\,}_{\mathrm{s}}+N^{\,}_{\mathrm{p}}>N^{\,}_{\mathrm{w}}$ 
tunneling vectors for each of the $M$ flavors of hypercubic stars 
and plaquettes, 
then precisely $N^{\,}_{\mathrm{s}}+N^{\,}_{\mathrm{p}}-N^{\,}_{\mathrm{w}}$ 
of these tunneling vectors are linearly dependent.  
This ensures that an array of quantum wires with 
$2M\,N^{\,}_{\mathrm{w}}$ 
gapless degrees of freedom has precisely 
$MN^{\,}_{\mathrm{w}}$ 
\textit{linearly independent} tunneling vectors of the form 
\eqref{arbitrary dimension tunneling vectors}.  
We will provide an example of this linear dependence in the next section, 
where we present an array of coupled wires of type $(3,1)$.

Once an appropriate set of interactions encoded by the tunneling vectors 
$\mathcal{T}^{(\texttt{j})}_{s}$
and 
$\mathcal{T}^{(\texttt{j})}_{p}$ 
has been chosen, for example by the construction outlined
in this section, the $d$-dimensional array of quantum wires 
becomes a gapped $D\equiv(d+1)$-dimensional state of matter.  
The excitations of this state of matter, as well as their
(possibly) fractional quantum numbers and any 
associated topological degeneracy on the $D$-torus,
can be studied using the methods of 
Sec.\ \ref{subsec: Fractionalization in the coupled wire array}.
As in that section, one identifies excitations with solitons in the pinned fields 
$\mathcal{T}^{(\texttt{j})\T}_{s}\,\mathcal{K}\Phi$ 
and 
$\mathcal{T}^{(\texttt{j})\T}_{p}\,\mathcal{K}\Phi$.
Depending on the values of $d$ and $k^{\,}_{0}$ 
that characterize the underlying hypercubic lattice,
these defects will be pointlike, stringlike, or membranelike in nature.  
When periodic boundary conditions are imposed, propagating these pointlike,
stringlike, or membranelike defects across the entire system defines nonlocal 
string and/or membrane operators, whose algebra can be used to determine 
the presence or absence of topological order in
the strongly-interacting, $D$-dimensional, gapped phase of matter.

\begin{figure}[t]
\begin{center}
\begin{flushleft}
\hspace{.5cm}(a)
\end{flushleft}
\includegraphics[width=.4\textwidth]{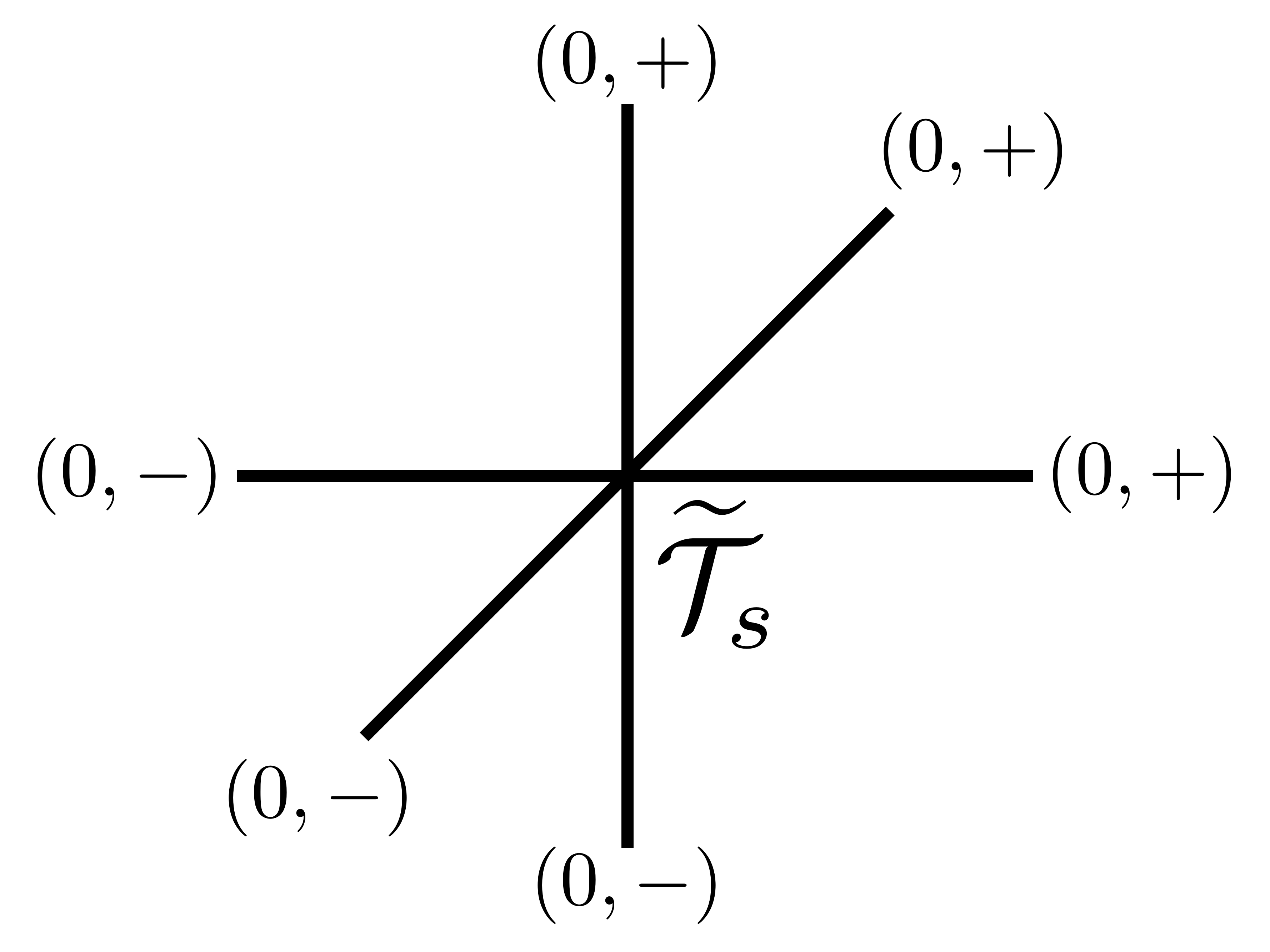}\\
\begin{flushleft}
\hspace{.5cm} (b)
\end{flushleft}
\includegraphics[width=.4\textwidth]{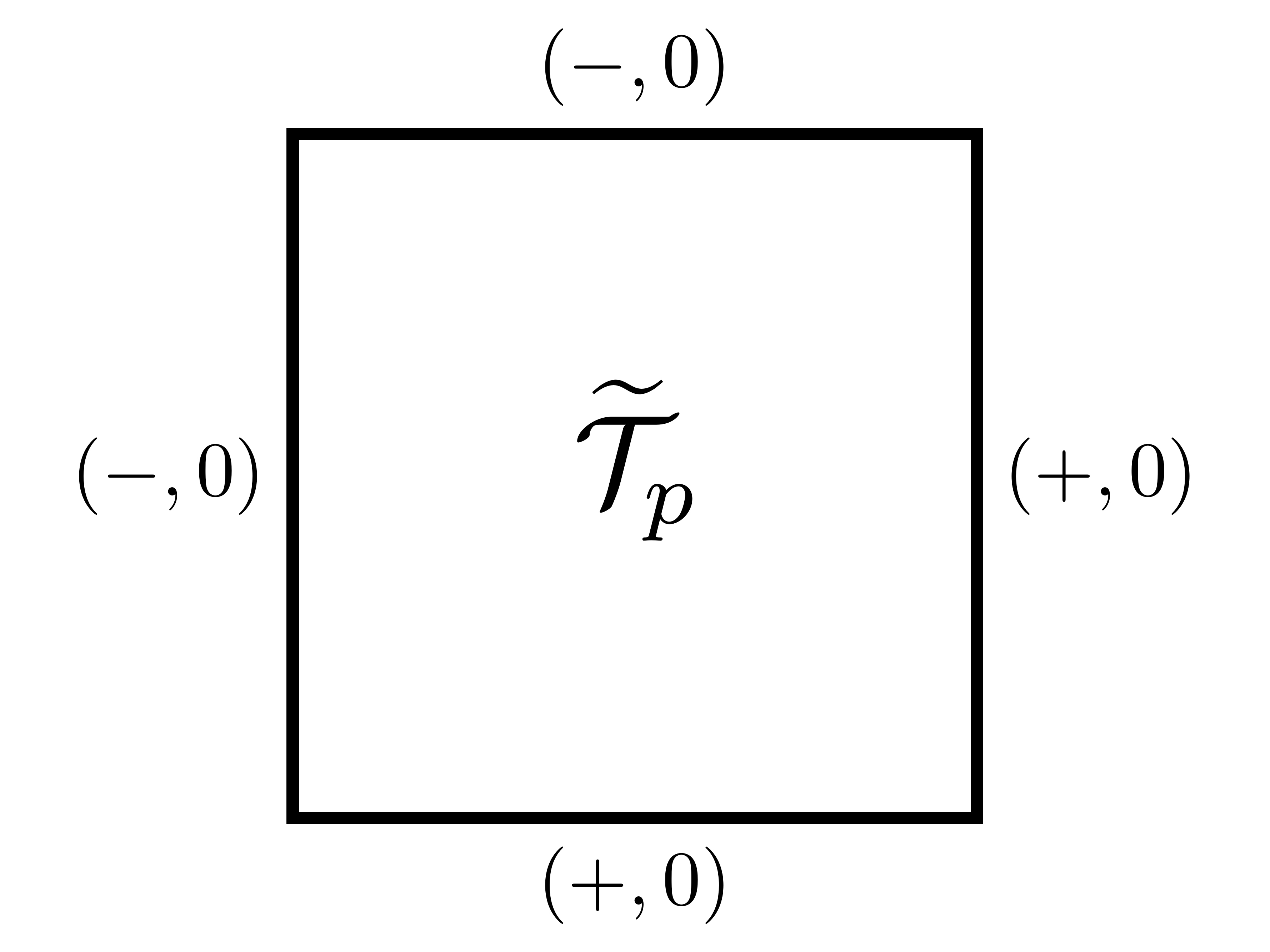}
\caption{
Pictorial representations of the tunneling vectors 
(a) $\tilde{\mathcal{T}}^{\,}_{s}$ 
and
(b) $\tilde{\mathcal{T}}^{\,}_{p}$,
defined in Eqs.\ \eqref{arbitrary dimension tunneling vectors},
for the array of quantum wires of type $(3,1)$.  
As in Fig.\ \ref{fig: Zm 2D tunneling vecs}, they are
built using the vectors $\tilde{v}$ and $\tilde{w}$ 
defined in Eqs.\ (\ref{eq: Zm 2D tunneling vecs}).
\label{fig: Zm 3D tunneling vecs (3,1)}
         }
\end{center}
\end{figure}

\begin{figure}[t]
\begin{center}
\begin{flushleft}
\hspace{.5cm}(a)
\end{flushleft}
\includegraphics[width=.36\textwidth]{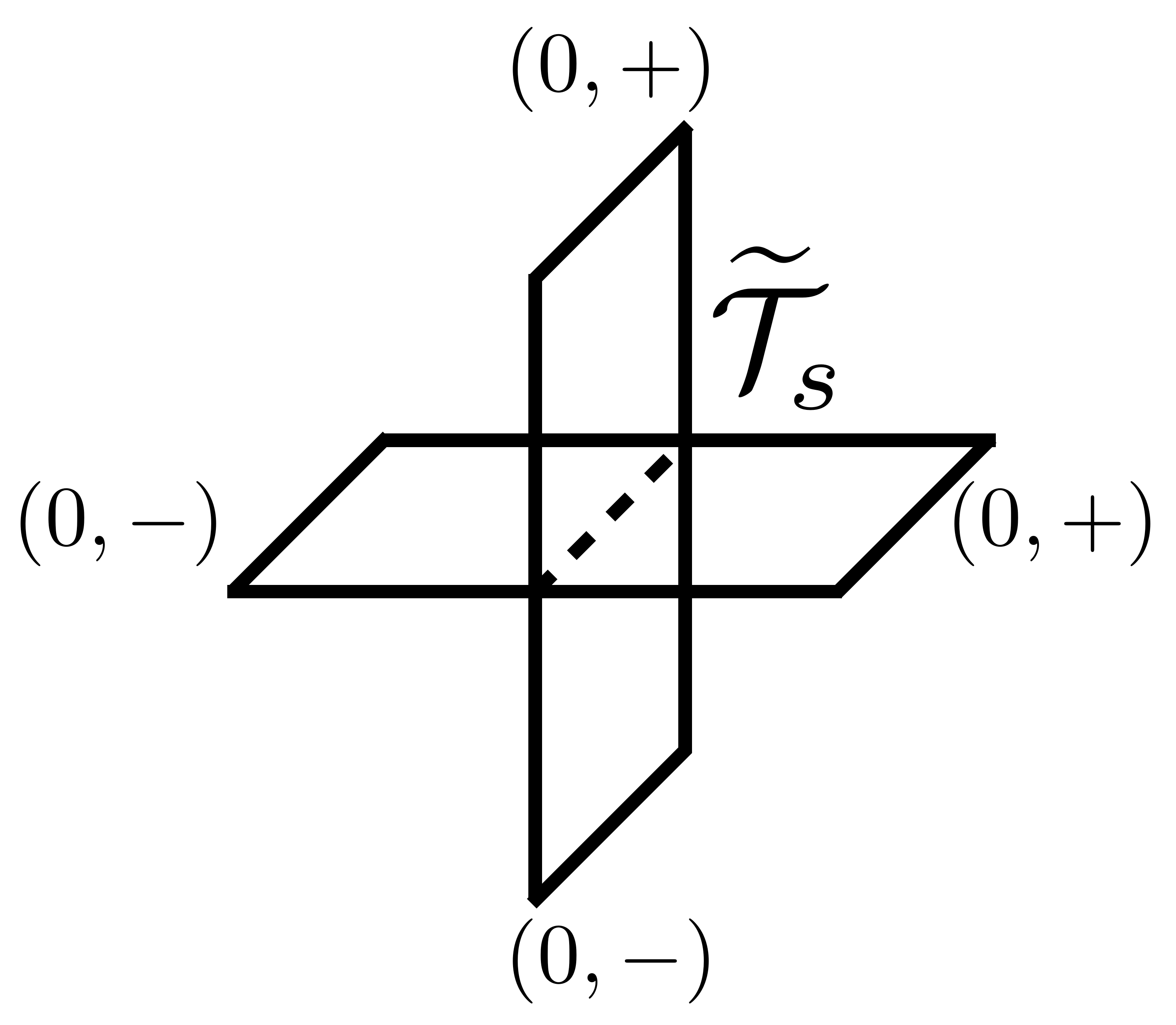}\\
\begin{flushleft}
\hspace{.5cm} (b)
\end{flushleft}
\includegraphics[width=.36\textwidth]{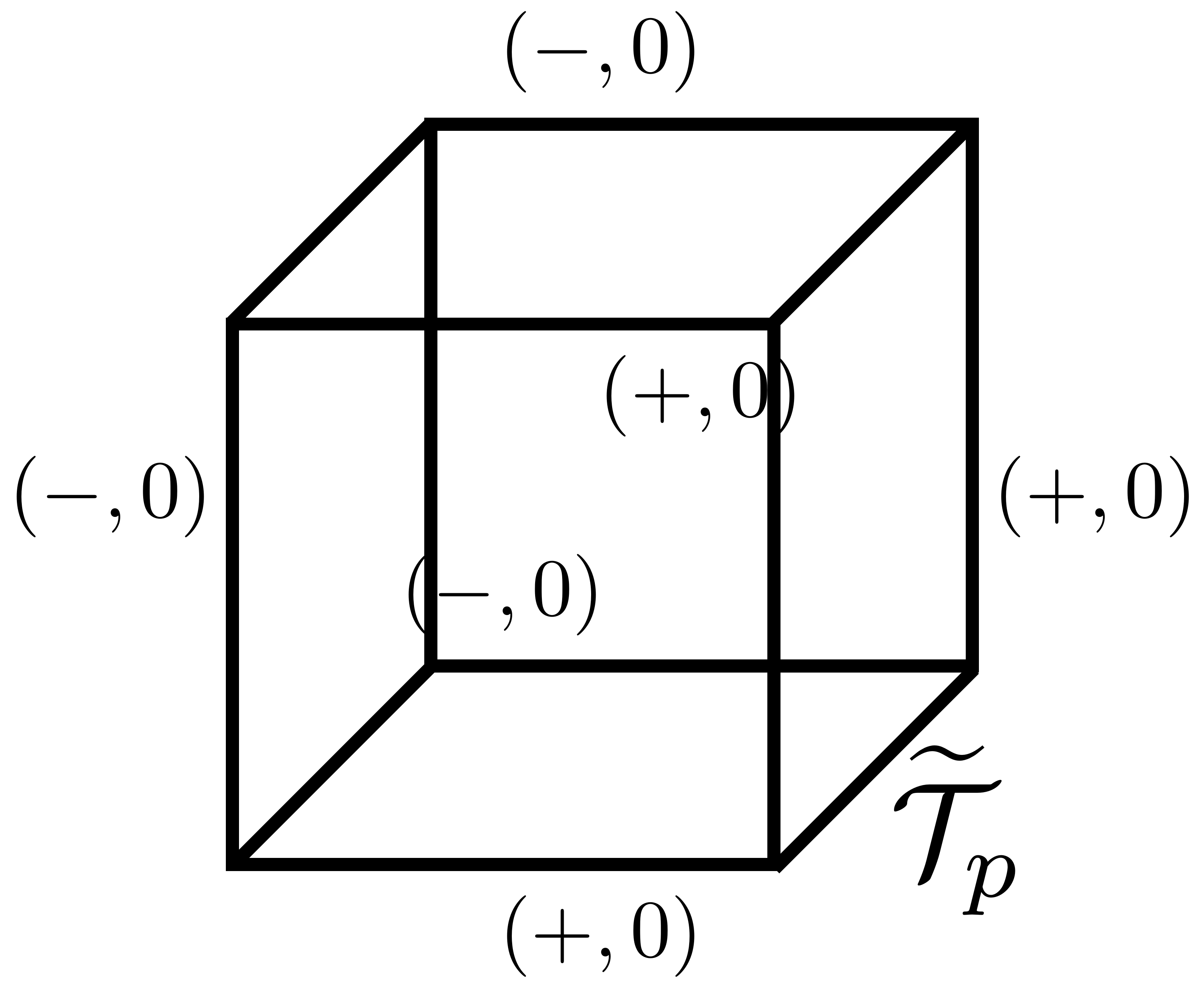}
\caption{
Pictorial representations of the tunneling vectors 
(a) $\tilde{\mathcal{T}}^{\,}_{s}$ 
and
(b) $\tilde{\mathcal{T}}^{\,}_{p}$,
defined in Eqs.~\eqref{arbitrary dimension tunneling vectors},
for the array of quantum wires of type $(3,2)$.  
As in Figs.\ \ref{fig: Zm 2D tunneling vecs}
and \ref{fig: Zm 3D tunneling vecs (3,1)}, they are
built using the vectors $\tilde{v}$ and $\tilde{w}$ 
defined in Eqs.\ 
\eqref{eq: Zm 2D tunneling vecs}.
\label{fig: Zm 3D tunneling vecs (3,2)}
         }
\end{center}
\end{figure}

\subsection{Example: $\mathbb{Z}^{\,}_{m}$ topological order in 
four-dimensional space from coupled wires}
\label{subsec: Example: 4D toric codes}

In this section, we provide a concrete example of how the construction 
of Abelian topological states of matter outlined in Sec.\
\ref{sec: Three-dimensional wire constructions} 
can be generalized to higher dimensions.  
In particular, we construct four-dimensional analogs of
the $\mathbb{Z}^{\,}_{m}$ topological states of matter explored in Sec.\
\ref{subsec: Zm example}.

Our starting point is a cubic array of quantum wires of type $(3,1)$, with 
periodic boundary conditions imposed in all four spatial directions from 
the outset. [We will also consider in parallel a related realization of 
$\mathbb{Z}^{\,}_{m}$ 
topological order that starts from the dual array of type $(3,2)$.]  
This array has the stars and plaquettes shown in Fig.\
\ref{fig: hypercubic star and plaquette}(b).
We use the coordinates $x,y,$ and $z$ to label directions 
within the cubic array, and $w$ to label the coordinate along each wire.

The initial Lagrangian of the system of decoupled wires is precisely 
the one described in Sec.\ 
\ref{subsubsec: Definitions and interwire couplings} 
for a system with $N^{\,}_{\mathrm{w}}=3N$
quantum wires, each containing $2M=2$ gapless degrees of freedom.  
In particular, starting from the free Lagrangian \eqref{L0 d=2} with
\begin{align}
\mathcal{K}\:=
\mathbbm{1}^{\,}_{3N}\otimes K^{\,}_{\mathrm{b}},
\end{align}
where the bosonic $K$-matrix $K^{\,}_{\mathrm{b}}$ is defined in Eq.\
\eqref{eq: def bosonic K b},
we perform the change of basis \eqref{eq: def W transformation} 
with the $2\times 2$ matrix $W$ given by Eq.\ \eqref{eq: Zm W to tilde}. 
In this way, we obtain a theory of decoupled wires with the $K$-matrix
$\tilde{K}^{\rm}_{m}$ given in Eq.~\eqref{eq: Zm K tilde}.
(It is worth pointing out here that this initial phase of the construction is, 
as we have seen in this paragraph, 
independent of the dimensionality of the array of quantum wires.)

Next, we couple the wires with the many-body interactions
$\tilde{\hat{L}}^{\,}_{\{\tilde{\mathcal{T}}\}}$,
defined as in Eq.\ \eqref{Lint}, 
for tunneling vectors $\tilde{\mathcal{T}}$ given by
Eqs.\ \eqref{arbitrary dimension tunneling vectors} 
with the two-dimensional vectors $\tilde{v}$ and $\tilde{w}$ 
defined in Eqs.\ \eqref{eq: Zm 2D tunneling vecs}.  
These tunneling vectors are shown in Fig.\ 
\ref{fig: Zm 3D tunneling vecs (3,1)} 
for the array of type (3,1), and in 
Fig.\ \ref{fig: Zm 3D tunneling vecs (3,2)} for the array of type (3,2).  
Using these pictorial representations of the tunneling vectors, 
one can verify that both sets of tunneling vectors satisfy
the Haldane criterion \eqref{Haldane criterion} with the $K$-matrix 
\eqref{eq: Zm K tilde}, as desired.  Furthermore, it is straightforward
to check that these tunneling vectors are charge-conserving. They
satisfy Eq.~\eqref{charge conservation} for any charge vector
$\mathcal Q$ of the form \eqref{charge vector definition}.

\begin{figure}[t]
\begin{center}
\includegraphics[width=.3\textwidth]{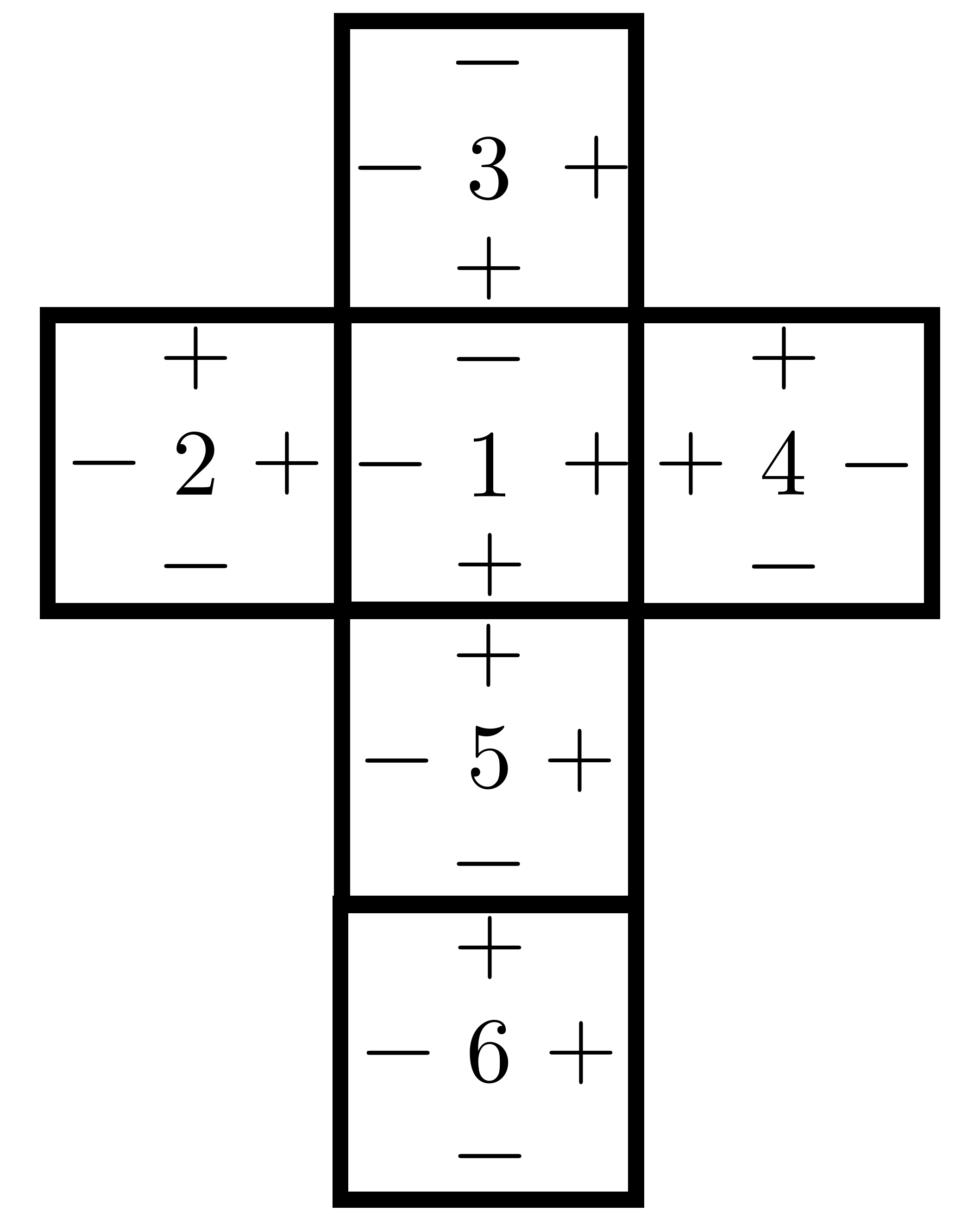}
\caption{
Pictorial representation of the six plaquette-centered tunneling vectors
$\mathcal{T}^{\,}_{p}$ surrounding a cubic cell of the array of quantum
wires labeled by (3,1). (Folding sides $2,\dots,5$ upwards out of the page
and placing side $6$ on top constructs the cubic cell.) 
The numbers $1,\dots,6$ label these
tunneling vectors in the order in which they appear in the Gram matrix $G$ in
Eq.\ \eqref{Gram matrix}.  The signs $\pm$ indicate whether a $\pm 1$
appears in the tunneling vector associated with that link.
\label{fig: Gram matrix}
         }
\end{center}
\end{figure}

We now verify that the tunneling vectors depicted in 
Figs.\ \ref{fig: Zm 3D tunneling vecs (3,1)} and
\ref{fig: Zm 3D tunneling vecs (3,2)} are sufficient in number to produce
a gapped four-dimensional state of matter.  We will focus here on the
array of coupled wires of type $(3,1)$, since the counting is identical
for the array of type $(3,2)$. To do this, we recall from 
Sec.\ \ref{subsec: Generalizing the results of Sec. ...} that the total number
of tunneling vectors is given by $N^{\,}_{\mathrm{s}}+N^{\,}_{\mathrm{p}}=4N$, while
the total number of gapless degrees of freedom in the array of decoupled 
quantum wires is $2N^{\,}_{\mathrm{w}}=6N$.  Since only $3N$ linearly independent
tunneling vectors are necessary to produce a fully gapped state of matter,
there must be a set of local constraints that removes $N$ tunneling vectors 
from the Haldane set $\mathbb{H}$. One can check that this is indeed the case, 
as the set of six tunneling vectors $\mathcal{T}^{\,}_{p}$ 
lining the surface of any cubic cell
of the three-dimensional cubic lattice are linearly dependent.  
One can verify this statement by computing the Gram matrix with elements
\begin{align}
G^{\,}_{p\pri{p}}\:=
\mathcal{T}^{\T}_{p}\,
\mathcal{T}^{\,}_{\pri{p}},
\end{align}
where the plaquettes $p$ and $\pri{p}$ border such a cubic cell.  
One finds (see Fig.~\ref{fig: Gram matrix} for guidance) that
\begin{align}\label{Gram matrix}
G = 
\begin{pmatrix}
4 & -1 & -1 & +1 & +1 & 0\\
-1 & 4 & -1 & 0 & +1 & +1\\
-1 & -1 & 4 & +1 & 0 & +1\\
+1 & 0 & +1 & 4 & -1 & -1\\
+1 & +1 & 0 & -1 & 4 & -1\\
0 & +1 & +1 & -1 & -1 & 4
\end{pmatrix},
\end{align}
which has vanishing determinant, indicating that this set of six tunneling
vectors is linearly dependent. Since the cubic lattice contains exactly $N$
such cubes when periodic boundary conditions are imposed, there are $N$
linearly dependent vectors that can be removed from the Haldane set 
$\mathbb{H}$.
[Note that a similar set of local constraints for \textit{stars} 
holds in the case of the array of wires of type (3,2).  
This is due to the duality between hypercubic arrays
of types $(d,k^{\,}_{0})$ and $(d,d-k^{\,}_{0})$ mentioned in 
Sec.\ \ref{subsec: Review of toric codes in arbitrary dimensions}.]

On the basis of the above arguments, we conclude that the cubic arrays of
types $(3,1)$ and $(3,2)$ both yield fully gapped, four-dimensional states of
matter when periodic boundary conditions are imposed.  The excitations
of both states of matter can be studied according to the methodology laid out
in Sec.~\ref{subsec: Fractionalization in the coupled wire array}, 
and by example in Sec.~\ref{subsubsec: Excitations}.  As in that section,
the building blocks of excitations in the array of coupled wires are 
the local operators
\begin{subequations}
\label{eq: 3D Zm defect hopping}
\begin{align}\label{eq: 3D Zm defect hopping a}
\hat{S}^{\dag}_{s^{\,}_{C^{\,}_{\mathrm{s}}}}(w)\:= 
\exp
\(
-\mathrm{i}\, 
\tilde{\hat{\phi}}^{\,}_{s^{\,}_{C^{\,}_{\mathrm{s}}},2}(w)
\)
\end{align}
and
\begin{align}\label{eq: 3D Zm defect hopping b}
\hat{P}^{\dag}_{p^{\,}_{C^{\,}_{\mathrm{p}}}}(w)\:= 
\exp
\(
+\mathrm{i}\, 
\tilde{\hat{\phi}}^{\,}_{p^{\,}_{C^{\,}_{\mathrm{p}}},1}(w)
\),
\end{align}
\end{subequations}
and the bilocal operators
\begin{subequations}
\label{eq: def 3D Zm w-string defs}
\begin{align}
\begin{split}
\hat{S}^{\dag}_{s^{\,}_{C^{\,}_{\mathrm{s}}}}(w^{\,}_{1},w^{\,}_{2})&\:=
\hat{S}^{\dag}_{s^{\,}_{C^{\,}_{\mathrm{s}}}}(w^{\,}_{2})\,
\hat{S}^{\phantom\dag}_{s^{\,}_{C^{\,}_{\mathrm{s}}}}(w^{\,}_{1})
\\
&=
\exp
\bigg(
-\mathrm{i}\, 
\int\limits^{w^{\,}_{2}}_{w^{\,}_{1}}\mathrm{d}w\ 
\partial^{\,}_{w}\tilde{\hat{\phi}}^{\,}_{s^{\,}_{C^{\,}_{\mathrm{s}}},2}(w)
\bigg),
\end{split}
\label{eq: def 3D Zm w-string defs a}
\end{align}
and
\begin{align}
\begin{split}
\hat{P}^{\dag}_{p^{\,}_{C^{\,}_{\mathrm{p}}}}(w^{\,}_{1},w^{\,}_{2})&\:=
\hat{P}^{\dag}_{p^{\,}_{C^{\,}_{\mathrm{p}}}}(w^{\,}_{2})\,
\hat{P}^{\phantom\dag}_{p^{\,}_{C^{\,}_{\mathrm{p}}}}(w^{\,}_{1})
\\
&=
\exp
\bigg(
+\mathrm{i}\, 
\int\limits^{w^{\,}_{2}}_{w^{\,}_{1}}\mathrm{d}w\ 
\partial^{\,}_{w}\tilde{\hat{\phi}}^{\,}_{p^{\,}_{C^{\,}_{\mathrm{p}}},1}(w)
\bigg).
\end{split}
\label{eq: def 3D Zm w-string defs b}
\end{align}
\end{subequations}
Here, we recall that the coordinate along each wire is now labeled by $w$,
and that the indices $C^{\,}_{\mathrm{s}}=1,\dots,2[d-(k^{\,}_{0}-1)]$ and
$C^{\,}_{\mathrm{p}}\:=1,\dots,2(k^{\,}_{0}+1)$ label the generalized cardinal
directions associated with each star or plaquette.  

The effects of the operators 
defined in Eqs.\
\eqref{eq: 3D Zm defect hopping}
and 
\eqref{eq: def 3D Zm w-string defs} 
on the pinned fields 
$\mathcal{T}^{\T}_{s}\,\tilde{\mathcal{K}}\,\tilde\Phi$ 
and 
$\mathcal{T}^{\T}_{p}\,\tilde{\mathcal{K}}\,\tilde\Phi$ 
can be computed analogously to Eqs.\ 
\eqref{eq: 2D pinning soliton star},
\eqref{eq: 2D pinning soliton star prime},
\eqref{eq: 2D pinning soliton plaquette},
and
\eqref{eq: 2D pinning soliton plaquette prime}.
As shown there, these operators give rise to the excitations 
of the array of coupled wires.
One can deduce whether the excitations created and propagated 
by these operators 
are pointlike, stringlike, or membranelike by first acting with one of these
operators on a single link.  This creates some number of defective stars 
or plaquettes (depending on the coordination number of that link 
and whether or not the operator acts along the direction of the wire).  
From there, one can grow a surface with excitations on its boundary 
by attempting to heal all defects created in this way with further applications
of the operators defined in Eqs.\ 
\eqref{eq: 3D Zm defect hopping}
or 
\eqref{eq: def 3D Zm w-string defs}.  
Processes analogous to this one are shown in Figs.\
\ref{fig: deconfined star defects} and
\ref{fig: xy line and membrane defects} 
for the array of type $(2,1)$ studied in Sec.\
\ref{sec: Three-dimensional wire constructions}.
As in Sec.~\ref{subsubsec: Excitations}, the electric
charge associated with these defects can be computed as in Eq.\
\eqref{eq: Zm charges for star and plaquette operators}.
The excitations of the array of type $(3,1)$ 
differ in character from those of the array of type $(3,2)$,
as we shall now see.

The excitations of the array of type $(3,1)$ can be pointlike, linelike, 
or membranelike in nature.  To see this, note that
applying the operator 
$\hat{S}^{\dag}_{s^{\,}_{C^{\,}_{\mathrm{s}}}}(w)$ 
in the wire labeled by 
$s^{\,}_{C^{\,}_{\mathrm{s}}}$
creates two defective stars, as there are two stars bordering each link 
in the array of wires.  These defective stars can be propagated away 
from one another, much as in Fig.\ \ref{fig: deconfined star defects}, 
by further applications of the operator $\hat{S}^{\dag}_{s^{\,}_{C^{\,}_{\mathrm{s}}}}(w)$.
Consequently, we may view the defective stars as pointlike
excitations with electric charge $\pm 2/m$ 
[recall Eq.\ \eqref{eq: Zm charges for star and plaquette operators}], 
connected by a ``string" of vertex operators.
One can also construct linelike excitations, for example by acting instead with
the bilocal operator 
$\hat{S}^{\dag}_{s^{\,}_{C^{\,}_{\mathrm{s}}}}(w^{\,}_{1},w^{\,}_{2})$,
similarly to Fig.\ \ref{fig: xy line and membrane defects}(a)-(b).

On the other hand, suppose that one applies the operator
$\hat{P}^{\dag}_{p^{\,}_{C^{\,}_{\mathrm{p}}}}(w)$ 
in the wire labeled by 
$p^{\,}_{C^{\,}_{\mathrm{p}}}$.  
In this case, one creates \textit{four} defective plaquettes, 
as each link is shared by four plaquettes.  
Attempting to heal these defects by subsequent applications 
of the operator $\hat{P}^{\dag}_{p^{\,}_{C^{\,}_{\mathrm{p}}}}(w)$ 
leads to a two-dimensional membrane of vertex operators
with linelike defects on its boundary,
much like in Fig.\ \ref{fig: xy line and membrane defects}(c)-(d).  
Furthermore, one can also create a 3-brane of
vertex operators with two-dimensional membranelike excitations 
on its terminating surfaces, by applying the operator 
$\hat{P}^{\dag}_{p^{\,}_{C^{\,}_{\mathrm{p}}}}(w^{\,}_{1},w^{\,}_{2})$ 
instead of 
$\hat{P}^{\dag}_{p^{\,}_{C^{\,}_{\mathrm{p}}}}(w)$.

A similar set of excitations can be constructed for the case of the array 
of type $(3,2)$.  The only difference is that, in this case,
the vertex operators associated with \textit{stars} naturally form membranes, 
similarly to the plaquettes in the array of type $(3,1)$.  This makes sense
in light of the duality between these two arrays of quantum wires, 
which exchanges stars and plaquettes, and therefore
also necessarily exchanges star and plaquette defects.

We now demonstrate that the gapped four-dimensional phases of matter associated
with the cubic arrays of types $(3,1)$ and $(3,2)$ are topologically ordered. 
calculating the minimal topological ground-state degeneracy on the four-torus,
\begin{equation}
\mathbb{T}^{4}\equiv 
S^{1}\times S^{1}\times S^{1}\times S^{1},
\end{equation}
by generalizing the analysis of Sec.\ 
\ref{subsubsec: Topological ground state degeneracy}, 
i.e., by presenting the algebra of nonlocal operators from which the degeneracy
is derived. We focus on the array of type $(3,1)$, 
as the degeneracy of the array of type $(3,2)$ 
is the same by the duality discussed in Sec.\
\ref{subsec: Review of toric codes in arbitrary dimensions}.
In both cases, the origin of the multidimensionality of the ground-state 
manifold is the nontrivial (for $m>1$) equal-time algebra
\begin{align}\label{vertex algebra in a wire}
\begin{split}
\hat{S}^{\dag}_{j}(w)\,\hat{P}^{\dag}_{j}(0,L)&=
\hat{P}^{\dag}_{j}(0,L)\,
\hat{S}^{\dag}_{j}(w)\, 
e^{-\mathrm{i}\, 2\pi/m},\\
\hat{P}^{\dag}_{j}(w)\, 
\hat{S}^{\dag}_{j}(0,L)&=
\hat{S}^{\dag}_{j}(0,L)\, 
\hat{P}^{\dag}_{j}(w)\, 
e^{+\mathrm{i}\, 2\pi/m},
\end{split}
\end{align}
which holds independently of dimensionality or lattice geometry 
as it is a property of operators defined in a single wire.
Consequently, there is no obstruction to repeating this analysis 
for any hypercubic array of type $(d,k^{\,}_{0})$.

The ground state degeneracy on $\mathbb{T}^{4}$ of the array of type $(3,1)$ 
is encoded in the algebra of the nonlocal operators
\begin{subequations}
\begin{align}
&
\hat{\mathcal{O}}^{\,}_{P,\Omega^{\,}_{\hat{x}\hat{y}\hat{z}}}(w)\:=
\prod_{p^{\,}_{C^{\,}_{\mathrm{p}}}\in\Omega^{\,}_{\hat{x}\hat{y}\hat{z}}} 
P^{\dag}_{p^{\,}_{C^{\,}_{\mathrm{p}}}}(w),
\\
&
\hat{\mathcal{O}}^{\,}_{P,\Omega^{\,}_{\hat{x}\hat{y}\hat{w}}}\:=
\prod_{p^{\,}_{C^{\,}_{\mathrm{p}}}\in\Omega^{\,}_{\hat{x}\hat{y}}}
P^{\dag}_{p^{\,}_{C^{\,}_{\mathrm{p}}}}(0,L),
\\
&
\hat{\mathcal{O}}^{\,}_{P,\Omega^{\,}_{\hat{x}\hat{z}\hat{w}}}\:=
\prod_{p^{\,}_{C^{\,}_{\mathrm{p}}}\in\Omega^{\,}_{\hat{x}\hat{z}}}
P^{\dag}_{p^{\,}_{C^{\,}_{\mathrm{p}}}}(0,L),
\\
&
\hat{\mathcal{O}}^{\,}_{P,\Omega^{\,}_{\hat{y}\hat{z}\hat{w}}}\:=
\prod_{p^{\,}_{C^{\,}_{\mathrm{p}}}\in\Omega^{\,}_{\hat{y}\hat{z}}}
P^{\dag}_{p^{\,}_{C^{\,}_{\mathrm{p}}}}(0,L),
\end{align}
which act along 3-branes, and
\begin{align}
&
\hat{\mathcal{O}}^{\,}_{S,\Gamma^{\,}_{x}}(w)\:=
\prod_{s^{\,}_{C^{\,}_{\mathrm{s}}}\in\Gamma^{\,}_{x}}
\hat{S}^{\dag}_{s^{\,}_{C^{\,}_{\mathrm{s}}}}(w), 
\\
&
\hat{\mathcal{O}}^{\,}_{S,\Gamma^{\,}_{y}}(w)\:=
\prod_{s^{\,}_{C^{\,}_{\mathrm{s}}}\in\Gamma^{\,}_{y}} 
\hat{S}^{\dag}_{s^{\,}_{C^{\,}_{\mathrm{s}}}}(w), 
\\
&
\hat{\mathcal{O}}^{\,}_{S,\Gamma^{\,}_{z}}(w)\:=
\prod_{s^{\,}_{C^{\,}_{\mathrm{s}}}\in\Gamma^{\,}_{z}} 
\hat{S}^{\dag}_{s^{\,}_{C^{\,}_{\mathrm{s}}}}(w), 
\\
&
\hat{\mathcal{O}}^{\,}_{S,\Gamma^{\,}_{w}}\:=
\hat{S}^{\dag}_{s^{\,}_{0}}(0,L),
\end{align}
\end{subequations}
which act along strings.  The volume 
$\Omega^{\,}_{\hat x \hat y \hat z}$, the surface
$\Omega^{\,}_{\hat x \hat y}$, the line $\Gamma^{\,}_{x}$, etc.\ are
defined analogously to their counterparts in Sec.\
\ref{subsubsec: Topological ground state degeneracy}.  
Their algebra is found to be
\begin{subequations}
\begin{align}
\hat{\mathcal{O}}^{\,}_{P,\Omega^{\,}_{\hat{x}\hat{y}\hat{z}}}(w)\, 
\hat{\mathcal{O}}^{\,}_{S,\Gamma^{\,}_{w}}&=
\hat{\mathcal{O}}^{\,}_{S,\Gamma^{\,}_{w}}\, 
\hat{\mathcal{O}}^{\,}_{P,\Omega^{\,}_{\hat{x}\hat{y}\hat{z}}}(w)\,
e^{+\mathrm{i}\,2\pi/m},
\\
\hat{\mathcal{O}}^{\,}_{P,\Omega^{\,}_{\hat{x}\hat{y}\hat{w}}}\, 
\hat{\mathcal{O}}^{\,}_{S,\Gamma^{\,}_{z}}(w)&=
\hat{\mathcal{O}}^{\,}_{S,\Gamma^{\,}_{z}}(w)\, 
\hat{\mathcal{O}}^{\,}_{P,\Omega^{\,}_{\hat{x}\hat{y}\hat{w}}}\,
e^{-\mathrm{i}\,2\pi/m},
\\
\hat{\mathcal{O}}^{\,}_{P,\Omega^{\,}_{\hat{x}\hat{z}\hat{w}}}\, 
\hat{\mathcal{O}}^{\,}_{S,\Gamma^{\,}_{y}}(w)&=
\hat{\mathcal{O}}^{\,}_{S,\Gamma^{\,}_{y}}(w)\, 
\hat{\mathcal{O}}^{\,}_{P,\Omega^{\,}_{\hat{x}\hat{z}\hat{w}}}\,
e^{-\mathrm{i}\,2\pi/m},
\\
\hat{\mathcal{O}}^{\,}_{P,\Omega^{\,}_{\hat{y}\hat{z}\hat{w}}}\, 
\hat{\mathcal{O}}^{\,}_{S,\Gamma^{\,}_{x}}(w)&=
\hat{\mathcal{O}}^{\,}_{S,\Gamma^{\,}_{x}}(w)\, 
\hat{\mathcal{O}}^{\,}_{P,\Omega^{\,}_{\hat{y}\hat{z}\hat{w}}}\,
e^{-\mathrm{i}\, 2\pi/m},
\end{align}
\end{subequations}
where we have made extensive use of Eq.\
\eqref{vertex algebra in a wire}.  
Similarly to what was found in Sec.\
\ref{subsubsec: Ground state degeneracy}, 
each line of the above algebra is independent from 
(i.e.,\ commutes with) the others, and 
contributes an $m$-fold topological degeneracy.  
We conclude that the total topological degeneracy on the four-torus of this
state of matter is
\begin{align}\label{4D degeneracy}
D^{\,}_{\mathbb{T}^{4}}= 
m^{4}.
\end{align}
For $m=2$, the 16-fold degeneracy coincides with the ground-state degeneracy 
of the four-dimensional toric code defined on the hypercubic lattice of type 
$(4,1)$\ \cite{Mazac12}.  For $m>1$, we have therefore arrived at
a state of matter whose low-lying excitations and topological ground-state 
degeneracy on the four-torus are consistent with a 
$\mathbb{Z}^{\,}_{m}$-topologically-ordered phase in four spatial dimensions.

From here, one could further generalize the discussion of 
Sec.~\ref{subsubsec: Surface theory}
in order to enumerate the possible gapped or gapless states of matter 
on the three-dimensional boundary of the four-dimensional bulk topological phase.  Terminating the cubic lattice of type $(3,1)$ in the $y$-direction, say, 
leads to a surface lattice of type $(2,1)$, i.e.,~a square lattice 
with wires on the links.  One is then free to impose any single-particle
tunneling or many-body interactions one wishes on the surface, 
so long as these surface terms are Haldane-compatible with the bulk.  
For example, one could search the space of tunneling
vectors like those defined in Sec.~\ref{subsec: 2D coupling discussion} 
to generate a set of allowed many-body interactions.
This method of studying the surface states can be readily generalized 
to any hypercubic array of quantum wires of type $(d,k^{\,}_{0})$,
like those studied in Sec.~\ref{sec: Higher-dimensional wire constructions}, 
to answer questions about higher-dimensional generalizations of the concept 
of surface topological order, for example.

The discussion of this section can be generalized to arrays of quantum wires 
of type $(d,k^{\,}_{0})$ to produce 
$\mathbb{Z}^{\,}_{m}$-topologically-ordered phases in higher dimensions.
Many of these higher-dimensional topological states of matter are particularly 
interesting in that they exhibit topological order at finite temperature%
~\cite{Mazac12}. 
The lowest-dimensional $\mathbb{Z}^{\,}_{m}$-topologically-ordered phase 
exhibiting topological order at finite temperature is the toric code 
of type (4,2). The discussion of this section demonstrates that one 
\textit{cannot} realize a topological state of matter in the universality class 
of the toric code of type (4,2) starting from an array of quantum wires of type 
(3,1) or (3,2). This is because both of these arrays yield topological states of 
matter whose degeneracy is consistent with the toric code of
type (4,1) [recall Eq.~\eqref{4D degeneracy}]. However, this does not preclude 
the possibility of designing arrays of quantum wires to yield
topological states of matter in $D=5$ or greater that have topological order 
at finite temperature. The detailed study of such
phases is beyond the scope of this work, 
but nevertheless a very interesting problem for future study.

\section{Conclusion}
\label{sec: Conclusion}

In this paper, we have outlined a general strategy for designing
Abelian topological phases of matter in $D$ spatial dimensions by
coupling an array of quantum wires in $d=D-1$ dimensions.  This
strategy hinges on the use of counting arguments introduced by
Haldane~\cite{Haldane95} to search for a set of compatible many-body
interactions that yields a gapped state of matter when the couplings
associated with these interactions are taken to infinity.  The
enumeration of the set of possible interactions, and the determination
of their compatibility, is aided by associating each interaction term
with one of the generalized stars and plaquettes of a $d$-dimensional
hypercubic lattice embedded in $D$-dimensional space.  In this sense,
the interactions that produce a gapped state of matter are arranged in
a manner reminiscent of the commuting projectors in a $d$-dimensional
toric code.

We found that many simplifications arise due to this similarity,
making these theories analytically tractable much as their forebears
in two dimensions.  In particular, the excitations of the arrays of
coupled wires can be studied thanks in part to analogies with similar
excitations in the $d$-dimensional toric code.  The fractional charge
and statistics of these excitations is readily accessible with
standard tools from Abelian bosonization.  Furthermore, when periodic
boundary conditions are imposed, the topological degeneracy (if any)
of the strongly-interacting, gapped, $D$-dimensional state of matter
can be determined with these tools.  Finally, when the array of
coupled wires is defined on a manifold with boundary, the stability of
gapless surface states on the $d$-dimensional boundary of the
$D$-dimensional topological phase can be addressed conveniently with
the same formalism in one less spatial dimension, provided that any
single-particle tunnelings or many-body interactions added to the
surface are compatible with those in the bulk.

There are many directions for future work in light of these findings.
First, it is important to note that the class of many-body
interactions introduced in Secs.~\ref{subsec: 2D coupling discussion}
and \ref{subsec: Generalizing the results of Sec. ...}  
are not the only ones possible, even when making use of the analogy to
$d$-dimensional toric codes; there are many other sets of compatible
tunneling vectors that can be associated with the stars and plaquettes
of hypercubic lattices.  Consequently, it would be instructive to map
out the set of all Abelian topological phases possible in three and
higher dimensions that are accessible with this approach.  Even in
three spatial dimensions, there are many possible topological field
theories beyond the BF-type theories explored in 
Sec.~\ref{subsubsec: Topological field theory}, 
such as those studied in
Refs.~\cite{Ye15a}, \cite{Ye15b}, and \cite{Chen15b}.  It would also
be interesting to determine whether other exactly solvable
commuting-projector Hamiltonians, besides toric codes, could be used
as bases for wire constructions like the ones undertaken in this work,
resulting in different classes of topological phases.  Second, it
would be interesting to study the surface states of these coupled-wire
arrays in more detail.  In particular, finding a useful way to
characterize gapless surfaces by extending the formalism presented in
this paper would be a very useful pursuit, as one might ask the
question of whether it is possible to find novel non-Fermi liquids or
conformal field theories on interacting surfaces of topological phases
in three or more dimensions.  In this pursuit, it would also be
crucial to make contact with existing work on the bulk-boundary
correspondence in three dimensions~\cite{Cho11,Chan13,Chen15b}.
Third, it is natural to ask how to extend this formalism to describe
non-Abelian topological states of matter.  This could be done by
investigating the possibility of using non-Abelian, rather than
Abelian, bosonization to describe the gapless wires and their
couplings to one another, as has been done in Refs.~\cite{Teo14} and
\cite{Huang15}.  Fourth, as was hinted at in this work, wire
constructions of topological phases in spatial dimensions greater than
two could prove useful in the study of surface topological order%
~\cite{Keyserlingk13,Vishwanath13,Wang13a,Wang13b,Burnell14,Chen14,%
Metlitski15,Mross15}.
In particular, it may be possible to use non-Abelian bosonization
techniques on the surfaces of Abelian topological phases to study
non-Abelian surface topological orders in a manner that treats the
surface and bulk physics simultaneously.

\begin{acknowledgments}
T.I. was supported by the National Science Foundation Graduate
Research Fellowship Program under Grant No.~DGE-1247312 and 
C.C. was supported by DOE Grant DEF-06ER46316.
\end{acknowledgments}

\appendix

\begin{widetext}

\section{Deconfinement of defects along the direction of a wire}
\label{appsec: Deconfinement of defects along the direction of a wire}

\setcounter{equation}{0}
\makeatletter
\renewcommand{\theequation}{A\arabic{equation}}

In this Appendix, we demonstrate that a pair of star defects in 
three dimensions, like the one shown in Fig.\
\ref{fig: deconfined star defects}(b), 
are deconfined from one another along the $z$-direction, 
despite the string of vertex operators connecting them.  
This is because a link in the string, 
which consists of two vertex operators applied on two legs of a star 
(see Fig.\ \ref{fig: deconfined star defects}), 
costs no additional energy if the vertex operators are displaced relative 
to one another along the $z$-axes of their respective wires.  
While we focus here on the specfic example of star defects in the 
$\mathbb{Z}^{\,}_{m}$-topologically-ordered state of matter constructed in Sec.\
\ref{subsec: Zm example}, 
the same analysis can be adapted to demonstrate that pointlike star and 
plaquette defects are deconfined in any dimension.

To see that this is the case, let us consider a star $s$ with two $2\pi/m$ 
solitons on the eastern and western legs.  We parameterize these solitons by 
decomposing the bosonic fields as
\begin{subequations}
\begin{align}
\begin{split}
\tilde{\hat{\phi}}^{\,}_{s^{\,}_{E},1}(z)&=
f^{\,}_{\mathrm{sol}}(z-z^{\,}_{E})
+
\tilde{\hat{\phi}}^{\prime}_{s^{\,}_{E},1}(z),
\\
\tilde{\hat{\phi}}^{\,}_{s^{\,}_{W},1}(z)&=
f^{\,}_{\mathrm{sol}}(z-z^{\,}_{W})
+
\tilde{\hat{\phi}}^{\prime}_{s^{\,}_{W},1}(z),
\end{split}
\end{align}
where the real-valued function
\begin{align}
f^{\,}_{\mathrm{sol}}(z-z^{\,}_{0})
&\:=
\frac{\pi}{m}\[\tanh\(\frac{z-z^{\,}_{0}}{\xi}\)+1\]
\end{align} 
is a fixed soliton profile centered
at $z^{\,}_{0}$, while it is the primed fields 
$\tilde{\hat{\phi}}^{\prime}_{s^{\,}_{C},1}(z)$ ($C=N,S,E,W$) that encode the
quantum fluctuations. 
\end{subequations}
On the one hand, if $z^{\,}_{E}=z^{\,}_{W}$, 
Eq.\ (\ref{eq: pinned fields if Zm topoorder a}) 
dictates that
\begin{subequations}
\begin{align}
\tilde{\mathcal{T}}^{\T}_{s}\,
\tilde{\mathcal{K}}^{\,}_m\,
\tilde{\hat{\Phi}}(z)=
m
\[
\tilde{\hat{\phi}}^{\,}_{s^{\,}_{N},1}(z)
-
\tilde{\hat{\phi}}^{\,}_{s^{\,}_{S},1}(z)
+
\tilde{\hat{\phi}}^{\prime}_{s^{\,}_{E},1}(z)
-
\tilde{\hat{\phi}}^{\prime}_{s^{\,}_{W},1}(z)
\]\equiv
\tilde{\mathcal{T}}^{\T}_{s}\,
\tilde{\mathcal{K}}^{\,}_m\,
\tilde{\hat{\Phi}}^{\prime}(z).
\end{align}
On the other hand, if $z^{\,}_{E}\neq z^{\,}_{W}$, 
Eq.\ (\ref{eq: pinned fields if Zm topoorder a}) 
dictates that
\begin{align}
\tilde{\mathcal{T}}^{\T}_{s}\,
\tilde{\mathcal{K}}^{\,}_m\,
\tilde{\hat{\Phi}}(z)=
m
\[
\tilde{\hat{\phi}}^{\,}_{s^{\,}_{N},1}(z)
-
\tilde{\hat{\phi}}^{\,}_{s^{\,}_{S},1}(z)
+
\tilde{\hat{\phi}}^{\prime}_{s^{\,}_{E},1}(z)
-
\tilde{\hat{\phi}}^{\prime}_{s^{\,}_{W},1}(z)
+
\delta\tilde{\phi}^{\,}_{s}(z)\],
\end{align}
where
\begin{align}
\delta\tilde{\phi}^{\,}_{s}(z)&=
\frac{\pi}{m}
\[
\tanh\(\frac{z-z^{\,}_{E}}{\xi}\)
-
\tanh\(\frac{z-z^{\,}_{W}}{\xi}\)
\].
\end{align}
\end{subequations}
In the limit $\xi\to0$ 
(i.e., the limit of perfectly sharp solitons), 
\begin{align}\label{difference of solitons}
\delta\tilde{\phi}^{\,}_{s}(z)&\longrightarrow
\frac{2\pi}{m}
\[
\Theta(z-z^{\,}_{E})
-
\Theta(z-z^{\,}_{W})
\].
\end{align}
Hence,
the difference in energy between the case with $z^{\,}_{E}\neq z^{\,}_{W}$ 
and the case with $z^{\,}_{E}=z^{\,}_{W}$ is given by
\begin{align}
\begin{split}
\delta E^{\,}_{s}&=
-U^{\,}_{s}\,
\int\mathrm{d}z\ 
\[
\cos
\(
\tilde{\mathcal{T}}^{\T}_{s}\,
\tilde{\mathcal{K}}^{\,}_m\,
\tilde{\hat{\Phi}}^{\prime}(z)
+
m\, 
\delta\tilde{\phi}^{\,}_{s}(z)
\)
-
\cos
\(
\tilde{\mathcal{T}}^{\T}_{s}\,
\tilde{\mathcal{K}}^{\,}_m\,
\tilde{\hat{\Phi}}^{\prime}(z)
\)
\]
\\
&=
-U^{\,}_{s}
\int\limits^{z^{\,}_{E}}_{z^{\,}_{W}}\mathrm{d}z\ 
\[
\cos
\(
\tilde{\mathcal{T}}^{\T}_{s}\,
\tilde{\mathcal{K}}^{\,}_m\,
\tilde{\hat{\Phi}}^{\prime}(z)
+
2\pi 
\)
-
\cos
\(
\tilde{\mathcal{T}}^{\T}_{s}\,
\tilde{\mathcal{K}}^{\,}_m\,
\tilde{\hat{\Phi}}^{\prime}(z)
\)
\]
\\
&=0.
\end{split}
\end{align}
Consequently, it costs no extra energy to move each vertex operator in a string 
up and down along each wire, as long as the solitons are sufficiently sharp.

\section{Discrete gauge symmetry and ground state in the limit 
of vanishing kinetic energy}
\label{appsec: Discrete gauge symmetry and ground state in the limit}

\setcounter{equation}{0}
\makeatletter
\renewcommand{\theequation}{B\arabic{equation}}

The goal of this Appendix is to make more explicit the connection
between the class of wire constructions considered in this paper and
well-known realizations of discrete lattice gauge theories, like the
toric code.  As in the previous Appendix, we will restrict ourselves
for the sake of concreteness to the case of the 
$\mathbb{Z}^{\,}_{m}$-topologically-ordered theories constructed in
Sec.\ \ref{subsec: Zm example}.  We further focus on the limit of
infinitesimal kinetic energy, which is discussed in
Sec.\ \ref{subsubsec: Energetics of pointlike and linelike defects}.

In this limit, the Hamiltonian of the coupled-wire theory 
(without disorder) is given by
\begin{subequations}
\label{Appendix B Hamiltonian}
\begin{align}
\tilde{\hat{H}}\approx 
\tilde{\hat{H}}^{\,}_{\{\tilde{\mathcal{T}}\}}\:=&\,
-
U^{\,}_{\mathrm{s}}
\int\limits^{L}_{0}
\mathrm{d}z\, 
\sum_{s}
\cos
\(
\tilde{\mathcal{T}}^{\T}_{s}\,
\tilde{\mathcal{K}}^{\,}_{m}\,
\tilde{\hat{\Phi}}(z)
\)
-
U^{\,}_{\mathrm{p}}
\int\limits^{L}_{0}
\mathrm{d}z\, 
\sum_{p}
\cos
\(
\tilde{\mathcal{T}}^{\T}_{p}\,
\tilde{\mathcal{K}}^{\,}_{m}\,
\tilde{\hat{\Phi}}(z)
\),
\label{Appendix B Hamiltonian a}
\end{align}
with
\begin{equation}
\tilde{\mathcal{T}}^{\T}_{s}\,
\tilde{\mathcal{K}}^{\,}_{m}\,
\tilde{\hat\Phi}(z)\:=
m
\[
\tilde{\hat{\phi}}^{\,}_{s^{\,}_{E},1}(z)
-
\tilde{\hat{\phi}}^{\,}_{s^{\,}_{W},1}(z)
+
\tilde{\hat{\phi}}^{\,}_{s^{\,}_{N},1}(z)
-
\tilde{\hat{\phi}}^{\,}_{s^{\,}_{S},1}(z)
\]
\label{Appendix B Hamiltonian b}
\end{equation}
and
\begin{equation}
\tilde{\mathcal{T}}^{\T}_{p}\,
\tilde{\mathcal{K}}^{\,}_{m}\,
\tilde{\hat\Phi}(z)\:=
m
\[\tilde{\hat{\phi}}^{\,}_{p^{\,}_{E},2}(z)
-
\tilde{\hat{\phi}}^{\,}_{p^{\,}_{W},2}(z)+
\tilde{\hat{\phi}}^{\,}_{p^{\,}_{S},2}(z)
-
\tilde{\hat{\phi}}^{\,}_{p^{\,}_{N},2}(z)
\].
\label{Appendix B Hamiltonian c}
\end{equation}
\end{subequations}

The theory defined in Eq.~\eqref{Appendix B Hamiltonian} 
possesses a set of discrete local symmetries, 
which we will call ``gauge symmetries."  
This set of gauge symmetries is generated by the unitary operators
\begin{subequations}
\label{eq: def generators Zm gauge trsf}
\begin{align}
\hat{A}^{\,}_{s}(z^{\,}_{s^{\,}_N},z^{\,}_{s^{\,}_S},z^{\,}_{s^{\,}_E},z^{\,}_{s^{\,}_W})\equiv
\hat{A}^{\,}_{s}
(\{z^{\,}_{s^{\,}_{C}}\})\:=
\exp
\Bigg(
\mathrm{i}
\[\tilde{\hat{\phi}}^{\,}_{s^{\,}_{E},1}(z^{\,}_{s^{\,}_{E}})
-
\tilde{\hat{\phi}}^{\,}_{s^{\,}_{W},1}(z^{\,}_{s^{\,}_{W}})
+
\tilde{\hat{\phi}}^{\,}_{s^{\,}_{N},1}(z^{\,}_{s^{\,}_{N}})
-
\tilde{\hat{\phi}}^{\,}_{s^{\,}_{S},1}(z^{\,}_{s^{\,}_{S}})
\]
\Bigg)
\label{eq: def generators Zm gauge trsf a}
\end{align}
and
\begin{align}
\hat{B}^{\,}_{p}(z^{\,}_{p^{\,}_N},z^{\,}_{p^{\,}_S},z^{\,}_{p^{\,}_E},z^{\,}_{p^{\,}_W})\equiv
\hat{B}^{\,}_{p}(\{z^{\,}_{p^{\,}_{C}}\})\:=
\exp
\Bigg(
\mathrm{i}
\[
\tilde{\hat{\phi}}^{\,}_{p^{\,}_{E},2}(z^{\,}_{p^{\,}_{E}})
-
\tilde{\hat{\phi}}^{\,}_{p^{\,}_{W},2}(z^{\,}_{p^{\,}_{W}})
+
\tilde{\hat{\phi}}^{\,}_{p^{\,}_{S},2}(z^{\,}_{p^{\,}_{S}})
-
\tilde{\hat{\phi}}^{\,}_{p^{\,}_{N},2}(z^{\,}_{p^{\,}_{N}})
\]
\Bigg).
\label{eq: def generators Zm gauge trsf b}
\end{align}
\end{subequations}
The operators $\hat{A}^{\,}_{s}$ and $\hat{B}^{\,}_{p}$ have the
physical interpretation of creating the smallest possible closed loop
of vertex operators containing the star $s$ or plaquette $p$.  When
$\{z^{\,}_{s^{\,}_{C}}\}=z$ for all cardinal directions 
$C=N,W,S,E$, 
the loop created by $\hat{A}^{\,}_{s}(\{z^{\,}_{s^{\,}_{C}}\})\equiv
\hat{A}^{\,}_{s}(z)$ is defined in a plane of constant $z$.  A
completely analogous statement is true of the operator
$\hat{B}^{\,}_{p}(\{z^{\,}_{p^{\,}_{C}}\})\equiv \hat{B}^{\,}_{p}(z)$
when $\{z^{\,}_{p^{\,}_{C}}\}=z$ for all $C=N,W,S,E$.  
Within this physical picture, the closed loop of vertex operators depicted in 
Fig.\ \ref{fig: deconfined star defects}(d) 
can be viewed as being created by the
product of all operators $\hat{B}^{\,}_{p}(z)$ for plaquettes $p$
contained within the perimeter of the loop.  When all $z$-points
within the set $\{z^{\,}_{s^{\,}_{C}}\}$ or $\{z^{\,}_{p^{\,}_{C}}\}$ are
different, the operators $\hat{A}^{\,}_{s}(\{z^{\,}_{s^{\,}_{C}}\})$ and
$\hat{B}^{\,}_{p}(\{z^{\,}_{p^{\,}_{C}}\})$ can be viewed as creating
``wavy" loops that are not confined to a single constant-$z$ plane.
By the arguments of Appendix A, such a ``wavy" loop is energetically
equivalent to a loop confined to a plane of constant $z$ in the limit
of vanishing kinetic energy.

To check that $\hat{A}^{\,}_{s}$ and $\hat{B}^{\,}_{p}$ are indeed
symmetries of the Hamiltonian \eqref{Appendix B Hamiltonian}, 
first note that
\begin{subequations}
\begin{align}
\hat{A}^{\,}_{s}(\{z^{\,}_{s^{\,}_{C}}\})\,\cos
\(
\tilde{\mathcal{T}}^{\T}_{s^{\prime}}\,
\tilde{\mathcal{K}}^{\,}_{m}\,
\tilde{\hat{\Phi}}^{\,}(z^{\prime})
\)\, \hat{A}^{\dag}_{s}(\{z^{\,}_{s^{\,}_{C}}\})=\cos
\(
\tilde{\mathcal{T}}^{\T}_{s^{\prime}}\,
\tilde{\mathcal{K}}^{\,}_{m}\,
\tilde{\hat{\Phi}}^{\,}(z^{\prime})
\)
\end{align}
and
\begin{align}
\hat{B}^{\,}_{p}(\{z^{\,}_{p^{\,}_{C}}\})\,\cos
\(
\tilde{\mathcal{T}}^{\T}_{p^{\prime}}\,
\tilde{\mathcal{K}}^{\,}_{m}\,
\tilde{\hat{\Phi}}^{\,}(z^{\prime})
\)\, \hat{B}^{\dag}_{p}(\{z^{\,}_{p^{\,}_{C}}\})=\cos
\(
\tilde{\mathcal{T}}^{\T}_{p^{\prime}}\,
\tilde{\mathcal{K}}^{\,}_{m}\,
\tilde{\hat{\Phi}}^{\,}(z^{\prime})
\)
\end{align}
\end{subequations}
can be seen to hold for all 
$s$, $s^{\prime}$, $p$, and $p^{\prime}$
if one observes that the bosonic fields entering
the right-hand sides of
Eqs.\ (\ref{eq: pinned fields if Zm topoorder a})
and
Eqs.\ (\ref{eq: pinned fields if Zm topoorder b})
are labeled exclusively by $\alpha=1$ (the charge 2 bosonic mode)
and $\alpha=2$ (the charge neutral bosonic mode),
respectively, whereas the $2\times2$ matrix $\tilde{K}^{\,}_{m}$ 
defined in Eq.\ \eqref{eq: Zm K tilde} is off diagonal,
i.e., any pair of bosonic fields carrying the same charge
from Eq.\ (\ref{eq: Zm Q tilde}) commute.
According to Eq.\ (\ref{eq: pinned fields if Zm topoorder a})
and
Eq.\ (\ref{eq: pinned fields if Zm topoorder b}), 
it is only when a star $s$ overlaps with a plaquette $p$
($s^{\,}_{N}=p^{\,}_{W}$ and $s^{\,}_{E}=p^{\,}_{S}$, say)
that either 
\begin{subequations}
\label{eq: dangerous terms for Zm gauge symmetry}
\begin{align}
\hat{A}^{\,}_{s}(\{z^{\,}_{s^{\,}_{C}}\})\,\cos
\(
\tilde{\mathcal{T}}^{\T}_{p}\,
\tilde{\mathcal{K}}^{\,}_{m}\,
\tilde{\hat{\Phi}}^{\,}(z^{\prime})
\)\, \hat{A}^{\dag}_{s}(\{z^{\,}_{s^{\,}_{C}}\})
\label{eq: dangerous terms for Zm gauge symmetry a}
\end{align}
or
\begin{align}
\hat{B}^{\,}_{p}(\{z^{\,}_{p^{\,}_{C}}\})\,\cos
\(
\tilde{\mathcal{T}}^{\T}_{s}\,
\tilde{\mathcal{K}}^{\,}_{m}\,
\tilde{\hat{\Phi}}^{\,}(z^{\prime})
\)\, \hat{B}^{\dag}_{p}(\{z^{\,}_{p^{\,}_{C}}\})
\label{eq: dangerous terms for Zm gauge symmetry b}
\end{align}
\end{subequations}
might transform nontrivially.

In order of increasing difficulty,
we shall assume first that
all $\{z^{\,}_{s^{\,}_{C}}\}$ and all $\{z^{\,}_{p^{\,}_{C}}\}$, respectively, 
lie in the same constant-$z$ plane and show that both
Eqs.\ (\ref{eq: dangerous terms for Zm gauge symmetry a})
and
(\ref{eq: dangerous terms for Zm gauge symmetry b})
transform trivially. To this end, we combine
Eqs.\ (\ref{eq: step 1 to define Zm}) 
and 
(\ref{eq: Zm K and Q tilde}) 
into the identities
\begin{subequations}
\label{eq: identities needed for proof Zm invariance}
\begin{align}
e^{
+\mathrm{i}\,
\tilde{\hat{\phi}}^{\,}_{j,1}(z)
  }\, 
\tilde{\hat{\phi}}^{\,}_{\pri j,2}(z^{\prime})\, 
e^{
-\mathrm{i}\, 
\tilde{\hat{\phi}}^{\,}_{j,1}(z)
  }=&\,
\tilde{\hat{\phi}}^{\,}_{\pri j,2}(z^{\prime})
+
\mathrm{i} 
\[
\tilde{\hat{\phi}}^{\,}_{j,1}(z),
\tilde{\hat{\phi}}^{\,}_{\pri j,2}(z^{\prime})
\]
\nonumber\\
=&\,
\tilde{\hat{\phi}}^{\,}_{\pri j,2}(z^{\prime})
-
\delta^{\,}_{j,j'}\,
\frac{2\pi}{m}\, 
\Theta(z-\pri z)
\label{eq: identities needed for proof Zm invariance a}
\end{align}
and 
\begin{align}
e^{
+\mathrm{i}\, 
\tilde{\hat{\phi}}^{\,}_{j,2}(z)}\, 
\tilde{\hat{\phi}
  }^{\,}_{\pri j,1}(z^{\prime})\, 
e^{
-\mathrm{i}\, 
\tilde{\hat{\phi}}^{\,}_{j,2}(z)
  }=&\,
\tilde{\hat{\phi}}^{\,}_{\pri j,1}(z^{\prime})
+
\mathrm{i} 
\[
\tilde{\hat{\phi}}^{\,}_{j,2}(z),
\tilde{\hat{\phi}}^{\,}_{\pri j,1}(z^{\prime})
\]
\nonumber\\
=&\,
\tilde{\hat{\phi}}^{\,}_{\pri j,1}(z^{\prime})
-
\delta^{\,}_{j,j'}\,
\frac{2\pi}{m}\, 
\Theta(z-\pri z).
\label{eq: identities needed for proof Zm invariance b}
\end{align}
\end{subequations}

When all $\{z^{\,}_{s^{\,}_{C}}\}$ 
lie in the same constant-$z$ plane
\begin{subequations}
\begin{equation}
\hat{A}^{\,}_{s}(\{z^{\,}_{s^{\,}_{C}}\})\,
\[
\tilde{\mathcal{T}}^{\T}_{p}\,
\tilde{\mathcal{K}}^{\,}_{m}\,
\tilde{\hat{\Phi}}^{\,}(z^{\prime})
\]\, \hat{A}^{\dag}_{s}(\{z^{\,}_{s^{\,}_{C}}\})\equiv
\hat{A}^{\,}_{s}(z)
\[
\tilde{\mathcal{T}}^{\T}_{p}\,
\tilde{\mathcal{K}}^{\,}_{m}\,
\tilde{\hat{\Phi}}^{\,}(z^{\prime})
\] 
\hat{A}^{\dag}_{s}(z)
\end{equation}
becomes
\begin{equation}
m
\[
\tilde{\hat{\phi}}^{\,}_{p^{\,}_{E},2}(z^{\prime})
-
e^{
\mathrm{i}\,\tilde{\hat{\phi}}^{\,}_{s^{\,}_{N},1}(z)
  }\, 
\tilde{\hat{\phi}}^{\,}_{p^{\,}_{W},2}(z^{\prime})\, 
e^{
-\mathrm{i}\,\tilde{\hat{\phi}}^{\,}_{s^{\,}_{N},1}(z)
  }
+
e^{
\mathrm{i}\,\tilde{\hat{\phi}}^{\,}_{s^{\,}_{E},1}(z)
  }\,
\tilde{\hat{\phi}}^{\,}_{p^{\,}_{S},2}(z^{\prime})\, 
e^{
-\mathrm{i}\, \tilde{\hat{\phi}}^{\,}_{s^{\,}_{E},1}(z)
  }
-
\tilde{\hat{\phi}}^{\,}_{p^{\,}_{N},2}(z^{\prime})
\].
\end{equation}
From Eqs.\ (\ref{eq: step 1 to define Zm})
and (\ref{eq: identities needed for proof Zm invariance a}),
we then conclude that
\begin{equation}
\begin{split}
\hat{A}^{\,}_{s}(z)
\[
\tilde{\mathcal{T}}^{\T}_{p}\,
\tilde{\mathcal{K}}^{\,}_{m}\,
\tilde{\hat{\Phi}}^{\,}(z^{\prime})
\] 
\hat{A}^{\dag}_{s}(z)
=&\,
m
\[
\tilde{\hat{\phi}}^{\,}_{p^{\,}_{E},2}(z^{\prime})
-
\tilde{\hat{\phi}}^{\,}_{p^{\,}_{W},2}(z^{\prime})
+
\tilde{\hat{\phi}}^{\,}_{p^{\,}_{S},2}(z^{\prime})
-
\tilde{\hat{\phi}}^{\,}_{p^{\,}_{N},2}(z^{\prime})
\]
\\
=&\,
\tilde{\mathcal{T}}^{\T}_{p}\,
\tilde{\mathcal{K}}^{\,}_{m}\,
\tilde{\hat{\Phi}}^{\,}(z^{\prime})
\end{split}
\end{equation}
\end{subequations}
transforms trivially. Similarly,
when all $\{z^{\,}_{p^{\,}_{C}}\}$ 
lie in the same constant-$z$ plane,
\begin{subequations}
\begin{align}
\hat{B}^{\,}_{p}(\{z^{\,}_{p^{\,}_{C}}\})\,
\[
\tilde{\mathcal{T}}^{\T}_{s}\,
\tilde{\mathcal{K}}^{\,}_{m}\,
\tilde{\hat{\Phi}}^{\,}(z^{\prime})
\]\, \hat{B}^{\dag}_{p}(\{z^{\,}_{p^{\,}_{C}}\})\equiv
\hat{B}^{\,}_{p}(z)
\[\tilde{\mathcal{T}}^{\T}_{s}\,
\tilde{\mathcal{K}}^{\,}_{m}\,
\tilde{\hat{\Phi}}^{\,}(z^{\prime})\] \hat{B}^{\dag}_{p}(z)
\end{align}
becomes
\begin{equation}
m
\[
e^{
+\mathrm{i}\,\tilde{\hat{\phi}}^{\,}_{p^{\,}_{S},2}(z)
  }\,
\tilde{\hat{\phi}}^{\,}_{s^{\,}_{E},1}(z^{\prime})\, 
e^{
-\mathrm{i}\,\tilde{\hat{\phi}}^{\,}_{p^{\,}_{S},2}(z)
  }
-
\tilde{\hat{\phi}}^{\,}_{s^{\,}_{W},1}(z^{\prime})
+
e^{
-\mathrm{i}\,\tilde{\hat{\phi}}^{\,}_{p^{\,}_{W},2}(z)
  }\, 
\tilde{\hat{\phi}}^{\,}_{s^{\,}_{N},1}(z^{\prime})\, 
e^{
+\mathrm{i}\,\tilde{\hat{\phi}}^{\,}_{p^{\,}_{W},2}(z)
  }
-
\tilde{\hat{\phi}}^{\,}_{s^{\,}_{S},1}(z^{\prime})
\].
\end{equation}
From Eqs.\ (\ref{eq: step 1 to define Zm})
and (\ref{eq: identities needed for proof Zm invariance b}),
we then conclude that
\begin{align}
\hat{B}^{\,}_{p}(z)
\[\tilde{\mathcal{T}}^{\T}_{s}\,
\tilde{\mathcal{K}}^{\,}_{m}\,
\tilde{\hat{\Phi}}^{\,}(z^{\prime})\] \hat{B}^{\dag}_{p}(z)=&\,
m
\[
\tilde{\hat{\phi}}^{\,}_{s^{\,}_{E},1}(z^{\prime})
-
\tilde{\hat{\phi}}^{\,}_{s^{\,}_{W},1}(z^{\prime})
+
\tilde{\hat{\phi}}^{\,}_{s^{\,}_{N},1}(z^{\prime})
-
\tilde{\hat{\phi}}^{\,}_{s^{\,}_{S},1}(z^{\prime})\]
\nonumber\\
=&\,
\tilde{\mathcal{T}}^{\T}_{p}\,
\tilde{\mathcal{K}}^{\,}_{m}\,
\tilde{\hat{\Phi}}^{\,}(z^{\prime})
\end{align}
\end{subequations}
also transforms trivially.

Repeating the above calculation for the case of general
$\{z^{\,}_{s^{\,}_{C}}\}$ and $\{z^{\,}_{p^{\,}_{C}}\}$, one finds
that the pinned fields $\tilde{\mathcal{T}}^{\T}_{p}\,
\tilde{\mathcal{K}}^{\,}_{m}\,\tilde{\hat{\Phi}}^{\,}(z^{\prime})$
and $\tilde{\mathcal{T}}^{\T}_{s}\,\tilde{\mathcal{K}}^{\,}_{m}\,
\tilde{\hat{\Phi}}^{\,}(z^{\prime})$ are not strictly invariant under
the action of $\hat{A}^{\,}_{s}$ and $\hat{B}^{\,}_{p}$, but that they
change by a difference of soliton profiles as in Eq.\
\eqref{difference of solitons}.  In Appendix A, 
it was shown that such a change in the
pinned fields does not affect the total energy in the limit of
vanishing kinetic energy, and we thus conclude that $\hat{A}^{\,}_{s}$
and $\hat{B}^{\,}_{p}$ commute with the Hamiltonian 
\eqref{Appendix B Hamiltonian} in this limit, 
for any pair $\{z^{\,}_{s^{\,}_{C}}\}$ and $\{z^{\,}_{p^{\,}_{C}}\}$.  

Having established that the Hamiltonian defined by Eqs.\
(\ref{Appendix B Hamiltonian}) 
displays a local $\mathbb{Z}^{\,}_{m}$ gauge symmetry,
we want to study its ground states. According to 
Elitzur's theorem~\cite{Elitzur75}, 
each ground state must be invariant
under the local $\mathbb{Z}^{\,}_{m}$ gauge symmetry.
Our goal is to verify this consequence of
Elitzur's theorem explicitly. To this end,
we observe that any argument
(\ref{Appendix B Hamiltonian b})
or
(\ref{Appendix B Hamiltonian c})
that appears in a cosine term from the Hamiltonian
(\ref{Appendix B Hamiltonian a})
is related to the generators 
(\ref{eq: def generators Zm gauge trsf a})
or
(\ref{eq: def generators Zm gauge trsf b})
of the local $\mathbb{Z}^{\,}_{m}$ symmetry through
\begin{subequations}
\begin{align}
\cos
\(
\tilde{\mathcal{T}}^{\T}_{s}\,
\tilde{\mathcal{K}}^{\,}_{m}\,
\tilde{\hat{\Phi}}^{\,}(z)
\)= 
\frac{
\left[\hat{A}^{\,}_{s}(z)\right]^{m}
+
\left[\hat{A}^{\dag}_{s}(z)\right]^{m}
     }
     {
2
     }
\end{align}
or
\begin{align}
\cos
\(
\tilde{\mathcal{T}}^{\T}_{p}\,
\tilde{\mathcal{K}}^{\,}_{m}\,
\tilde{\hat{\Phi}}^{\,}(z)
\) = 
\frac{
\left[\hat{B}^{\,}_{p}(z)\right]^{m}
+
\left[\hat{B}^{\dag}_{p}(z)\right]^{m}
     }
     {
2 
     },
\end{align}
\end{subequations}
respectively.

We shall ignore the issue of topological degeneracy 
by assuming a unique ground state when
the lattice of wires spans a manifold of vanishing genus,
or by restricting to one topological sector of the theory
defined on the torus.  We demand that the ground state 
satisfy two properties.  

First, in order to be a ground state, a state must
consist of a superposition of field configurations that minimize both
cosine terms in Eq.\ \eqref{Appendix B Hamiltonian} simultaneously.
(This is possible because these two sets of terms are Haldane
compatible, and because we work in the strong-coupling limit
$U^{\,}_{\rm s},U^{\,}_{\rm p}\to\infty$.)  Since the 
charge-2 bosonic fields do not commute with
the charge-neutral bosonic fields,
we can use either set of bosonic
fields to label the full set of classical field configurations
minimizing both sets of cosine terms. We denote
the eigenfunctionals of the fields
$\tilde{\hat{\phi}}^{\,}_{j,1}(z)$
by 
$\ket{\{\tilde{\phi}^{\,}_{j,1}(z)\}}$,
i.e.,
\begin{equation}
\tilde{\hat{\phi}}^{\,}_{j,1}(z)\,
\ket{\{\tilde{\phi}^{\,}_{j,1}(z)\}}=
\tilde{\phi}^{\,}_{j,1}(z)\,
\ket{\{\tilde{\phi}^{\,}_{j,1}(z)\}}.
\end{equation}
Among all these eigenfunctionals, we select the eigenfunctionals
for which
\begin{equation}
\cos
\Big(
\tilde{\mathcal{T}}^{\T}_{s}\,
\tilde{\mathcal{K}}^{\,}_{m}\,
\tilde{\Phi}^{\,}(z) 
\Big)=1
\end{equation}
holds for all $s$.
We now define (up to normalization) a ``reference state"
\begin{align}
\ket{\varphi}\:= 
\int\limits^{L}_{0}\mathrm{d}\pri z\,
\sum_{\{n^{\,}_{j}\in\,\mathbb{Z}\}}
\Ket{
\left\{
\tilde{\phi}^{\,}_{j,1}(z)
+
2\pi\,n^{\,}_{j}\,\Theta(z-\pri z)
\right\}
    },
\end{align}
where $j=1,\dots,2N$. The sum 
over the integers  $n^{\,}_{j}\in\mathbb{Z}$
accounts for the fact that classical field configurations differing
from one another by a soliton with an integer charge are equivalent
from the point of view of the cosines. According to
Eq.\
\eqref{eq: identities needed for proof Zm invariance b}, 
we must have
\begin{subequations}\label{Bp action}
\begin{align}
&
\left[\hat{B}^{\,}_{p}(\{z^{\,}_{p^{\,}_{C}}\})\right]^{n}\, 
\hat{\tilde\phi}^{\,}_{j,1}(z^{\prime})\,
\left[\hat{B}^{\dag}_{p}(\{z^{\,}_{p^{\,}_{C}}\})\right]^{n}=
\hat{\tilde\phi}^{\,}_{j,1}(z^{\prime})
-
\delta^{\,}_{j\in p}\, 
\frac{2\pi\,n}{m}\, 
\Theta(z^{\,}_{j}-\pri z),
\\
&
\left[\hat{B}^{\dag}_{p}(\{z^{\,}_{p^{\,}_{C}}\})\right]^{n}\, 
\hat{\tilde\phi}^{\,}_{j,1}(z^{\prime})\,
\left[\hat{B}^{\,}_{p}(\{z^{\,}_{p^{\,}_{C}}\})\right]^{n}=
\hat{\tilde\phi}^{\,}_{j,1}(z^{\prime})
+
\delta^{\,}_{j\in p}\, 
 \frac{2\pi\,n}{m}\, 
\Theta(z^{\,}_{j}-\pri z),
\end{align}
\end{subequations}
where $\delta^{\,}_{j\in p}$ is a function that returns
1 if $j\in p$ and 0 otherwise, for any $n\in\mathbb{Z}$.  
Thus, one concludes that
\begin{align}
\cos
\(
\tilde{\mathcal{T}}^{\T}_{p}\,
\tilde{\mathcal{K}}^{\,}_{m}\,
\tilde{\hat{\Phi}}^{\,}(z)
\)\ket{\varphi} = 
\left\{
\frac{
\left[\hat{B}^{\,}_{p}(z)\right]^{m} 
+
\left[\hat{B}^{\dag}_{p}(z)\right]^{m} 
     }
     {
2
     }
\right\}
\ket{\varphi}=
\ket{\varphi}.
\end{align}
Therefore, the state $\ket{\varphi}$ minimizes both cosine terms simultaneously.

The second constraint to be imposed on the ground state of
Hamiltonian (\ref{Appendix B Hamiltonian}) is that it is also
invariant under the symmetry group generated by the operators
$\hat{B}^{\,}_{p}(\{z^{\,}_{p^{\,}_{C}}\})$. The state
$\ket{\varphi}$ is not up to the task, as Eq.\ \eqref{Bp action}
demonstrates that $\ket{\varphi}$ is not invariant under applications
of the operator 
$\left[\hat{B}^{\,}_{p}(\{z^{\,}_{p^{\,}_{C}}\})\right]^{n}$
for $1\leq n<m$.  For a single
plaquette $p$ at fixed $\{z^{\,}_{p^{\,}_{C}}\}$, 
however, one can check using Eq.~\eqref{Bp action} and the fact that
$\left[\hat{B}^{\,}_{p}(\{z^{\,}_{p^{\,}_{C}}\})\right]^{m}\,
\ket{\varphi}=\ket{\varphi}$
that the state
\begin{align}
\left\{
1
+
\sum^{m-1}_{n=1}
\left[
\hat{B}^{\,}_{p}(\{z^{\,}_{p^{\,}_{C}}\})
\right]^{n}
\right\}
\ket{\varphi}
\label{eq: intermediary construction gauge invariant GS}
\end{align}
is.  We must therefore extend the construction
(\ref{eq: intermediary construction gauge invariant GS})
to all plaquettes $p$ and all
points $\{z^{\,}_{p^{\,}_{C}}\}$ along the wires.  This is
accomplished by the (unnormalized) state
\begin{align}
\ket{\rm GS}\:=
\exp
\left(
\sum_{p}
\int\limits_{0}^{L} 
\mathrm{d}\{z^{\,}_{p^{\,}_{C}}\}\, 
\log
\left(
1
+
\sum^{m-1}_{n=1}
\left[\hat{B}^{\,}_{p}(\{z^{\,}_{p^{\,}_{C}}\})\right]^{n}
\right)
\right)
\ket{\varphi}.
\end{align}
The operator $\exp(\dots)$ 
on the right-hand side
applies all possible products of
operators 
$\left[\hat{B}^{\,}_{p}(\{z^{\,}_{p^{\,}_{C}}\})\right]^{n}$
over all plaquettes $p$ and all points $\{z^{\,}_{p^{\,}_{C}}\}$.

The ground state $\ket{\rm GS}$ is a phase-coherent and
equal-amplitude superposition of all possible configurations of 
closed loops of vertex operators.  These closed loops can be like the one depicted in
Fig.~\ref{fig: deconfined star defects}(d), i.e.,
restricted to a single plane of constant $z$, 
or more general configurations involving
``wavy" closed loops.  Thus, the ground state $\ket{\rm GS}$ can be
viewed as a quasi-two-dimensional
``soup of loops" in which the closed loops are
further allowed to fluctuate in the $z$-direction.  This is a direct
generalization of the ground state of Kitaev's toric code
\cite{Kitaev03} to the context of the coupled-wire systems considered
in this work.  Furthermore, it explicitly demonstrates in the limit of
vanishing kinetic energy that the string and membrane operators built
using the vertex operators defined in Eqs.\
\eqref{eq: 2D Zm defect hopping} 
and the bilocal operators defined in Eqs.\
\eqref{eq: def 2D Zm z-string defs}, 
respectively, do not lead to an observable change
in the ground states except at their ends and edges.

\medskip
\end{widetext}

\bibliographystyle{apsrev}

\bibliography{bib_3d_wire}

\begin{thebibliography}{71}
\bibitem[{foo()}]{footnote_kappa_tilde}
\bibinfo{howpublished}{We henceforth assume that the vectors
  $\tilde{v}^{(\texttt{j})}$ and $\tilde{w}^{(\texttt{j})}$ can be chosen in
  such a way that the matrix $\varkappa^{-1}$ defined in Eq.~\eqref{kappa def}
  has nonzero determinant, so that $\varkappa$ is guaranteed to exist.}
\end{thebibliography}

\end{document}